%% file: main.tex
\documentclass[%
pof,
 amsmath,amssymb,
preprint,%
]{revtex4-1}

\usepackage{graphicx}
\usepackage{caption}
\usepackage{comment}
\usepackage{enumitem}
\usepackage{subcaption}
\usepackage{dcolumn}
\usepackage{bm}

\usepackage[utf8]{inputenc}
\usepackage[T1]{fontenc}
\usepackage{mathptmx}
\usepackage{etoolbox}
\usepackage{xcolor}
\usepackage[colorinlistoftodos]{todonotes}
\usepackage{soul}
\usepackage{multirow}
\usepackage{etoolbox}
\usepackage{pifont}
\usepackage{booktabs}
\usepackage{float}
\usepackage{threeparttable}

\usepackage{natbib}
\setcitestyle{authoryear,open={(},close={)},citesep={;},aysep={,},yysep={,}}
\begin{document}

\title{Self-Tuning Dynamic Explicit Modal Filtering Based on Local Flow Characteristics for Large-Eddy Simulation}

\author{Mohammadmahdi Ranjbar}
\author{Ali Mostafavi}
\author{Farzad Mashayek}
\email[Corresponding author: ]{mashayek@arizona.edu}
\affiliation{Department of Aerospace and Mechanical Engineering, University of Arizona, Tucson, Arizona 85719, USA}
\renewcommand{\abstractname}{}
\begin{abstract}
This work improves upon our previously introduced explicit dynamic modal filter (DEMF) within the framework of the discontinuous Galerkin spectral element method (DGSEM) by introducing a mechanism for self-tuning of the model parameters. The new self-tuning dynamic explicit modal filter (STDEMF) also extends the methodology for obtaining modal values from nodal values beyond Chebyshev grids and polynomials to general collocation points and orthogonal polynomial bases by leveraging orthogonality. The generated modes are used to remove the built-up energy due to unresolved sub-grid scales (SGS) in large-eddy simulation (LES) of turbulent flows. The STDEMF improves the performance of DEMF in two ways. First, the filter kernel applied to the modes is adapted from a cut-off kernel to a hyperbolic tangent shape, which automatically adjusts the model for different polynomial orders. Second, the cut-off mode is computed dynamically for each element as a function of local flow characteristics, including the local Kolmogorov length scale and the second invariants of the strain and rotation rate tensors. The suggested formulation for the cut-off mode treats unresolved elements distinctly and improves performance by avoiding under- or over-dissipation. Moreover, the cut-off mode evolves over time within the same element as turbulent characteristics vary. The model is evaluated on three flows, homogeneous isotropic decaying, the Taylor-Green vortex, and periodic channel flow, each with distinct turbulent characteristics. Comparisons of the results show that the STDEMF model outperforms the DEMF model and the Smagorinsky eddy viscosity model.
\end{abstract}
\maketitle
\input{include/introduction.tex}

\input{include/methodology}
\input{include/results.tex}
\input{include/conclusion.tex}

\input{include/acknowledgements.tex}
\input{include/authordeclaration.tex}
\input{include/dataavailability.tex}

\bibliography{cfd}
\end{document}

%% file: include/introduction.tex
\section{\label{sec:intro}Introduction}

Among the numerical schemes used to solve multi-dimensional nonlinear conservation laws, including the Navier-Stokes (NS) equations, the high-order discontinuous Galerkin (DG) method has gained increasing popularity in recent years. The DG method has been implemented to simulate various flows in complex geometries, including incompressible \citep{Liu_2024, Zhang_2022, Shahbazi_2007, Ferrer_2011}, compressible \citep{Li_2021,Li_2021b,Zhang_2023,Ghiasi_2018,Qi_2024}, and multiphase \citep{Gao_2018, Pochet_2013, Orlando_2024} flows. This growing popularity of the method can be attributed to two key factors. First, it combines the geometric flexibility of finite elements \citep{HONG2025,Mostafavi_2024, Mostafavi_2025,XI2025,MOSTAFAVI2025sh,mostafavi2024mass} with the high accuracy of spectral methods \citep{SPALART1991,Okamoto_2025}, allowing high accuracy even on unstructured and distorted meshes \citep{Kopriva_1998}. Second, it offers improved robustness compared to the continuous Galerkin counterpart by incorporating additional numerical dissipation through upwind Riemann solvers \citep{Kirby_2003}.

Although the dissipation introduced by the upwind Riemann solver increases the robustness of the method compared to the continuous Galerkin version, it remains insufficient for 
under-resolved turbulent flow simulations. The spectral vanishing viscosity (SVV) is amidst the methods used to introduce numerical dissipation in DG methods. SVV functions by adding an artificial viscous flux combined with a high-pass modal filter. It was initially introduced by \cite{Tadmor_1989} for Fourier and continuous Galerkin discretization to regularize the solution and enhance the convergence of the one-dimensional (1D) nonlinear conservation laws, such as the inviscid Burger equations. \cite{Chen_1993} later extended the application of SVV to multidimensional scalar conservation laws with periodic boundary conditions using the Fourier spectral method. Subsequent studies further expanded the scope of SVV from Fourier-based implementations to non-periodic cases by employing Chebyshev and Legendre Polynomials \citep{Andreassen_1994, Kaber_1996}. \cite{Andreassen_1994} employed SVV with Chebyshev polynomials to simulate an atmospheric gravity wave governed by the nonlinear, inviscid, two-dimensional (2D) Euler equations with gravity. In the absence of a dissipation mechanism, nonlinear energy transfer to higher frequencies caused the breakdown of long-time simulations, making SVV essential. \cite{Kaber_1996} employed Legandre-based SVV to stabilize non-linear conservation laws and tested the method on 1D and 2D Euler equations. Additionally, spectral vanishing super/hyper-viscosity methods were developed for both periodic and non-periodic cases, using higher-order derivatives in the flux formulation rather than the second-order form employed in SVV \citep{Ma_1998, Tadmor_1989, Guermond_2003}.

\cite{Karamanos_2000} implemented SVV in the NS equations to dissipate small-scale turbulent energy in under-resolved cases and enable large-eddy simulations (LES). The artificial viscous flux followed the formulation of \cite{Tadmor_1989}, and their results showed that SVV outperformed eddy-viscosity spectral LES in turbulent channel flow.

Over the past two decades, researchers have continuously implemented and refined SVV for LES. \cite{Kirby_2002} introduced a modified version of the SVV approach by \cite{Karamanos_2000} for the continuous Galerkin method by adaptively computing the viscosity amplitude based on the local strain. \cite{PASQUETTI_2005} conducted a sensitivity study of SVV parameters for LES of a cylinder wake, showing that larger cut-off modes and smaller artificial viscosity improved performance. Extending this work, \cite{Pasquetti_2006} integrated SVV into his Defiltering–Transport–Filtering algorithm within a semi-Lagrangian LES framework, successfully simulating the confined cylinder wake. Later, \cite{Pasquetti_2007} applied the SVV–LES approach to thermally stratified cylinder wakes and rotor–stator flows, achieving strong agreement with DNS and experiments. \cite{Kirby_2006} developed a positive semi-definite formulation of SVV, well-suited for stabilizing spectral/hp element methods in hybrid-shaped domains for incompressible NS equations. \cite{Severac_2007} applied SVV to LES of turbulent rotating-cavity flows by incorporating it into the cylindrical NS equations using a Chebyshev–Fourier pseudo-spectral method, achieving good agreement with DNS and experiments. \cite{Koal_2012} extended the SVV formulation in cylindrical coordinates to include the axis region using a spectral element–Fourier discretization and introduced two modified kernels to enhance near-axis stability. \cite{Moura_2016} developed an adaptive SVV approach that dynamically scales with local advection speed and mesh spacing and incorporated a power kernel to further improve accuracy and stability. \cite{Chen_2022} developed an SVV stabilization for non-standard spectral element methods with triangular and mixed meshes. Their tests showed that while SVV reduces accuracy for linear elliptic equations, it significantly enhances stability and accuracy for high Reynolds number incompressible flows, where standard spectral methods fail. 

The SVV framework can also complement explicit subgrid-scale modeling by replacing the artificial flux in its formulation with a turbulence eddy-viscosity model and applying a high-pass modal filter. \cite{Manzanero_2020} introduced a Smagorinsky-based SVV model for high-order DG methods by modifying the artificial flux formulation of classical SVV. Their approach applies a high-pass modal filter to the Smagorinsky LES model and demonstrates that, with a suitable filter kernel, SVV enhances stability, preserves spectral convergence, and effectively controls dissipation in turbulent regions while vanishing in the laminar limit. This Smagorinsky-SVV LES model was later extended to an entropy-stable formulation for DG methods by \cite{Mateo_Gab_n_2022}.

The common feature of SVV studies discussed above is the use of an artificial viscous flux combined with a high-pass modal filter to dissipate energy at high-wavenumber motions. However, adding an extra flux term increases the mathematical complexity and alters the structure of the governing equations, necessitating separate analyses, such as Von Neumann stability or dispersion/dissipation studies, to evaluate the scheme’s numerical behavior \citep{Manzanero_2020, Moura_2015, Moura_2016}. Another, more straightforward alternative to SVV for stabilizing DG methods and enabling LES is to introduce a mechanism that directly filters solution variables to remove energy from the large resolved scales without relying on artificial viscosity. Turbulent flows exhibit a multi-scale nature where energy is transferred from large to small scales and dissipated through viscosity \citep{Tikhomirov_1991,Bozorgpour2024}. A hierarchical expansion of flow variables using orthogonal functions allows the extraction of spectral information. In addition to the nodal representation in DG spectral element method, a modal expansion can be used, where the coefficients, referred to as modes, capture turbulence characteristics and the associated scales \citep{Ranjbar_2023, Ranjbar_2024POF}.

The studies by \cite{Ranjbar_2024AIAA, Ranjbar_2023, Ranjbar_2024POF, Ranjbar2024APS} showed that high-wavenumber modes correspond to small-scale motions, while low-wavenumber modes represent large-scale structures. They also demonstrated that energy transfer from large to small scales manifests in the variations of mode magnitudes \citep{Ranjbar_2024POF}. Furthermore, their findings revealed that energy accumulation in turbulent flows is reflected in increased magnitudes of high-wavenumber modes \citep{Ranjbar_2024POF}. By applying a low-pass modal filter to dissipate energy from high-wavenumber modes, the accumulation of energy in under-resolved turbulent flows was prevented and the solution was stabilized. This demonstrated that modal filtering mimics the effects of SVV, serving as an alternative or complement to subgrid-scale modeling in LES of turbulent flows. To distinguish laminar and well-resolved turbulent regions from under-resolved areas, they developed a sensor integrated with the modal filtering approach \citep{Ranjbar_2024POF}. The sensor assesses turbulence resolvedness by comparing the local Kolmogorov length scale and average grid spacing.

The dynamic explicit modal filter (DEMF) method of \cite{Ranjbar_2024POF} employs Chebyshev polynomials for the modal expansion of the solution. Since their spectral element method uses a Chebyshev grid, the transformation to obtain the modes, the discrete Chebyshev transform, simplifies to the discrete cosine transform. This enables the use of the discrete cosine transform from the Fastest Fourier Transform in the West (FFTW) library for efficient computations of the modes \citep{Frigo_2005}.

In this study, we extend the use of orthogonal expansions by exploiting their orthogonality to extract modal coefficients, enabling the DEMF approach to be applied to any set of grid and orthogonal polynomials. While Legendre polynomials are employed here, the method is readily adaptable to other polynomial bases. Furthermore, the modal filtering approach of \cite{Ranjbar_2024POF} is refined by incorporating a low-pass hyperbolic tangent filter kernel, which self-tunes most of the previous empirical parameters to determine the amount of energy removal in unresolved regions through the cut-off mode parameter. The cut-off mode is proportional to the polynomial order within each element and is evaluated dynamically for each element based on local flow characteristics, including shear, rotation, and the degree of turbulence unresolvedness. The paper is structured as follows to introduce and assess the new self-tuning dynamic explicit modal filter (STDEMF) method. The next section introduces the governing equations and numerical scheme. Next, we describe the transformation between nodal and modal spaces and detail the modal filtering approach. The results section evaluates the proposed method through three test cases, followed by concluding remarks summarizing the key findings.

%% file: include/methodology.tex
\section{\label{sec:method}Formulation and Methodology}

This section presents the governing equations, the numerical scheme used for the discretizations, the methodology for obtaining modal coefficients, and the filtering approach.

\subsection{Governing Equations and the Numerical Solution Approximation}{\label{subsection:method}}

The non-dimensional 3D compressible NS equations in conservative form
\begin{equation}
 \label{eq:NS_Conservative}
\frac{\partial{\vec{Q}}}{\partial{t}}+ \frac{\partial{\vec{F_j^a}}}{\partial{x_j}}=
\frac{1}{\mathrm{Re}_{c}}\frac{\partial{\vec{F_j^v}}}{\partial{x_j}}  ,
\end{equation}
where
\begin{equation}
\vec{Q}=
\begin{bmatrix}
    \rho, \ \
    \rho v_1, \ \
    \rho v_2, \ \
    \rho v_3, \ \
    \rho e
\end{bmatrix}^T
\end{equation}
is the solution vector, and
\begin{equation}
\vec{F_j^a}=
\begin{bmatrix}
    \rho v_j, \ \
    \rho v_1v_j+P\delta_{1j}, \ \
    \rho v_2v_j+P\delta_{2j}, \ \
    \rho v_3v_j+P\delta_{3j}, \ \
    v_j\left(\rho e+P\right)
\end{bmatrix}^T
\end{equation}
are inviscid fluxes, and
\begin{equation}
\vec{F_j^v}=
\begin{bmatrix}
    0, \ \
    \tau_{1j}, \ \
    \tau_{2j}, \ \
    \tau_{3j}, \ \
    \sum_{k=1}^3 v_k \tau_{jk} + \frac{1}{\left(\gamma-1\right)\mathrm{Ma}_{c}^{2}\mathrm{Pr}_{c}} \frac{\partial T}{\partial x_j}
\end{bmatrix}^T
\end{equation}
are viscous fluxes, are spatially discretized using the discontinuous Galerkin spectral element method (DGSEM), a nodal version of DG schemes \citep{Kopriva_2009}. Here,  $\rho$  denotes the dimensionless density, while $v_1$, $v_2$, and $v_3$ represent the dimensionless velocity components along the $x$, $y$, and $z$ directions, respectively, and $\delta_{ij}$ is the Kronecker delta function. $P$ is the dimensionless pressure, $T$ denotes the dimensionless temperature, and $\gamma$ represents the ratio of specific heats. The quantity $e$ represents the dimensionless internal energy and is computed as
\begin{equation}
\rho e = \frac{P}{\gamma-1} + \frac{\rho v_iv_i}{2}.
\end{equation}
The dimensionless shear stress components acting on various planes in different directions are denoted by $\tau$ and are defined as 
\begin{equation}
\tau_{ij} = \left( \frac{\partial v_j}{\partial x_{i}} + \frac{\partial v_i}{\partial x_j} \right) - \frac{2}{3} \left( \nabla \cdot \vec{v} \right) \delta_{ij},
\end{equation}
where $\vec{v}$ is the velocity vector. To close the system of the governing equations, the non-dimensional equation of state \citep{Jacobs_2003}
\begin{equation}
P=\frac{\rho T }{\gamma \mathrm{Ma}_c}
\end{equation} is used to solve for the pressure.

The dimensionless numbers emerging from the non-dimensionalization of the NS equations include the reference Reynolds number (\(\mathrm{Re}_c\)), Prandtl number (\(\mathrm{Pr}_c\)), and Mach number (\(\mathrm{Ma}_c\)), defined as \citep{Jacobs_2003}
\begin{equation}
    \mathrm{Re}_c=\frac{U_c^* L_c^* \rho_c^*}{\mu_c^*}, \quad \mathrm{Pr}_c = \frac{c_p^* \mu_c^*}{\kappa^*}, \quad \mathrm{Ma}_c=\frac{U_c^*}{\sqrt{\gamma R^* T_c^*}},
\end{equation}
where $U_c^*$, $L_c^*$, $\rho_c^*$, $\mu_c^*$, and $T_c^*$ represent the reference dimensional values for velocity, length, density, dynamic viscosity, and temperature, respectively. In addition, $c_p^*$, $\kappa^*$, and $R^*$ are the dimensional specific heat at constant pressure, thermal conductivity, and gas constant, respectively.

The DGSEM implemented for the spatial discretization of NS equations employs non-overlapping curvilinear hexahedral elements to discretize the physical domain. A polynomial trasnfinite mapping method is implemented to transform physical elements, ${\cal E}$, onto a standard reference element, ${\cal E}_{f} = [-1,1]^3$. This mapping  establishes the relation between the physical coordinates $\vec{x} = (x,y ,z)^T \in {\cal E}$ and the reference coordinates $\vec{\xi}=(\xi, \eta, \zeta)^T \in {\cal E}_{f}$, through the transformation $\vec{x} = \vec{\mathcal{X}}\left(\vec{\xi}\right)$, where $\vec{\mathcal{X}}$ denotes the transformation operator. The transformation modifies the differential operators, and as a result, Eq. (\ref{eq:NS_Conservative}) becomes
\begin{equation}
 \label{eq:NS_Transformed}
\frac{\partial{\widetilde{Q}}}{\partial{t}} + \frac{\partial{\widetilde{F_i^a}}}{\partial{\xi_i}}=
\frac{1}{\mathrm{Re}_{c}}\frac{\partial{\widetilde{F_i^v}}}{\partial{\xi_i}} ,
\end{equation}
where
\begin{equation}
 \label{eq:NSDSEM}
\widetilde{Q} = \left|J\right|\vec{Q}, \quad \widetilde{F_i^a} = \left|J\right| \frac{\partial{\xi_i}}{\partial{x_j}} \vec{F_j^a}, \quad \widetilde{F_i^v} = \left|J\right| \frac{\partial{\xi_i}}{\partial{x_j}} \vec{F_j^v},
\end{equation}
with $\sim$ denoting the transformed quantities and $J$ representing the Jacobian of the transformation matrix. Here, $\widetilde{F^a}$ and $\widetilde{F^v}$ are the contravariant convective and viscous fluxes, respectively, \citep{Kopriva_2009}. 

Locally smooth test functions, $\phi$, are introduced and multiplied by Eq. (\ref{eq:NS_Transformed}) and the weak form is subsequently obtained by integrating the resulting expression over the reference element
 \begin{equation}
 \label{eq:integral form}
\int_{{{\cal E}_{f}}} \frac{\partial{\widetilde{Q}}}{\partial{t}} \phi \, 
+ \int_{{{\cal E}_f}} \frac{\partial{\widetilde{F_i^a}}}{\partial{\xi_i}} \phi \,
= \frac{1}{\mathrm{Re}_{c}}\int_{{{\cal E}_{f}}}\frac{\partial{\widetilde{F_i^v}}}{\partial{\xi_i}} \phi.
\end{equation}
The second term, associated with the inviscid fluxes, is rewritten as the sum of two integrals by applying integration by parts, separating the boundary fluxes of the elements from the interior fluxes,
\begin{equation}
\int_{{{\cal E}_{f}}} \frac{\partial{\widetilde{Q}}}{\partial{t}} \phi  
+ \int_{\partial{{\cal E}_f}} \widetilde{F_i^a} \hat{n_i} \phi - \int_{{{\cal E}_f}} {\widetilde{F_i^a}}  \frac{\partial{\phi}}{\partial{\xi_i}}
=  \frac{1}{\mathrm{Re}_{c}}\int_{{{\cal E}_{f}}}\frac{\partial{\widetilde{F_i^v}}}{\partial{\xi_i}} \phi,
\end{equation}
where $\hat{n}$ represents the unit normal vector on the faces of the reference element, $\partial {\cal E}_f$. Since the solution approximation with collocation polynomials in the DGSEM method is performed independently within each element, discontinuities in the fluxes appear at the interfaces between adjacent elements. To resolve this and preserve conservation, the discontinuous inviscid flux at the element interfaces is replaced with a numerical inviscid flux, ${\widetilde{F^{a\star}}}$. This yields 
\begin{equation} \label{eq:transformed and Riemann Flux}
\int_{{{\cal E}_{f}}} \frac{\partial{\widetilde{Q}}}{\partial{t}} \phi  
+ \int_{\partial{{\cal E}_f}} \widetilde{F_i^{a\star}} \hat{n_i} \phi - \int_{{{\cal E}_f}} {\widetilde{F_i^a}}  \frac{\partial{\phi}}{\partial{\xi_i}}
=  \frac{1}{\mathrm{Re}_{c}}\int_{{{\cal E}_{f}}}\frac{\partial{\widetilde{F_i^v}}}{\partial{\xi_i}} \phi.
\end{equation}
The numerical flux $\widetilde{F^{a\star}}$ is computed by applying the Riemann Solver to the inter-element fluxes \citep{PEYVAN2021} and is one of the methods for introducing dissipation into DG schemes \citep{Wang_2013,de_la_Llave_Plata_2018, Beck_2014}. A similar procedure can be applied to the term involving viscous fluxes in Eq. (\ref{eq:integral form}) by performing integration by parts and treating the boundary contributions separately. Common approaches for handling the viscous flux term include the Bassi–Rebay (BR1 and BR2) schemes \citep{Bassi_1996,Quaegebeur_2019} and the interior penalty method \citep{Arnold_2002,FERRER2011}.

The DGSEM employs approximation functions based on Lagrange collocating polynomials of order $\mathcal{P}$ within each element for the transformed solution variables, $\widetilde{Q}$, and the contravariant convective and viscous fluxes, $\widetilde{F^a}$ and $\widetilde{F^v}$. The solution variables within each element are represented as
\begin{equation}\label{eq:Q_nodal}
    \widetilde{Q}(\vec{\xi}, t)=\sum_{i=0}^{\mathcal{P}} \sum_{j=0}^{\mathcal{P}} \sum_{k=0}^{\mathcal{P}} \widetilde{Q}_{i,j,k}(t)l_{i}(\xi) l_{j}(\eta) l_{k}(\zeta),
\end{equation}
where
\begin{equation}
l_{i}\left(\xi\right) = \prod_{n=0, n \neq i}^{\mathcal{P}}\frac{\xi-\xi_{n}}{\xi_{i}-\xi_{n}},
\end{equation} 
are the Lagrange polynomials associated with the collocation points, and $\widetilde{Q}_{i,j,k}(t)$ denote the nodal solution values at the collocation points. The contravariant fluxes are represented in a similar manner through polynomial expansions
\begin{equation}\label{eq:F_nodal}
    \widetilde{F}(\vec{\xi}, t)=\sum_{i=0}^{\mathcal{P}} \sum_{j=0}^{\mathcal{P}} \sum_{k=0}^{\mathcal{P}} {\widetilde{F}_{i,j,k}}(t)l_{i}(\xi) l_{j}(\eta) l_{k}(\zeta),
\end{equation}
with $\widetilde{F}_{i,j,k}(t)$ representing the nodal flux values (applicable to both $\widetilde{F}^a$ and $\widetilde{F}^v$) at the collocation points. Various collocation points can be used, including Gauss-Chebyshev, Gauss-Legendre, and Gauss-Lobatto. Gauss-Lobatto points are often preferred for constructing energy- or entropy-stable schemes \citep{Ferrer_2023,Manzanero_2020_b} by leveraging the Summation-By-Parts Simultaneous Approximation Term (SBP-SAT) property \citep{Manzanero_2017, Fisher_2013, Chen_2017, Carpenter_2014}. They also enable split-form formulations with two-point fluxes, enhancing stability at the cost of higher computational effort \citep{Chandrashekar_2013, Winters_2018, Gassner_2016}.

Substituting the collocation equations for the solution and the fluxes in Eq. (\ref{eq:transformed and Riemann Flux}) and by selecting the test functions to be the same as the Lagrange interpolating functions and employing Gauss quadrature rule for both volume and surface integrals, the DGSEM formulation simplifies to a set of decoupled ordinary differential equations (ODEs), where each nodal solution value evolves independently in time.

For temporal integration, the low-storage explicit Runge-Kutta scheme of third order \citep{Williamson_1980} is used. To optimize computational efficiency, parallel implementations are performed for all computations, combining MPI and OpenMP on multiprocessor CPUs. The use of an explicit time integration scheme enhances parallelization efficiency when compared to implicit solvers, such as Rosenbrock-type Runge-Kutta methods \citep{Bassi_2015,Ferrer_2023}.

\subsection{Self-tuning Dynamic Explicit Modal Filtering}

In this section, the orthogonality of the approximation functions are used to obtain modal coefficients from the calculated nodal solution values. Then, a low-pass filter is applied to the modes before reconstructing the filtered solution.

In 1D problems, the solution within each element can be expressed as
\begin{equation} \label{eq:1DNodal&Modal}
\widetilde{Q}(\xi,t) = \sum_{i=0}^{\mathcal{P}}\widetilde{Q}_{i}(t){l}_{i}(\xi) = \sum_{i=0}^{\mathcal{P}}\hat{Q}_{i+1}(t)P_{i}(\xi), \
\end{equation}
where $\hat{Q}_{i+1}$ are the modal coefficients, and $P_{i}(\xi)$ are the orthogonal polynomials of degree $i$. The orthogonality property gives
\begin{equation} \label{eq:orthonality}
\int_{-1}^{1} P_{n}\left(\xi\right) P_{m}\left(\xi\right) w\left(\xi\right)  d\xi = \int_{-1}^{1} P_{n}^2\left(\xi\right) w(\xi) d\xi \ \delta_{nm},
\end{equation}
where $w(\xi)$ is the corresponding weight function. Multiplying Eq. (\ref{eq:1DNodal&Modal}) by one orthogonal polynomial $P_k\left(\xi\right)$ and the weight function $w(\xi)$, and integrating over the reference element,
\begin{equation}
\begin{split}
\int_{-1}^{1} \widetilde{Q}\left(\xi,t\right)P_k\left(\xi\right) w\left(\xi\right) d\xi &= \sum_{i=0}^{\mathcal{P}} \widetilde{Q}_{i}\left(t\right) \int_{-1}^{1} l_i\left(\xi\right)P_k\left(\xi\right)w\left(\xi\right) d\xi  \\&=\sum_{i=0}^{\mathcal{P}} \hat{Q}_{i+1}\left(t\right) \int_{-1}^{1} P_i\left(\xi\right)P_k\left(\xi\right) w\left(\xi\right) d\xi\\&= \hat{Q}_{k+1}\left(t\right)\int_{-1}^{1} P_{k}^2\left(\xi\right) w\left(\xi\right) d\xi,
\end{split}
\end{equation}
yields the modal coefficients
\begin{equation}\label{eq:modes}
\hat{Q}_{k+1}\left(t\right) = \sum_{i=0}^{\mathcal{P}} \frac{\int_{-1}^{1} l_i\left(\xi\right)P_k\left(\xi\right)w\left(\xi\right) d\xi}{\int_{-1}^{1} P_{k}^2\left(\xi\right)w\left(\xi\right) d\xi} \widetilde{Q}_i\left(t\right) = \sum_{i=0}^{\mathcal{P}} \mathcal{M}_{ki} \widetilde{Q}_{i}\left(t\right),  
\end{equation}
with
\begin{equation}\label{ModalMatrix}
\mathcal{M}_{ki} = \frac{\int_{-1}^{1} l_i\left(\xi\right) P_k\left(\xi\right)w\left(\xi\right) d\xi}{\int_{-1}^{1} P_{k}^2\left(\xi\right) w\left(\xi\right) d\xi},
\end{equation}
defined as the forward matrix evaluating modes from nodal values. Similarly, multiplying Eq. (\ref{eq:1DNodal&Modal}) by a Lagrange polynomial $l_i\left(\xi\right)$ and integrating over the reference element gives
\begin{equation}
\begin{split}
\int_{-1}^{1} \widetilde{Q}\left(\xi,t\right)l_i\left(\xi\right) d\xi &= \sum_{i=0}^{\mathcal{P}} \widetilde{Q}_i\left(t\right) \int_{-1}^{1}l_i\left(\xi\right)l_k\left(\xi\right) d\xi =\widetilde{Q}_k\left(t\right)\int_{-1}^{1} l_{k}^2\left(\xi\right) d\xi \\&= \sum_{i=0}^{\mathcal{P}} \hat{Q}_{i+1}\left(t\right) \int_{-1}^{1} P_i\left(\xi\right)l_k\left(\xi\right) d\xi,
\end{split}
\end{equation}
yielding 
\begin{equation}\label{eq:nodalvalues}
\widetilde{Q}_k\left(t\right) = \sum_{i=0}^{\mathcal{P}} \frac{\int_{-1}^{1} P_i\left(\xi\right)l_k\left(\xi\right)  d\xi}{\int_{-1}^{1} l_{k}^2\left(\xi\right)  d\xi} \hat{Q}_{i+1}\left(t\right) = \sum_{i=0}^{\mathcal{P}} \mathcal{N}_{ki} \hat{Q}_{i+1}\left(t\right), 
\end{equation}
where the backward transformation matrix, evaluating nodal values from modes, is given by
\begin{equation}
\quad \mathcal{N}_{ki} = \frac{\int_{-1}^{1} P_i\left(\xi\right)l_k\left(\xi\right) d\xi}{\int_{-1}^{1} l_{k}^2\left(\xi\right) d\xi}.
\end{equation}
It is important to note that while continuous Lagrange polynomials are not orthogonal over $[-1,1]$, their discrete counterparts, associated with Gauss-Lobatto or Gauss-Legendre quadrature points, exhibit orthogonality under the respective quadrature rule used for integration \citep{Mateo_Gab_n_2022}. To efficiently exploit these properties within the numerical framework, the matrices $\mathcal{M}$ and $\mathcal{N}$ are computed once at the beginning of the simulation and stored for later use.

For 3D problems, the tensor product of  1D formulas constructs the nodal and modal representations
\begin{equation} \label{eq:3DNodal&Modal}
\widetilde{Q}(\vec{\xi},t) = \sum_{i,j,k=0}^{\mathcal{P}}\widetilde{Q}_{i,j,k}(t){l}_{i}(\xi){l}_{j}(\eta){l}_{k}(\zeta) = \sum_{i,j,k=0}^{\mathcal{P}}\hat{Q}_{i+1,j+1,k+1}(t)P_{i}(\xi)P_{j}(\eta)P_{k}(\zeta). \
\end{equation}
Multiplying the nodal and modal collocation equations in Eq.~(\ref{eq:3DNodal&Modal}) by the tensor product of orthogonal polynomials and the weight function $P_{lmn}=P_{l}\left(\xi\right)P_{m}\left(\eta\right)P_{n}\left(\zeta\right)w\left(\xi\right)w\left(\eta\right)w\left(\zeta\right)$, and integrating over the reference element, ${\cal E}_{f}=[-1,1]^3$, yields the modal coefficients. The tensor-product structure allows the integral over the reference element to be separated into three independent integrals over the reference interval $L = [-1,1]$. Consequently, the transformation from nodal values to modal coefficients in three dimensions can be performed through three successive matrix–vector multiplications using the one-dimensional forward transformation matrix $\mathcal{M}$. The resulting modal coefficients are then given by
\begin{equation}\label{eq:3Dmodes}
\hat{Q}_{l+1,m+1,n+1}(t) = 
\sum_{i=0}^{\mathcal{P}} \sum_{j=0}^{\mathcal{P}} \sum_{k=0}^{\mathcal{P}} 
\mathcal{M}_{li}\, \mathcal{M}_{mj}\, \mathcal{M}_{nk}\, \widetilde{Q}_{i,j,k}(t).
\end{equation}
Similarly, the inverse transformation back to the nodal space in 3D problems is given by 
\begin{equation}\label{eq:3Dnodes}
\widetilde{Q}_{l,m,n}(t) = 
\sum_{i=0}^{\mathcal{P}} \sum_{j=0}^{\mathcal{P}} \sum_{k=0}^{\mathcal{P}} 
\mathcal{N}_{li}\, \mathcal{N}_{mj}\, \mathcal{N}_{nk}\, \hat{Q}_{i+1,j+1,k+1}(t).
\end{equation}

The orthogonality property can be used to extract modal coefficients for any collocation grid and orthogonal polynomial basis. In this study, Gauss–Lobatto points with Legendre polynomials and a weight function of $w\left(\xi\right)=1$ are employed. Certain grid–polynomial combinations can simplify the nodal-to-modal transformation. For example, \cite{Ranjbar_2024POF} used Chebyshev polynomials on a staggered Chebyshev grid, which allows efficient transformations via the Discrete Cosine Transform and its inverse using the FFTW library \citep{Frigo_2005}. This approach, however, is limited to Chebyshev collocation points and polynomials.

Compared to the modes associated with artificial flux in SVV studies, the modes corresponding to solution variables capture physics-based information and can be used for local flow characterization and assessing flow resolvedness \citep{Ranjbar_2024POF}. Following \citet{Ranjbar_2024POF}, the 3D modal representation is mapped to a 1D format, with modes grouped according to their polynomial characteristics. The aggregated values of these mode groups, denoted as ``energy levels,'' $\Gamma_n$, are given by
\begin{equation} \label{eq:energy levels}
\Gamma_{n} = \left(\sum_{i=1}^{\mathcal{P}+1}\sum_{j=1}^{\mathcal{P}+1}\sum_{k=1}^{\mathcal{P}+1}\hat{Q}_{i,j,k}^{2}\theta(i,j,k)\right)^{\frac{1}{2}},  \quad\quad 2 \le n \le \operatorname{round}\left(\sqrt{3\left(\mathcal{P}+1\right)^2}\right)
\end{equation}
where 
\begin{equation}
     \theta(i,j,k) = 
     \left\{
       \begin{array}{cc}
         1 & \quad \quad \quad \sqrt{i^2 + j^2 + k^2} \in [n - 0.5, n + 0.5) \\
         0 & \quad \quad \text{otherwise}
       \end{array}
     \right..
\end{equation}

A 1D low-pass filter kernel is then applied to the grouped modes and provides a dissipation mechanism. Building on our previous DEMF method \citep{Ranjbar_2024POF}, which employed a cut-off filter kernel, the present study introduces a hyperbolic tangent filter kernel, and the filtered modes are computed as
\begin{equation}
\hat{Q}^f_{i,j,k}\left(t\right) = \mathcal{F}_{n}\hat{Q}_{i,j,k}\left(t\right),
\end{equation}
where
\begin{equation} \label{eq:filterkernel}
\mathcal{F}_n =
\begin{cases} 
1 & n \leq M \\ 
-\tanh \left[ -3\frac{(n - N)^2}{(M - N)^2} \right] & n > M
\end{cases},
\end{equation}
with $N$ being the total number of ``mode groups'' (not the total number of modes) defined as
\begin{equation}
N = \operatorname{round}\left(\sqrt{3\left(\mathcal{P}+1\right)^2}\right),
\end{equation}
and $M$ is the cut-off mode, which is the main feature of the present-study model and is evaluated dynamically based on turbulent characteristics. Strictly speaking, $M$ represents a group of modes in the 1D representation rather than a single mode; however, for simplicity and consistency with the SVV literature, it is referred to as the cut-off mode instead of cut-off mode group. The index $n$ in the filtering formula is determined for each mode, $\hat{Q}_{i,j,k}$, using the relation
\begin{equation}
\sqrt{i^2 + j^2 + k^2} \in [n - 0.5, n + 0.5). \\
\end{equation}
$\hat{Q}^f_{i,j,k}\left(t\right)$ is then used in Eq. (\ref{eq:3Dnodes}) to convert back to the nodal space, yielding $\widetilde{Q}^f_{i,j,k}\left(t\right)$.

Modal-based filters have been widely studied in the SVV context, with various formulations proposed. The simplest is the cut-off filter, which sets modes below or above a threshold to zero and others to one \citep{Tadmor_1989}. However, smoother transitions between zero and one have shown improved performance \citep{Tadmor_1994}. Alternatives such as exponential \citep{Kirby_2006, Karamanos_2000}, power \citep{Moura_2016, Manzanero_2017}, and other functions \citep{Koal_2012} have also been explored, along with studies on optimal parameter tuning \citep{Tadmor_1989, Maday_1993, Karamanos_2000, PASQUETTI_2005}. This work implements a hyperbolic tangent filter kernel preserving modes below the cut-off mode while exponentially reducing the magnitude of the modes above it, reaching zero for the highest mode. The kernel also adapts to different polynomial orders via the embedded parameters $N$ and $M$. While SVV methods in 3D typically construct filter kernels using tensor products of 1D shapes \citep{Karamanos_2000, Kirby_2006, Manzanero_2020, Mateo_Gab_n_2022}, this work instead maps modes into a 1D representation and directly applies a 1D kernel.

The cut-off mode, $M$, controls the manipulation of modes and the removal of energy. Extensive studies have been conducted on the appropriate selection of $M$ in SVV frameworks, with different researchers arriving at varying values depending on the application \citep{Tadmor_1989, Karamanos_2000, PASQUETTI_2005, Kirby_2006}. Following the recommendation of \cite{Karamanos_2000}, which suggests that the cut-off mode formulation can be adapted to couple with flow dynamics by incorporating local variables such as the strain field within each element \citep{Kirby_2002}, this study defines the cut-off mode based on the degree to which turbulence is unresolved, as well as on the strain and rotational rate tensors.

The second invariant of the strain rate tensor is defined as \citep{Tlales_2024}
\begin{equation}
Q_{S} = \frac{1}{2}\left(\left(tr\left(S\right)\right)^2-tr\left(S^2\right)\right), \\
\end{equation}
where $S$ is the strain rate tensor. The rotational rate tensor $\Omega$ has a similar invariant given by
\begin{equation}
Q_{\Omega} = -\frac{1}{2}\left(tr\left(\Omega^2\right)\right). \\
\end{equation}
The quantity $Q_S$ identifies regions with high local viscous dissipation rate of the kinetic energy, where more negative values indicate stronger viscous effects \citep{Zhou_2015}. In contrast, $Q_{\Omega}$ is proportional to the enstrophy density and highlights rotational structures in the flow, where more positive values correspond to stronger rotational motion \citep{Zhou_2015}. These variables were employed in the study by \cite{Tlales_2024} for machine learning–based mesh adaptation in turbulent regions. By analyzing the probability of having viscous and rotational flow features using $Q_S$ and $Q_{\Omega}$, the mesh was refined accordingly \citep{Tlales_2024}. In this study, the cut-off mode formulation incorporates these variables to regulate filtering in unresolved elements.

In this work, the cut-off mode $M$ is determined from the ratio of the local Kolmogorov length scale $\eta$ to the average grid spacing $\Delta$, along with a measure of shear and rotation, and is expressed as
\begin{equation} \label{eq:cutoff mode}
M = N \times \left(\frac{\Delta}{\eta}\right)^{-c}\left(1 - X\right)^{c},
\end{equation}
where $c$ is a positive constant and $X$ quantifies the shear and rotation within each element. The variable $X$ is defined as
\begin{equation} \label{eq:xprobability}
X = X_S X_{\Omega},
\end{equation}
where
\begin{equation} \label{eq:xviscous}
X_S = 1 - \exp\left(\frac{-|\bar{Q}_S|}{\langle |\bar{Q}_S| \rangle + \kappa}\right),
\end{equation}
and
\begin{equation} \label{eq:xrotation}
X_{\Omega} = 1 - \exp\left(\frac{-\bar{Q}_{\Omega}}{\langle \bar{Q}_{\Omega} \rangle + \kappa}\right),
\end{equation}
with $\bar{Q}_S$ and $\bar{Q}_{\Omega}$ being the nodal averages of $Q_S$ and $Q_{\Omega}$ within each element, and $\langle \ \rangle$ shows averaging over unresolved elements that are along the same homogeneous direction(s) in space and/or time. For periodic box turbulence, averaging can be performed over all unresolved elements. For wall-bounded flows, averaging is performed along homogeneous directions, or, for stationary flows, over time for each element once the flow becomes statistically steady. An alternative to using averaging in the denominator is to apply a threshold. Mesh adaptation studies based on $\Omega$ and $S$ have shown that the optimal threshold is problem-dependent \citep{PANG2021}. Moreover, even a small change in the threshold significantly alters the regions detected for mesh refinement \citep{Tlales_2024}. The proposed formulations for $X_S$ and $X_{\Omega}$ yield values in $[0,1]$, providing a relative measure of shear and rotation. A small constant $\kappa$ is included in the denominators to prevent division by zero.

According to Eq.~(\ref{eq:cutoff mode}), as the flow becomes increasingly unresolved, indicated by smaller Kolmogorov length scales relative to the average grid spacing, the cut-off mode shifts toward lower modal indices, thereby increasing the removal of energy from high-frequency modes. Bigger magnitudes of $Q_S$ and $Q_{\Omega}$ increase the values of $X_S$ and $X_\Omega$, which in turn reduces the cut-off mode $M$ and intensifies filtering.

Unresolved elements are detected by coupling the modal filtering with the sensor developed by \cite{Ranjbar_2024POF}. This sensor evaluates the local Kolmogorov length scale for each element and compares it with the average grid spacing of the same element. The non-dimensional local Kolmogorov length scale is computed from the non-dimensional turbulent kinetic energy dissipation rate,
\begin{equation}\label{eq:turbulent energy dissipation} \epsilon = \frac{2}{\mathrm{Re}_c}S':S'-\frac{2}{3\,\mathrm{Re}_c}\left(\frac{\partial v'_k}{\partial x_k}\right)^2, \end{equation}
where $S'$ is the fluctuating strain-rate tensor, $:$ denotes the double contraction, and $v'_k$ represents velocity fluctuations. The dissipation rate is then volumetrically averaged over the nodes within each element, $\bar{\epsilon}$, and used to compute the non-dimensional local Kolmogorov length scale
\begin{equation}\label{eq:kolmogorovlengthscale} \eta = \left(\frac{1}{\mathrm{Re}_c}\right)^{3/4} \left( \frac{1}{\bar{\epsilon}} \right)^{1/4}. \end{equation}
The average grid spacing for each element, accounting for non-uniform node distribution, is defined as \citep{Komperda_2020}
\begin{equation} \label{eq:gridspacing} \Delta = \frac{J^\frac{1}{3}}{\prod_{i=1}^{3}\left(\mathcal{P}_{i}+1\right)^\frac{1}{3}},
\end{equation}
and filtering is applied only in elements where $\eta < \Delta$. The rate of change of kinetic energy used for comparison of the results in the homogeneous TGV case is computed as
\begin{equation}\label{eq:energy dissipation rate}
-\frac{dK}{dt} = \frac{1}{V} \frac{d}{dt} \int_V \frac{1}{2} \rho v_iv_i \: dv,
\end{equation}
where $V$ is the total volume of the domain and $K$ is the kinetic energy.

In contrast to \cite{Ranjbar_2024POF} and the commonly used formulation $M = 5\sqrt{N}$ \citep{Karamanos_2000}, which apply a fixed cut-off mode uniformly across all elements, the present model offers several advantages. First, filtering adapts to each unresolved element based on its local turbulence characteristics, including the degree of unresolvedness, shear, and rotation. Second, for unresolved elements, the cut-off mode evolves dynamically over time, unlike in \cite{Ranjbar_2024POF,Karamanos_2000,Manzanero_2020,Mateo_Gab_n_2022}, where it remains constant. Third, unlike the integer-only cut-offs in \cite{Ranjbar_2024POF}, the filter kernel here allows continuous, real-valued cut-off values, and naturally adjusts to different polynomial orders through the parameters $N$ and $M$.

\begin{figure}[!ht]
    \centering
    \includegraphics[width=0.85\textwidth]{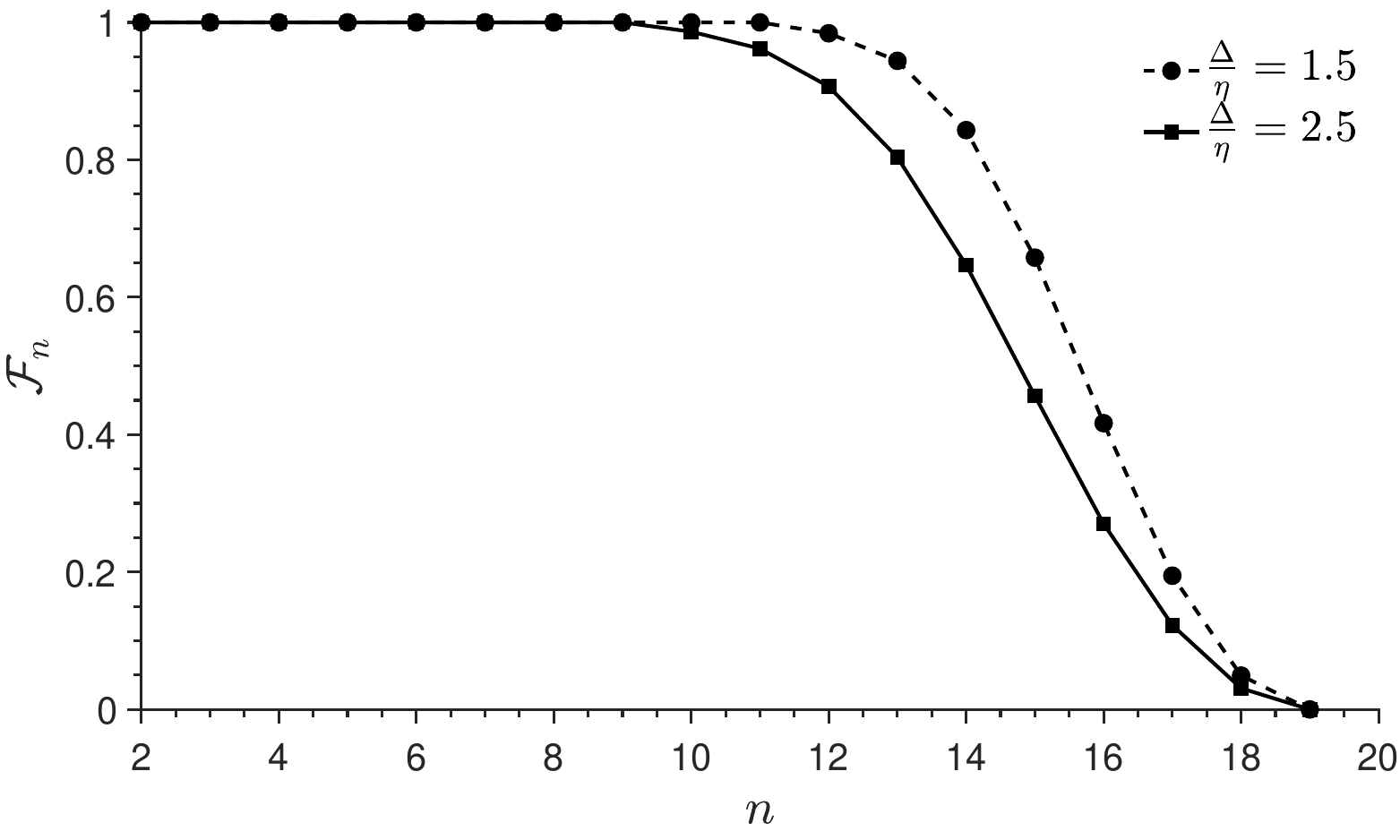}
    \caption{\label{fig:kernels} Filter kernels for two elements with $\frac{\Delta}{\eta}=1.5$ and $\frac{\Delta}{\eta}=2.5$ ($N=19$, $X=0.6$, $c=0.4$ in Eq.~(\ref{eq:cutoff mode})).}
\end{figure}

As an example, Fig.~\ref{fig:kernels} presents the filter kernels for two elements with the same average grid spacing and $X=0.6$, but different local Kolmogorov length scales. The polynomial order in these elements is $\mathcal{P}=10$, which yields $11$ modes in each direction, giving $N=19$. The constant $c=0.4$ is used for the cut-off mode calculation in Eq. (\ref{eq:cutoff mode}). As seen in Fig. \ref{fig:kernels}, the element with the smaller $\eta$ has a lower cut-off mode of 9.12 as opposed to 11.2 for the element with larger $\eta$, thereby influencing more modes and removing more energy from the higher modes.

\begin{figure}[!ht]
    \centering
    \includegraphics[height=0.47\textheight]{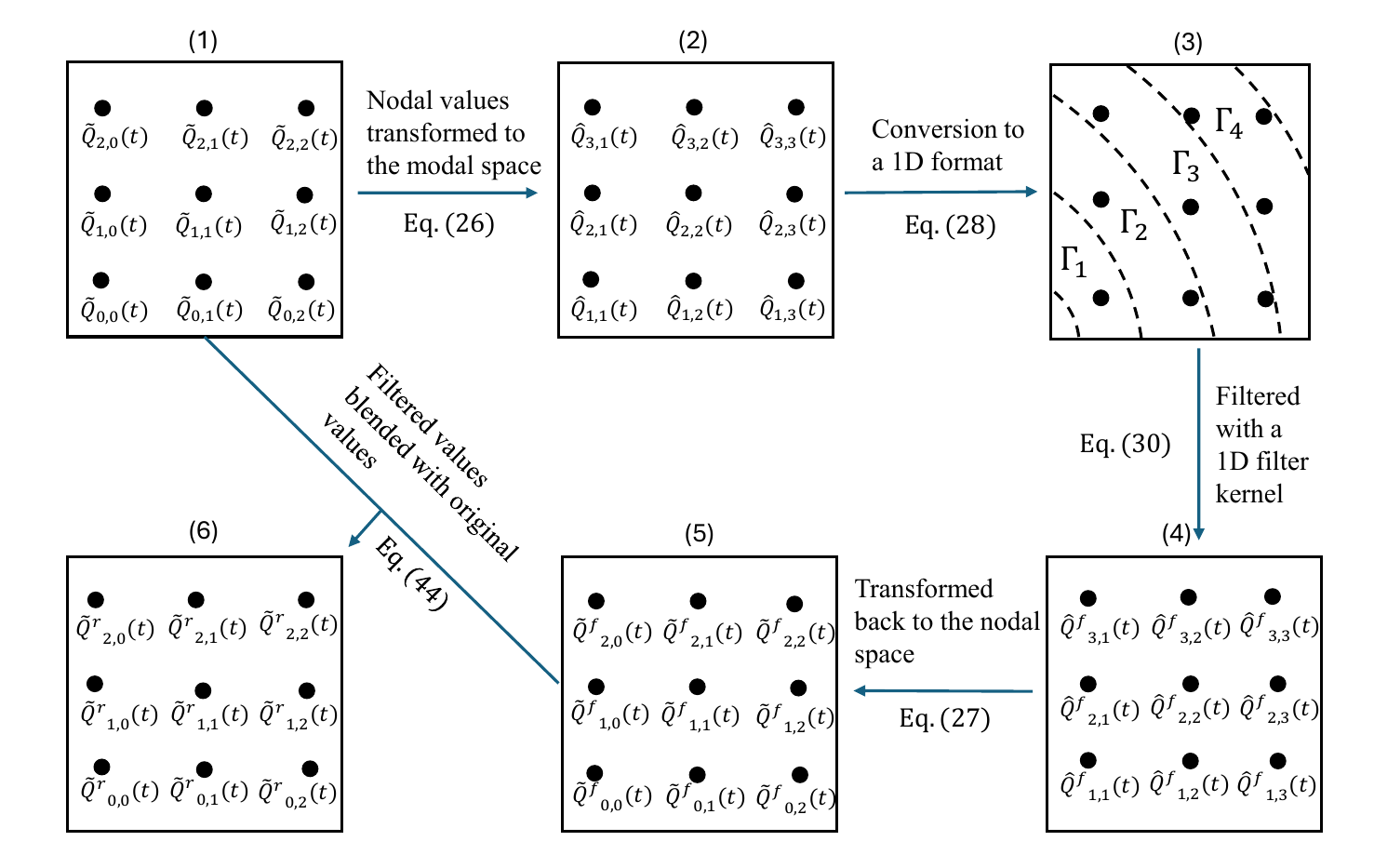}
    \caption{\label{fig:filtering} Schematic of modal filtering for a 2D problem with $\mathcal{P}=2$ inside each element.}
\end{figure}

The main parameter in the filtering formulation is the cut-off mode, which controls the removal of energy. To enhance control and flexibility over filtering, two additional parameters are introduced. First, the filtering frequency $F$ is defined as the inverse of the number of timesteps between consecutive filtering operations. Second, the combination factor $0\leq \alpha \leq 1$ blends the filtered values with the original values. The blending is expressed as
\begin{equation}
\widetilde{Q}^r_{i,j,k}\left(t\right) = \alpha \widetilde{Q}^f_{i,j,k}\left(t\right) +(1-\alpha)\widetilde{Q}_{i,j,k}\left(t\right),
\end{equation}
where $\widetilde{Q}^r_{i,j,k}$ is the blended value used to advance the simulation, $\widetilde{Q}^f_{i,j,k}$ is the filtered value, and $\widetilde{Q}_{i,j,k}$ is the original unfiltered value. \cite{Ghiasi_2019} demonstrated that different combinations of $\alpha$ and $F$ yielding the same $\alpha \times F$ produce identical results. In this study, the values $\alpha = 0.01$ and $F = 1$ are used in all cases. To summarize all the steps of the filtering process, Fig. \ref{fig:filtering} illustrates the modal filtering procedure for a 2D problem with polynomial order $\mathcal{P}=2$ within the elements.

\subsection{Software}

The simulations were performed using the open-source Fortran code HORSES3D \citep{Ferrer_2023}. HORSES3D is an object-oriented solver originally designed to solve the compressible NS equations using DG discretizations and explicit time marching. The code has since been extended to handle a variety of physical phenomena, including incompressible flows, multiphase flows, acoustics, and particle dynamics. Over time, numerous techniques, such as shock capturing, turbulence modeling, energy/entropy-stable schemes, polynomial adaptation, and immersed boundary methods, have been incorporated into the code. The code is actively maintained and is available on GitHub under the MIT license at the developer’s repository: \href{https://github.com/loganoz/horses3d}{https://github.com/loganoz/horses3d}.

%% file: include/results.tex
\section{Results and Discussions}

This section presents numerical results obtained by applying the STDEMF model to three benchmark flows. The results are compared with the unfiltered DGSEM, the DEMF model of \citet{Ranjbar_2024POF}, and the DGSEM combined with the Smagorinsky eddy-viscosity turbulence model. Some considerations about the model are presented at the end.

\subsection{ Homogeneous Isotropic Decaying (HID)}

The proposed model is developed as an alternative for sub-grid scale modeling, aiming to remove unresolved turbulent energy of the sub-grid scales. Therefore, small-scale turbulence serves as the first case to evaluate the model. The HID flow, studied in several references \citep{george1992,skrbek2000,batchelor1948,Blaisdell_1993}, is used as the test case for evaluating the model's performance on small-scale turbulence.

\begin{figure}[htbp]
    \centering
    \begin{subfigure}[t]{0.48\textwidth} 
        \includegraphics[width=\textwidth,height=1\textwidth,keepaspectratio]{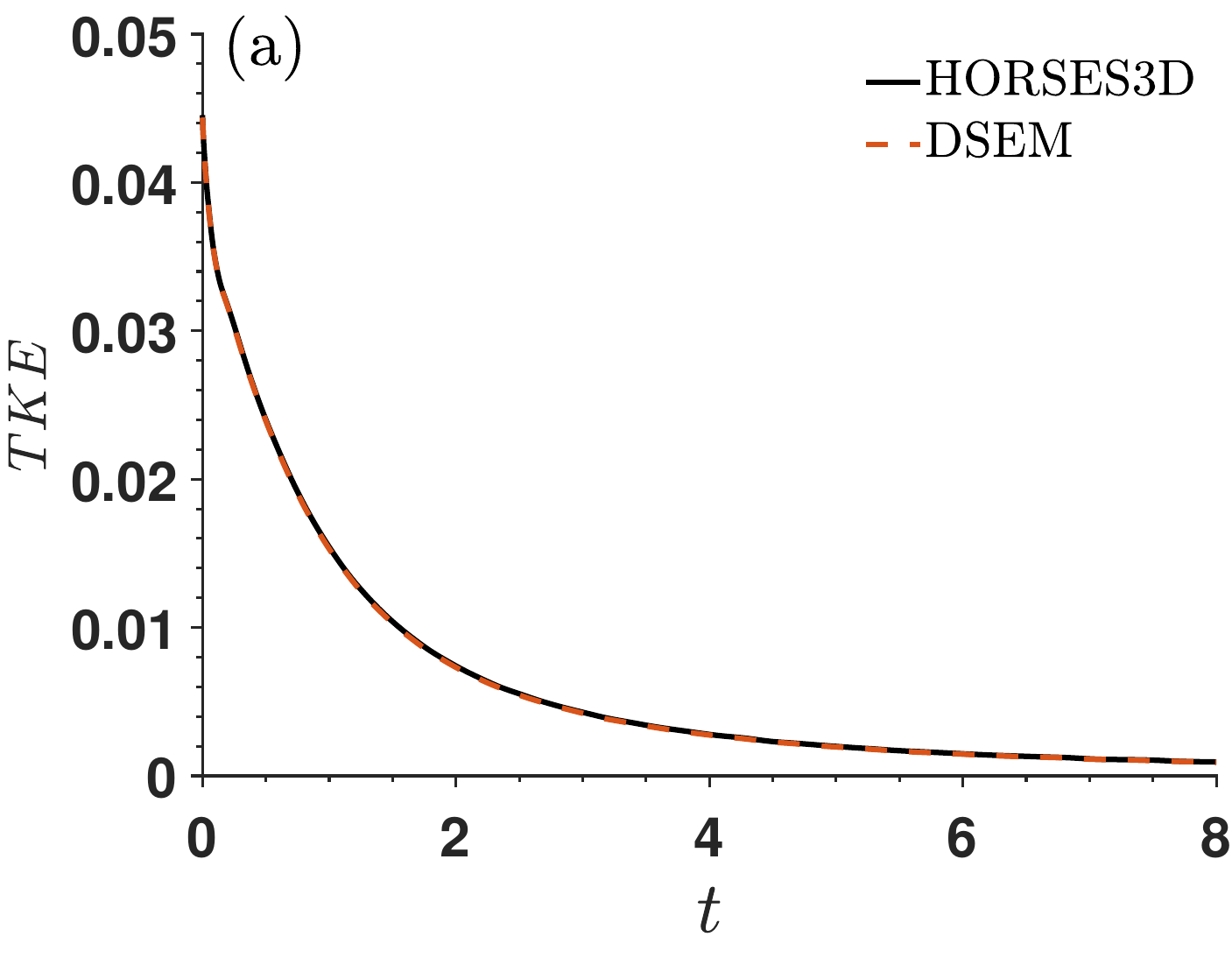}
    \end{subfigure}
    \hspace{0.02\textwidth}
    \begin{subfigure}[t]{0.48\textwidth}
        \includegraphics[width=\textwidth,height=1\textwidth,keepaspectratio]{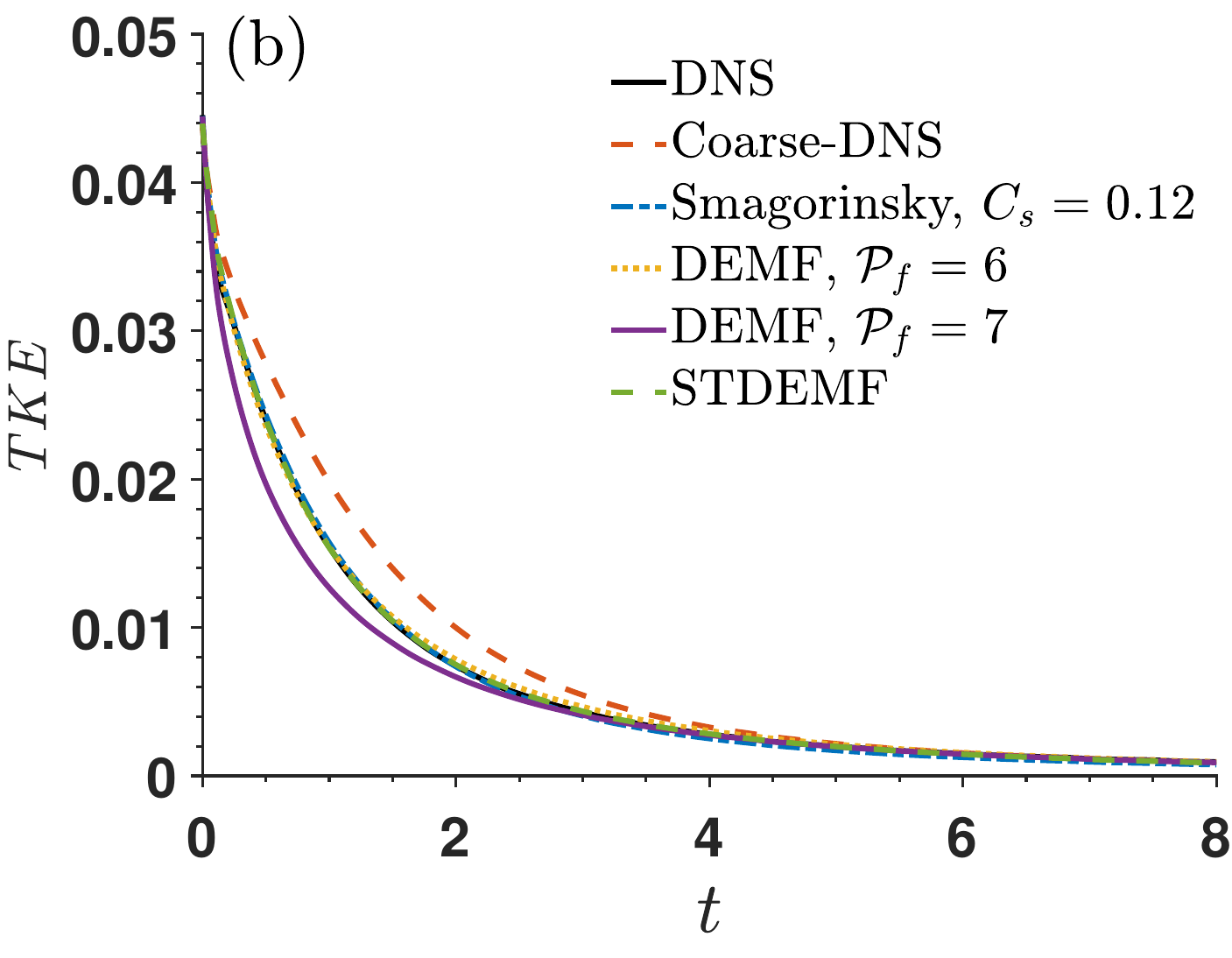}
    \end{subfigure}
    \vspace{1em}
    \begin{subfigure}[t]{0.48\textwidth}
        \centering
        \includegraphics[width=\textwidth,height=1\textwidth,keepaspectratio]{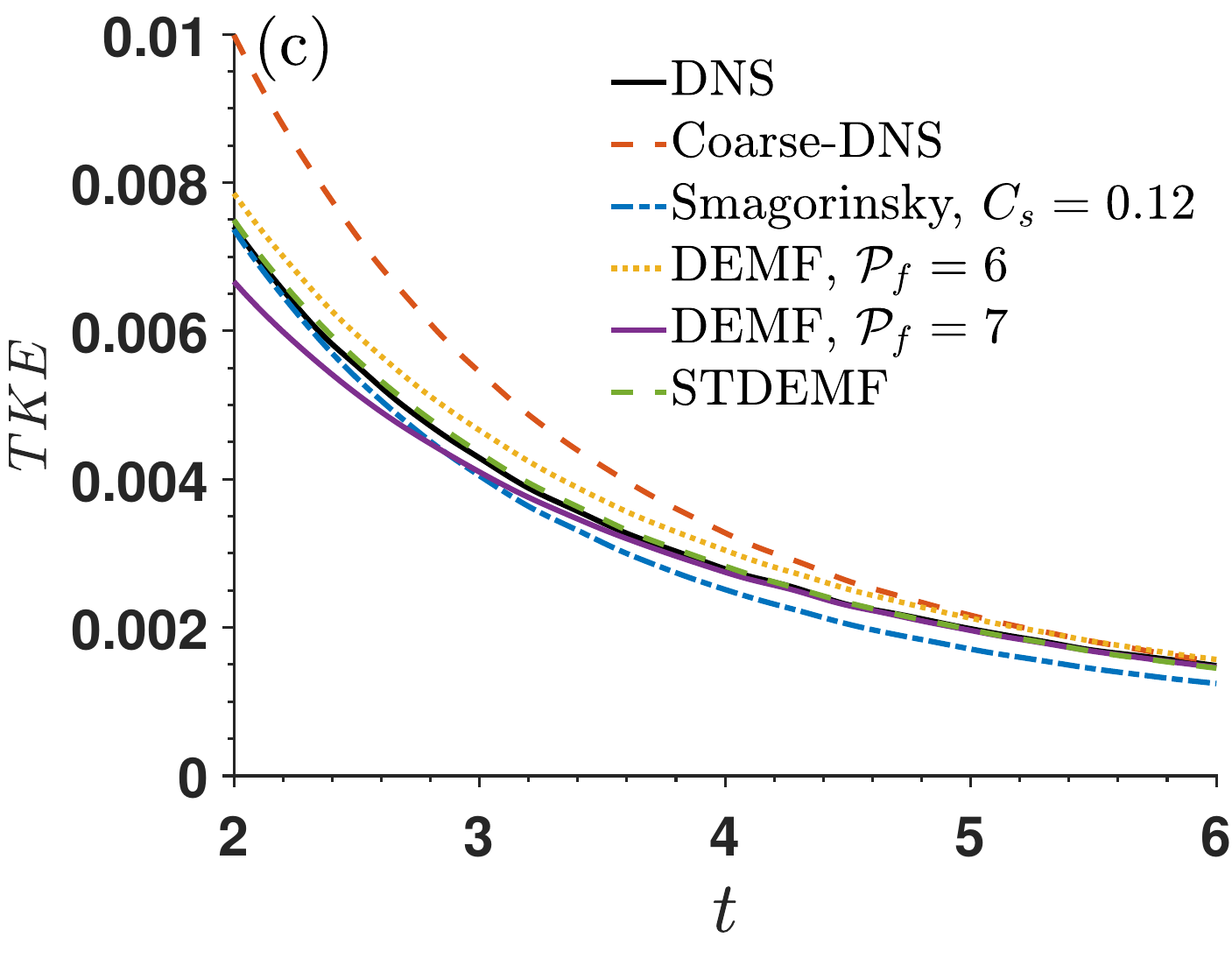}
    \end{subfigure}
    \caption{Comparison of the turbulent kinetic energy in the HID flow: (a) DNS results between the HORSES3D and DSEM solvers; (b) DNS, coarse-DNS, Smagorinsky, DEMF, and STDEMF model results for $t=[0,8]$; and (c) a zoomed view of DNS, coarse-DNS, Smagorinsky, DEMF, and STDEMF model results for $t=[2,6]$.}
    \label{fig:isoitropictke}
\end{figure}

The HID flow is initialized following the $idc96$ case of \cite{Blaisdell_1993}, where a top-hat initial energy spectrum distribution with contributing wavenumbers $8 \le k \le 16$ is used to initialize the velocity and temperature fields. The flow is simulated in a periodic cubic domain $V=[0, 2\pi]^3$ at a Reynolds number of $\mathrm{Re_c}=450$, with air as the working fluid characterized by $\mathrm{Pr_c}=0.72$ and $\gamma=1.4$. Validation is performed by comparing DNS results from the present solver, HORSES3D, with those from the discontinuous spectral element method (DSEM) code implemented by \cite{Ranjbar_2024POF}, both based on high-order spectral element methods. For DNS at this Reynolds number, the domain is discretized using $12^3$ elements with a polynomial order of $\mathcal{P}=8$ inside each element. Figure \ref{fig:isoitropictke}(a) shows the turbulent kinetic energy (TKE) obtained from both solvers, demonstrating good agreement.

After performing DNS and validating the case, the resolution was reduced to create an under-resolved flow, on which the model was then applied for assessment. Table \ref{tab:isocases} summarizes the resolutions used for both the DNS and the coarse-DNS cases. 
\begin{table}[h!]
    \centering
    \caption{Details of the grids used for different cases of the HID flow at $\mathrm{Re_c}=450$.}
    \begin{tabular}{c@{\hspace{1cm}}c@{\hspace{1cm}}c@{\hspace{1cm}}c@{\hspace{1cm}}c}
        \toprule
        Case& Case ID& $\mathcal{P}$ & Number of elements & Number of grid points   \\
        \hline
        HID - DNS & 12$\mathcal{P}$8 & 8 &  $12\times12\times12$ & 1,259,712 \\
        HID - Coarse DNS & 6$\mathcal{P}$7 & 7 &  $6\times6\times6$ & 110,592 \\
        \hline
        \hline
    \end{tabular}
    \label{tab:isocases}
\end{table}
In addition to the STDEMF and  DEMF models, the Smagorinsky eddy-viscosity model is also applied to the unresolved case. The Smagorinsky eddy-viscosity model introduces an artificial turbulence viscosity, $\mu_t$, based on the gradients of the resolved velocity and local grid size, following the formulation
\begin{equation}
    \mu_t = (C_s\Delta)^2 |S|
\end{equation}
where $C_s$ is the Smagorinsky constant.

Figure \ref{fig:isoitropictke}(b) and (c) present a comparison of the non-filtered coarse-DNS results against DNS, DEMF, STDEMF, and the Smagorinsky model. All the cases simulated using HORSES3D employ the standard form with the low-dissipation Roe Riemann solver for the convective fluxes \citep{Ferrer_2023}. For the viscous flux, the BR1 method with a penalty parameter of zero is used, which corresponds to averaging the viscous flux at element interfaces. Reducing the resolution without applying any model results in higher turbulent kinetic energy values, which is attributed to the missing subgrid-scale dissipation mechanism. When the Smagorinsky model is applied, the added artificial viscosity dissipates the excess unremoved energy, and the results show agreement with the DNS data. The constant value of $C_s = 0.12$ yields the optimal results in this case. Although the Smagorinsky results are closer to DNS, a closer inspection of the plot in Fig. \ref{fig:isoitropictke}(c) shows that at later times ($t > 2.4$), as turbulence decreases substantially, the Smagorinsky model exhibits overdissipation. For the DEMF model, the user must decide on the number of mode groups, $\mathcal{P}_f$, from which energy should be removed. It is seen in the figure that $\mathcal{P}_f = 6$ and $\mathcal{P}_f = 7$ are the cases between which the DNS results lie. The cut-off filter kernel that removes the last six groups of modes ($\mathcal{P}_f = 6$) in unresolved elements shows better agreement with DNS at earlier times ($t < 2$) compared to the case where the last seven groups of modes ($\mathcal{P}_f = 7$) are removed. The $\mathcal{P}_f = 7$ case shows excessive overdissipation at earlier times. However, at later times ($t > 3.6$), the DEMF case with $\mathcal{P}_f = 7$ shows better agreement compared to the case with $\mathcal{P}_f = 6$. Lastly, the STDEMF model, which implements a tangent hyperbolic filter kernel and self-selects the cut-off mode for each unresolved element separately, shows the best agreement with DNS results among all the implemented models. The results shown for the new STDEMF model correspond to a constant value of $c=0.65$ in Eq. (\ref{eq:cutoff mode}).

\begin{figure}[htbp]
    \centering
    \begin{subfigure}[t]{0.48\textwidth} 
        \includegraphics[width=\textwidth,height=1\textwidth,keepaspectratio]{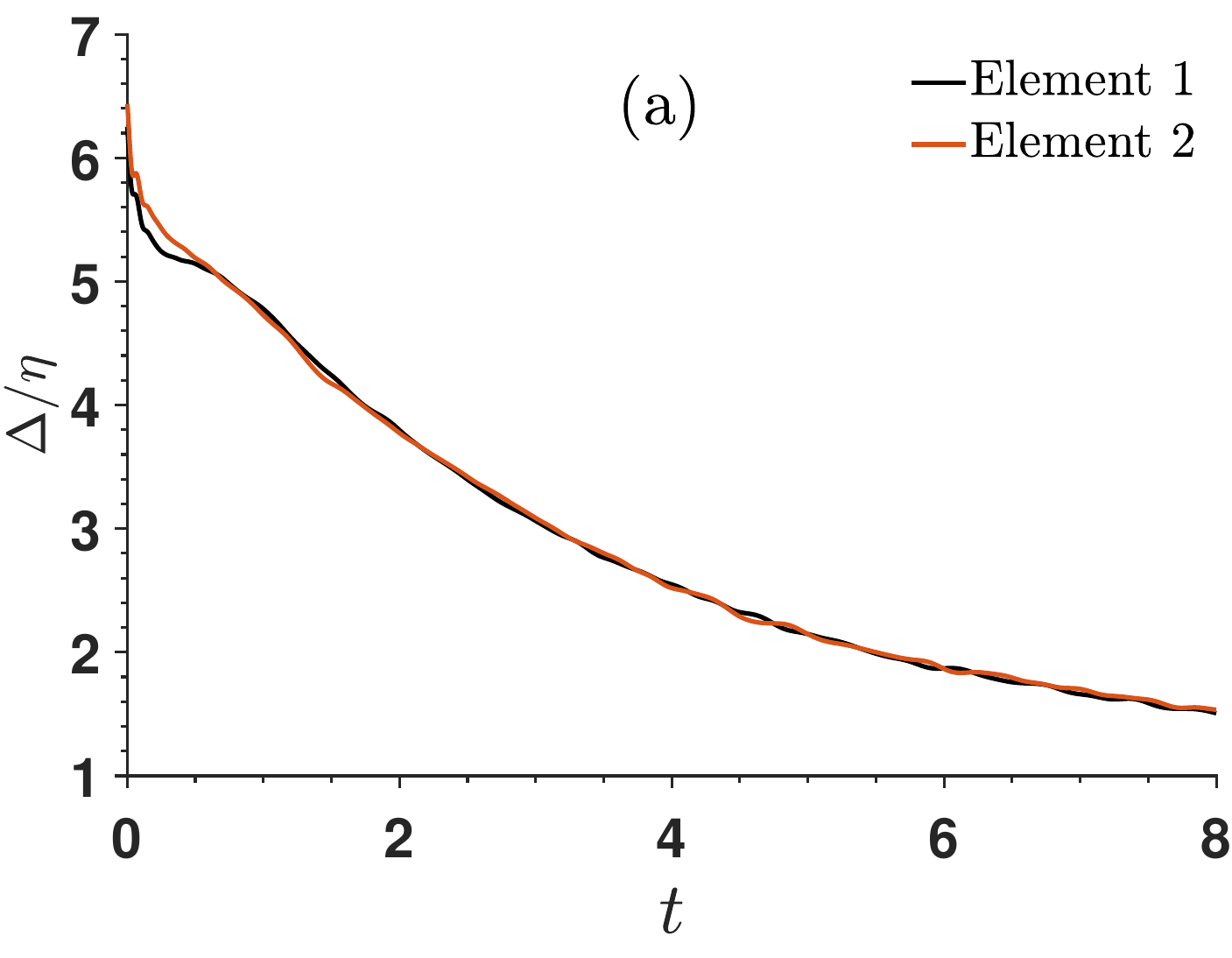}      
    \end{subfigure}
    \hspace{0.01\textwidth}
    \begin{subfigure}[t]{0.48\textwidth}
        \includegraphics[width=\textwidth,height=1\textwidth,keepaspectratio]{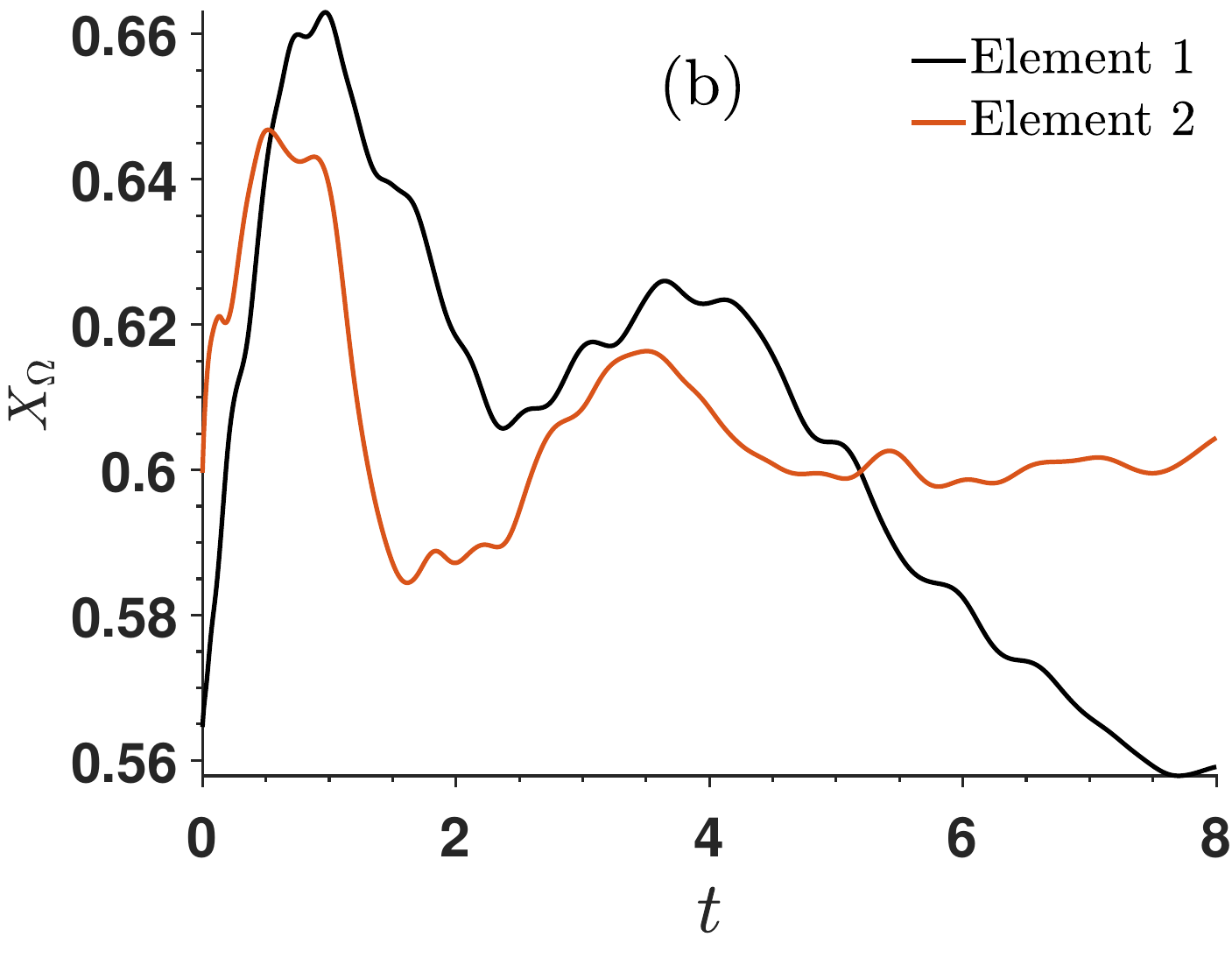}
    \end{subfigure}
    \vspace{1em}
    \begin{subfigure}[t]{0.48\textwidth} 
        \includegraphics[width=\textwidth,height=1\textwidth,keepaspectratio]{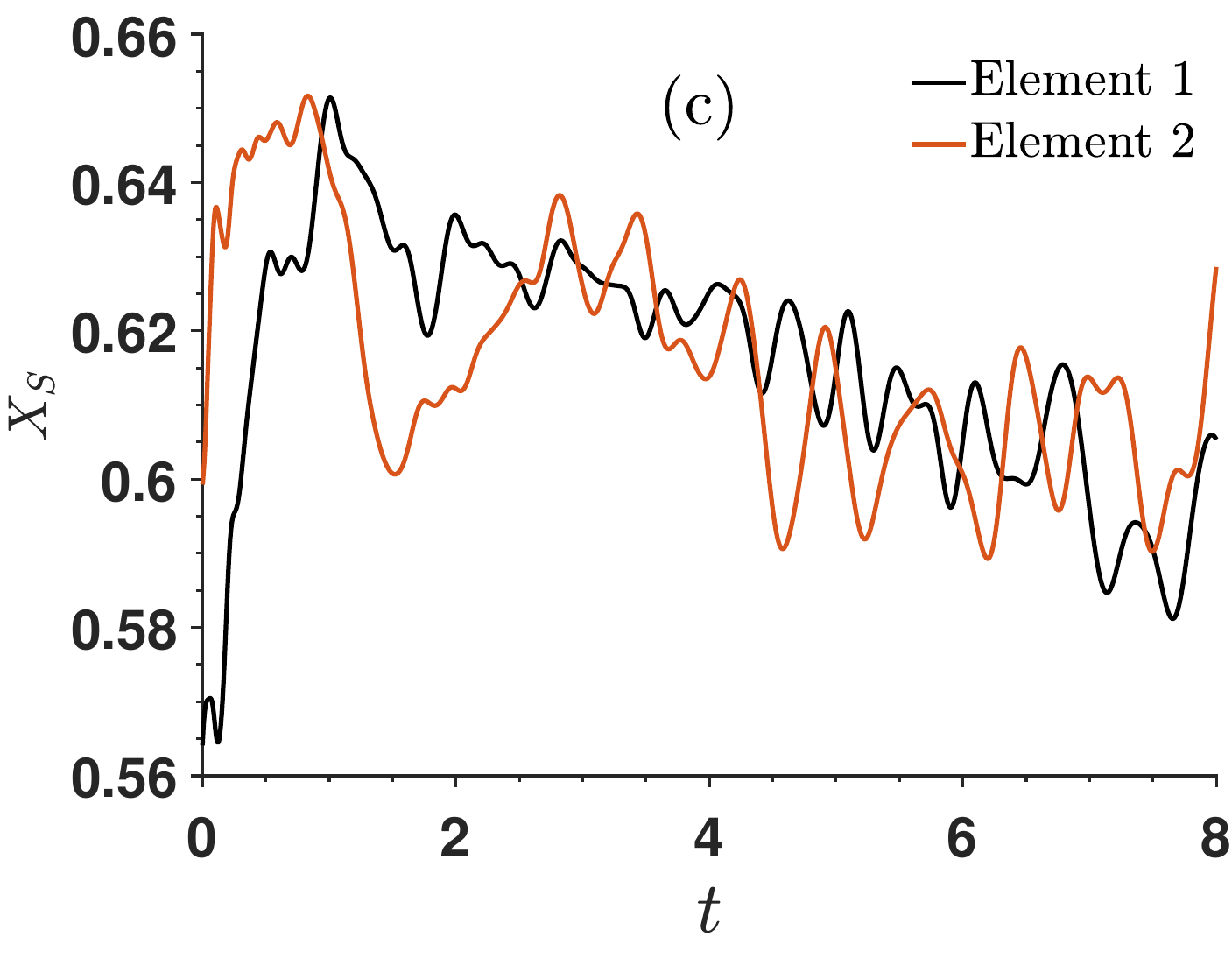}
    \end{subfigure}
    \hspace{0.02\textwidth}
    \begin{subfigure}[t]{0.48\textwidth}
        \includegraphics[width=\textwidth,height=1\textwidth,keepaspectratio]{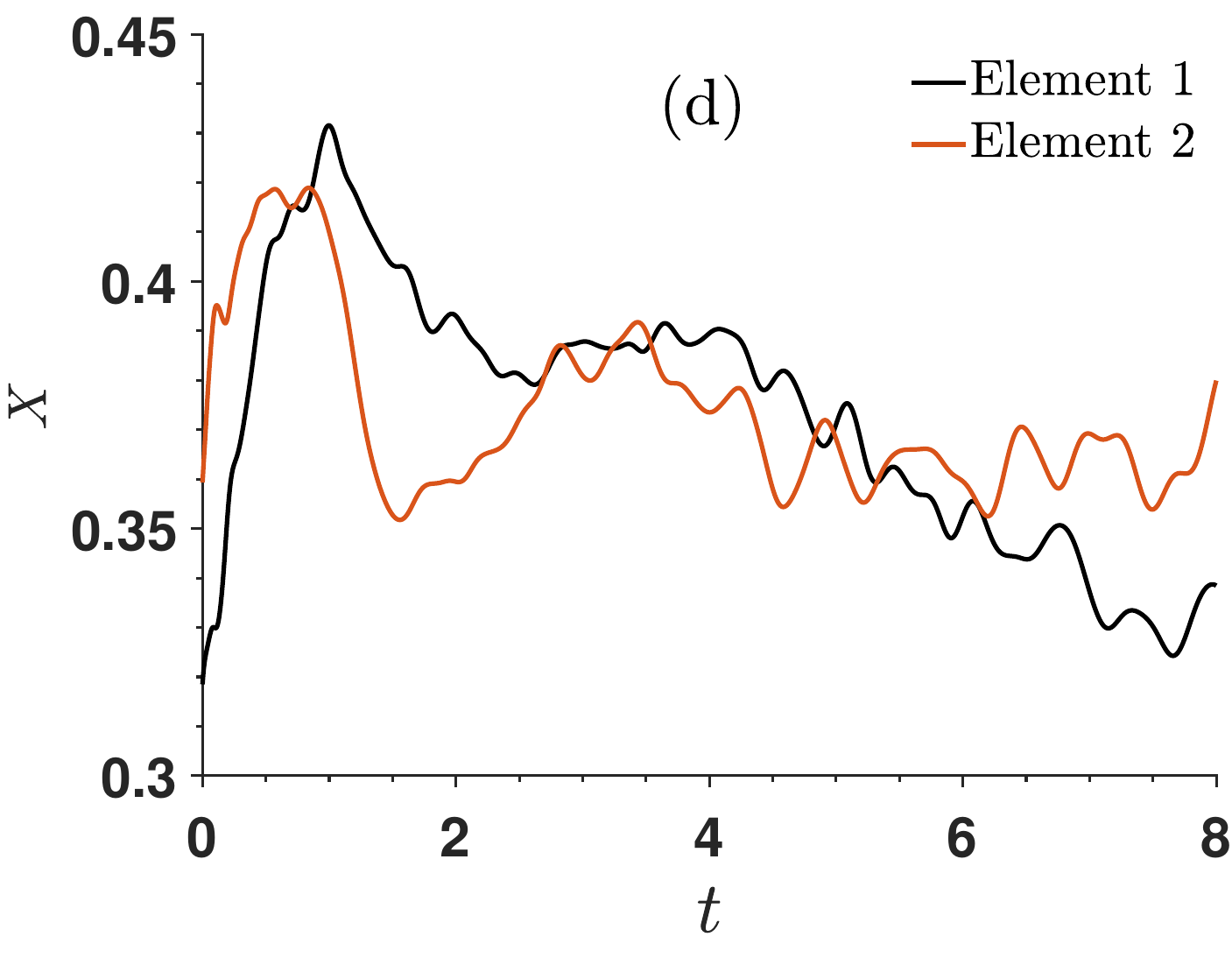}
    \end{subfigure}
    \vspace{1em}
    \begin{subfigure}[t]{0.48\textwidth}
        \centering
        \includegraphics[width=\textwidth,height=1\textwidth,keepaspectratio]{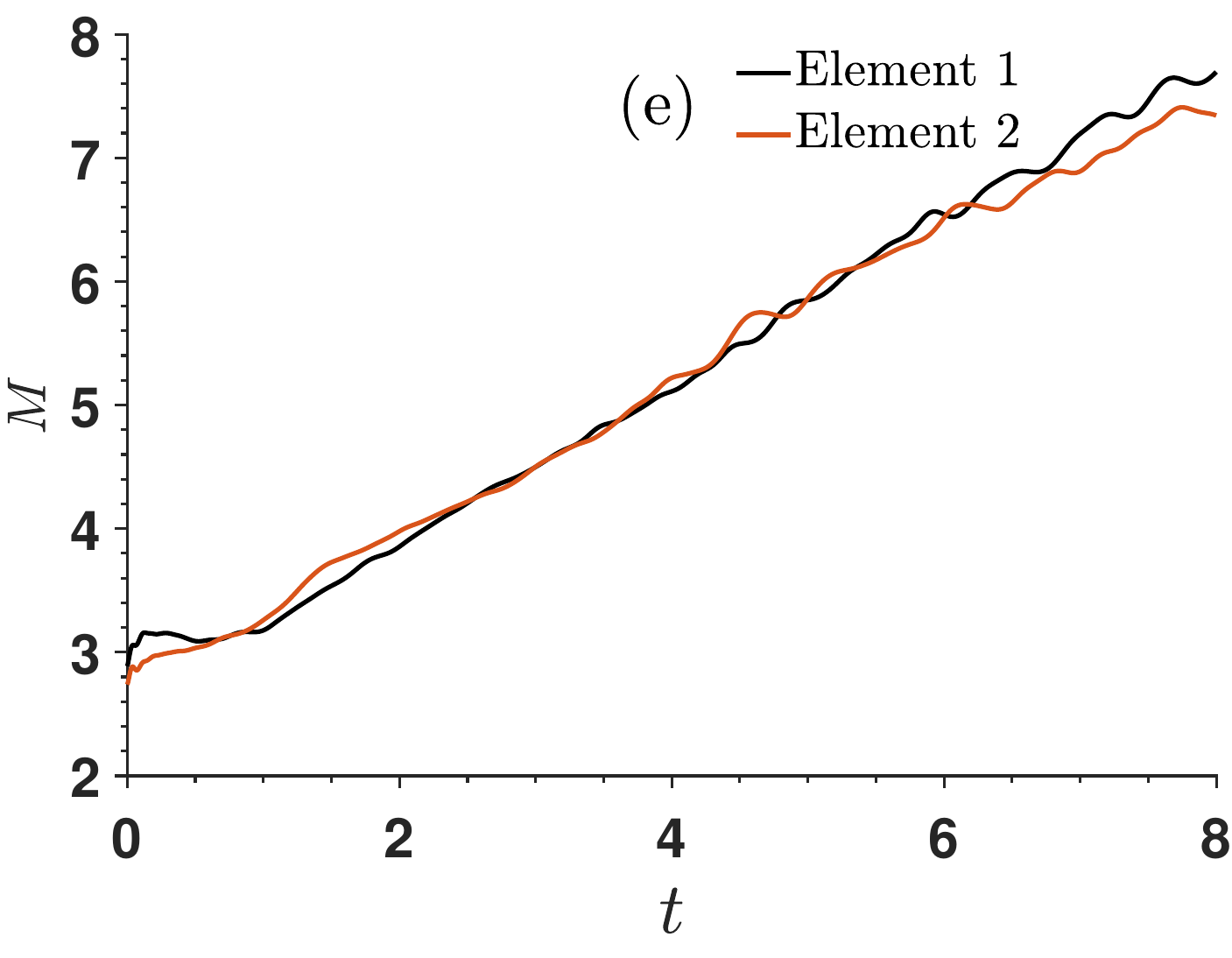}
    \end{subfigure}
    \caption{Temporal evolution of the STDEMF model parameters for two distinct elements in the $6\mathcal{P}7$ case of HID flow: (a) $\Delta/\eta$, (b) $X_\Omega$, (c) $X_S$, (d) $X$, and (e) the cut-off mode $M$.}
    \label{fig:isomodelparameters}
\end{figure}

The advantages of the STDEMF model stem from its ability to self-select the cut-off mode and, consequently, the amount of dissipation for different unresolved elements, preventing both under-dissipation and over-dissipation. As an example of how the model evolves over time for different elements, Fig. \ref{fig:isomodelparameters} illustrates the dynamic behavior of the cut-off mode and the associated parameters for two arbitrarily-chosen elements in the coarse-DNS case. As shown in Fig. \ref{fig:isomodelparameters}(a), at early times, the ratio $\Delta/\eta$ is large for the two locations, indicating a higher degree of flow unresolvedness. This requires greater artificial dissipation, resulting in lower cut-off modes, $M$, at initial times. As the flow develops, $\eta$ increases and the ratio $\Delta/\eta$ decreases. Additionally, $X_S$, $X_\Omega$, and $X$ are shown in Fig. \ref{fig:isomodelparameters}(b)–(d). Although $\eta$ exhibits similar behavior for both elements, $X_S$, $X_\Omega$, and consequently $X$, which captures the combined effects of shear and rotation, evolve differently over time. Considering the temporal variations of $\Delta/\eta$ and $X$, the cut-off mode $M$ shows an overall increasing trend, causing reduced energy removal over time, while still displaying differences between the elements. This indicates that even when the Kolmogorov length scales become similar at later times, variations in $X$ continue to adjust the cut-off mode and, therefore, the amount of energy removed.

\begin{figure}[!ht]
    \centering
    \includegraphics[width=0.8\textwidth]{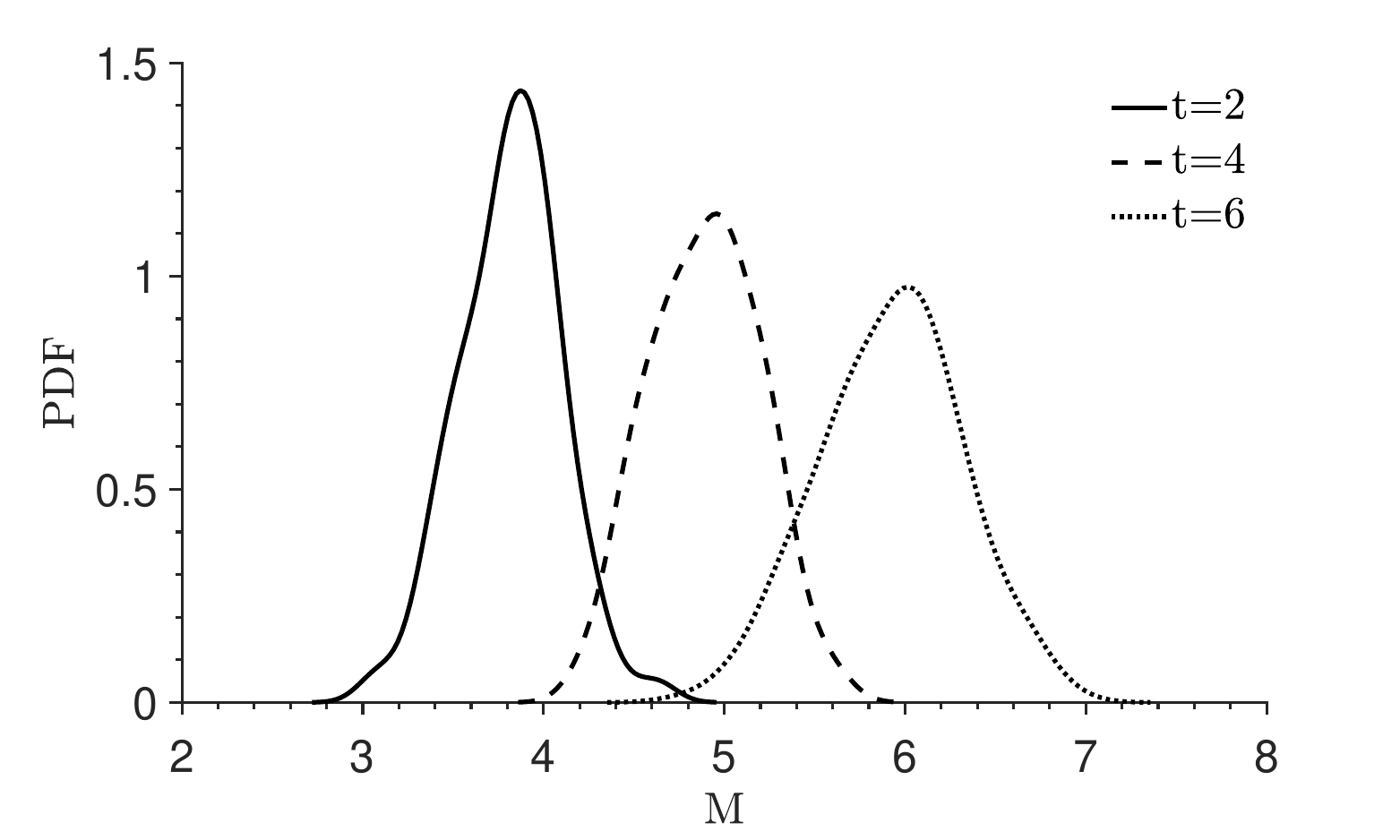}
    \caption{\label{fig:isoMPDF} Probability density function of the cut-off mode for the $6\mathcal{P}7$ case of HID flow at different times.}
\end{figure}

Figure \ref{fig:isoMPDF} shows the probability density function (PDF) of the cut-off mode $M$ for all elements of the domain in the $6\mathcal{P}7$ case of HID flow at $t=2$, $t=4$, and $t=6$. As turbulence decays, the distribution of $M$ shifts toward higher values, leading to reduced energy removal. The mean (standard deviation) of $M$ increases from $3.815$ (0.078) at $t=2$ to $4.895$ (0.0913) at $t=4$ and $5.923$ (0.153) at $t=6$. These results indicate that spatial variations between elements remain small while the flow evolves in time, with dissipation decreasing as $M$ increases. The lower cut-off modes observed in the STDEMF model, compared to the DEMF model arise from the use of a tangent hyperbolic kernel. This kernel exponentially attenuates the magnitudes of the modes above the cut-off, reducing them to zero at the highest mode. In contrast, the cut-off filter kernel employed in the DEMF model sets all modes above the threshold abruptly to zero.

\begin{figure}[htbp]
    \centering
    \begin{subfigure}[t]{0.48\textwidth} 
        \includegraphics[width=\textwidth,height=1\textwidth,keepaspectratio]{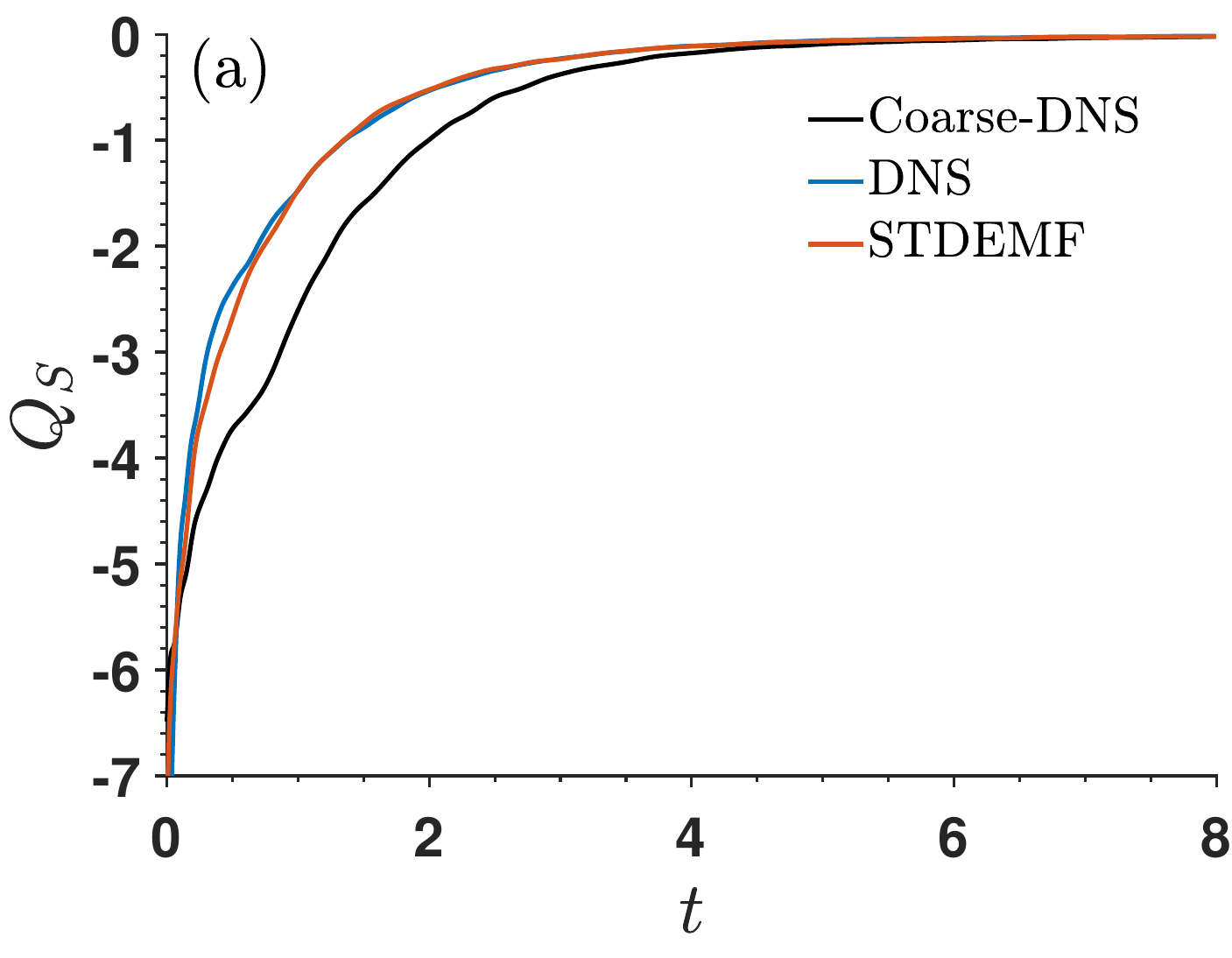}
    \end{subfigure}
    \hspace{0.02\textwidth}
    \begin{subfigure}[t]{0.48\textwidth}
        \includegraphics[width=\textwidth,height=1\textwidth,keepaspectratio]{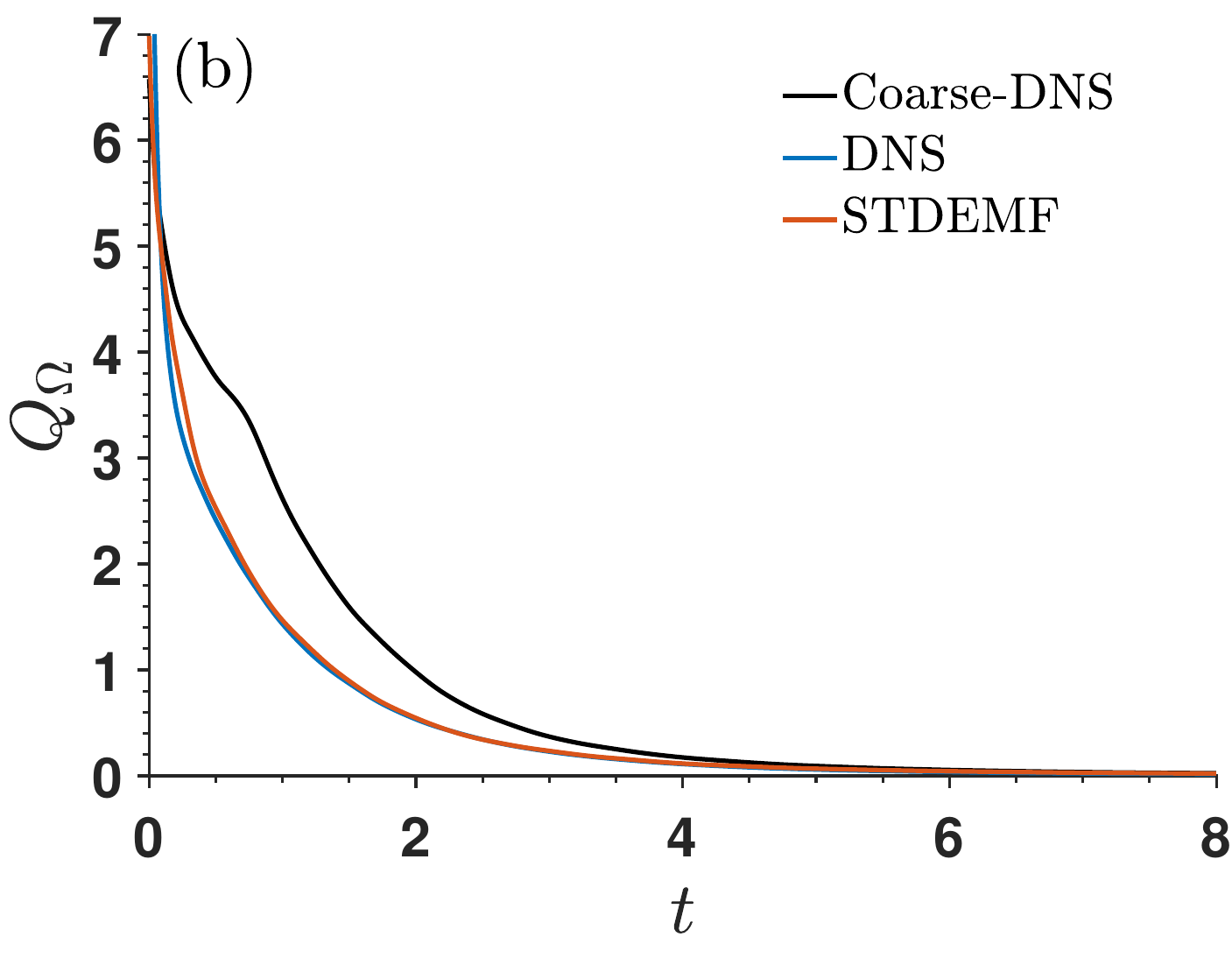}
    \end{subfigure}
    \vspace{1em}
    \begin{subfigure}[t]{0.48\textwidth} 
        \includegraphics[width=\textwidth,height=1\textwidth,keepaspectratio]{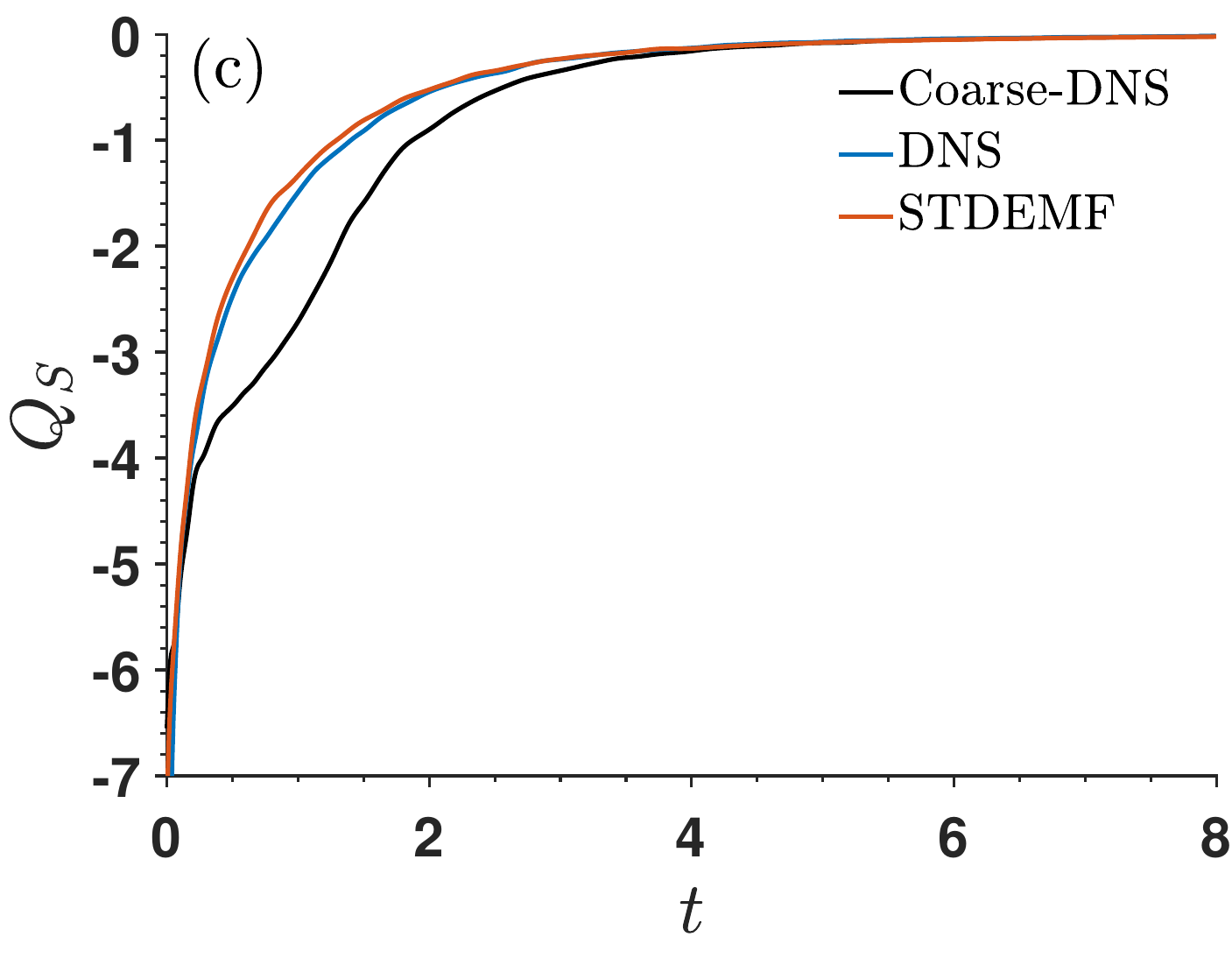}
    \end{subfigure}
    \hspace{0.02\textwidth}
    \begin{subfigure}[t]{0.48\textwidth}
        \includegraphics[width=\textwidth,height=1\textwidth,keepaspectratio]{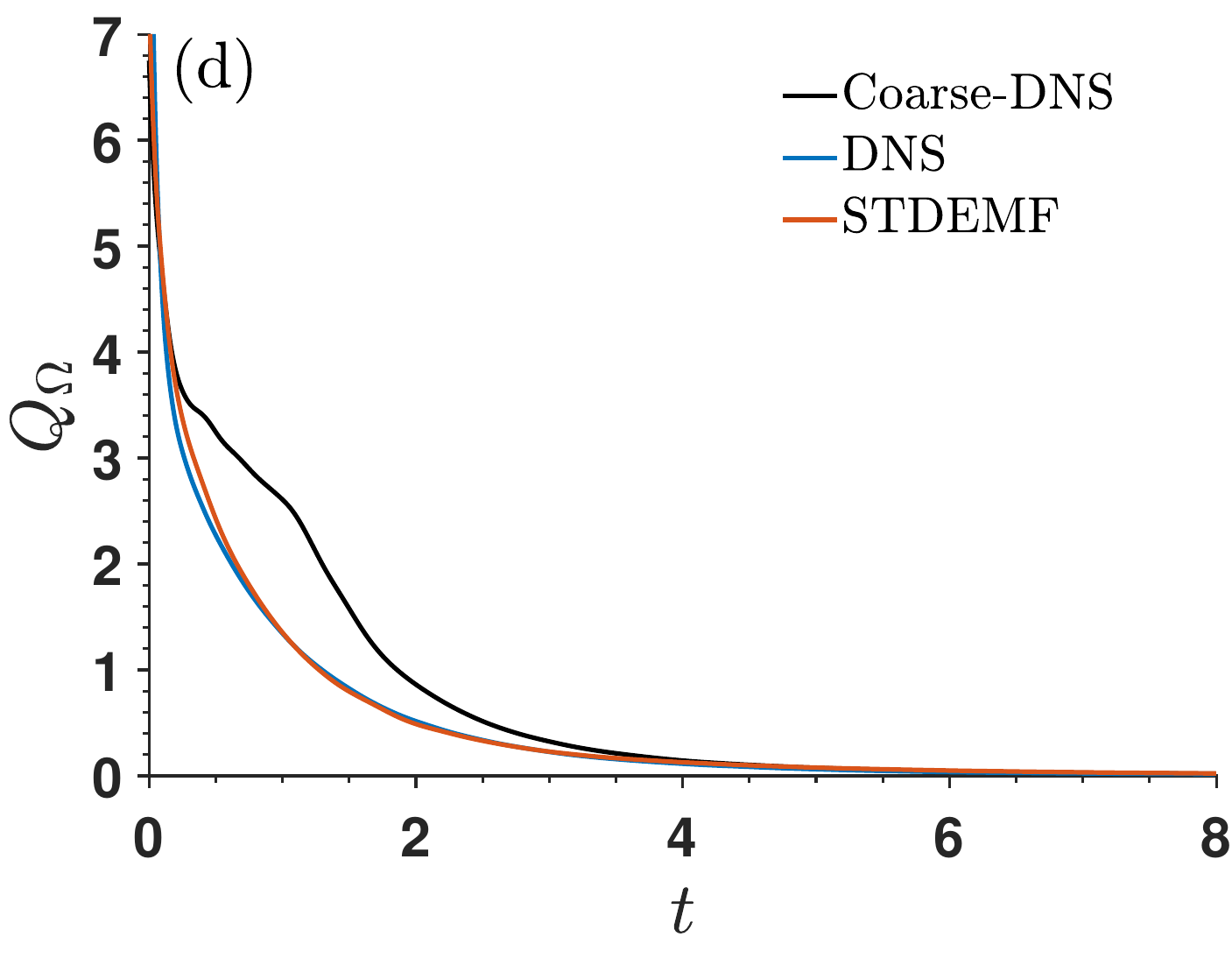}
    \end{subfigure}
    \caption{Element-level comparison of DNS, coarse-DNS, and the STDEMF model for (a) $Q_S$ in element $1$, (b) $Q_{\Omega}$ in element $1$, (c) $Q_S$ in element $2$, and (d) $Q_{\Omega}$ in element $2$, with the same elements of Fig. \ref{fig:isomodelparameters}.}
    \label{fig:element-level model assessment}
\end{figure}

To further examine how the proposed model operates within individual elements, element-level results are provided for the two elements studied in Fig. \ref{fig:isomodelparameters}. Figure \ref{fig:element-level model assessment} compares the model parameters $Q_S$ and $Q_\Omega$ from the DNS, the coarse-DNS, and the STDEMF model. It shows that in the coarse-DNS case, due to the lack of small-scale mechanisms, bigger magnitudes are obtained for both $Q_S$ and $Q_{\Omega}$ in both elements, particularly before $t=5$. In contrast, although the coarse-mesh resolution is one order of magnitude less than the DNS mesh resolution, the model parameters calculated with the STDEMF model are very close to the DNS results. 

The HID flow considered here is initialized with small-scale isotropic turbulence and, as discussed above, evolves primarily in time, with only minor spatial variations across the domain. To further evaluate the model’s performance, it is also applied to a flow that undergoes transition to turbulence and exhibits significant spatial variations in turbulent properties.

\subsection{ Taylor-Green Vortex (TGV)}

The TGV flow, studied extensively in the literature \citep{taylor_1937, Fehn_2021, Manrique_de_Lara_2023, Manzanero_2020}, represents a flow that initially exhibits a laminar state, transitions to turbulence, and eventually decays into an isotropic turbulence state that dissipates due to viscosity. The flow evolution introduces spatial discrepancies in turbulence properties, which, on a coarse mesh, lead to resolved and unresolved regions \citep{Ranjbar_2024POF}. Consequently, the TGV serves as an ideal test case for evaluating the STDEMF model across laminar, transitional, and turbulent regimes, ensuring that the model is deactivated in laminar and well-resolved turbulent regions through the sensor. 

\begin{figure}[!ht]
    \centering
    \includegraphics[width=0.85\textwidth]{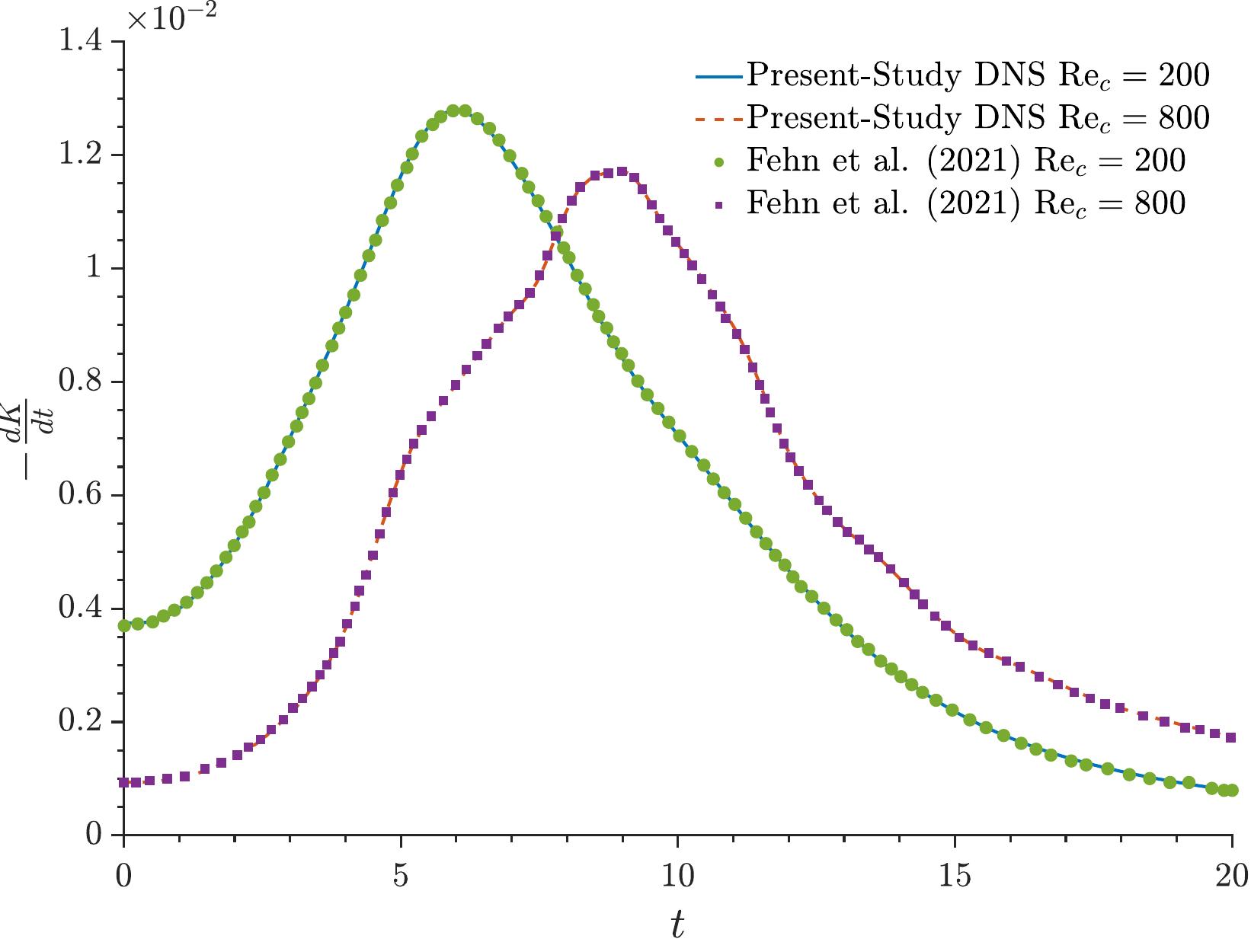}
    \caption{\label{fig:tgvvalid} Comparison of the temporal evolution of the kinetic energy dissipation rate between the present-study DNS and the results of \cite{Fehn_2021} in the TGV flow.}
\end{figure}

The TGV is initialized with a periodic flow field within a computational domain of $V=[0, 2\pi]^3$ using the initial conditions 
\begin{equation}
\begin{aligned}
    &\rho = 1, \\
    &u = \sin\left(x\right) \cos\left(y\right) \cos\left(z\right), \\
    &v = -\cos\left(x\right) \sin\left(y\right) \cos\left(z\right), \\
    &w = 0, \\
    &P = \frac{100}{\gamma} + \frac{1}{16}\left[\cos\left(2x\right) + \cos\left(2y\right)\right]\left[\cos\left(2z\right) + 2\right].
\end{aligned}
\end{equation}
Two Reynolds numbers, $\mathrm{Re_c} = 200$ and $\mathrm{Re_c} = 800$, are simulated using the resolutions reported by \cite{Fehn_2021}. For $\mathrm{Re_c} = 200$, the suggested resolutions of $128^3$ and $256^3$ are employed, while for $\mathrm{Re_c} = 800$, the suggested resolutions of $256^3$ and $512^3$ are used. The details of the DNS and coarse-DNS grids used for model validation are provided in Table \ref{tab:tgvcases}. Consistent with \cite{Fehn_2021}, the two resolutions in each case yield identical results for the evolution of the kinetic energy and its dissipation rate. Validation against the results of \citet{Fehn_2021} demonstrates that the temporal evolution of the kinetic energy dissipation rate at both Reynolds numbers is in close agreement with the reference results (Fig. \ref{fig:tgvvalid}).
 \begin{table}[h]
    \centering
    \caption{Details of the grids used for different cases of TGV flow.}
    \begin{tabular}{c@{\hspace{0.7cm}}c@{\hspace{0.7cm}}c@{\hspace{0.7cm}}c@{\hspace{0.7cm}}c@{\hspace{0.7cm}}c}
        \toprule
        Case& $\mathrm{Re_c}$ &  Case ID& $\mathcal{P}$ & Number of elements & Number of grid points   \\
        \hline
         TGV - DNS & 200& 18$\mathcal{P}$7 & 7 & $18\times18\times18$ & 2,985,984  \\

         TGV - Coarse DNS & 200 & 11$\mathcal{P}$4 & 4 &  $11\times11\times11$ & 166,375 \\
         
         TGV - DNS & 800&  20$\mathcal{P}$12 &  12 &  $20\times20\times20$& 17,576,000\\
         
         TGV - Coarse DNS & 800& 7$\mathcal{P}$10 & 10 &  $7\times7\times7$ & 456,533\\
        
        \hline
        \hline
    \end{tabular}
    \label{tab:tgvcases}
\end{table}
 
\begin{figure}[htbp]
    \centering
    \begin{subfigure}[t]{0.48\textwidth}
        \includegraphics[width=\textwidth,height=0.95\textwidth]
        {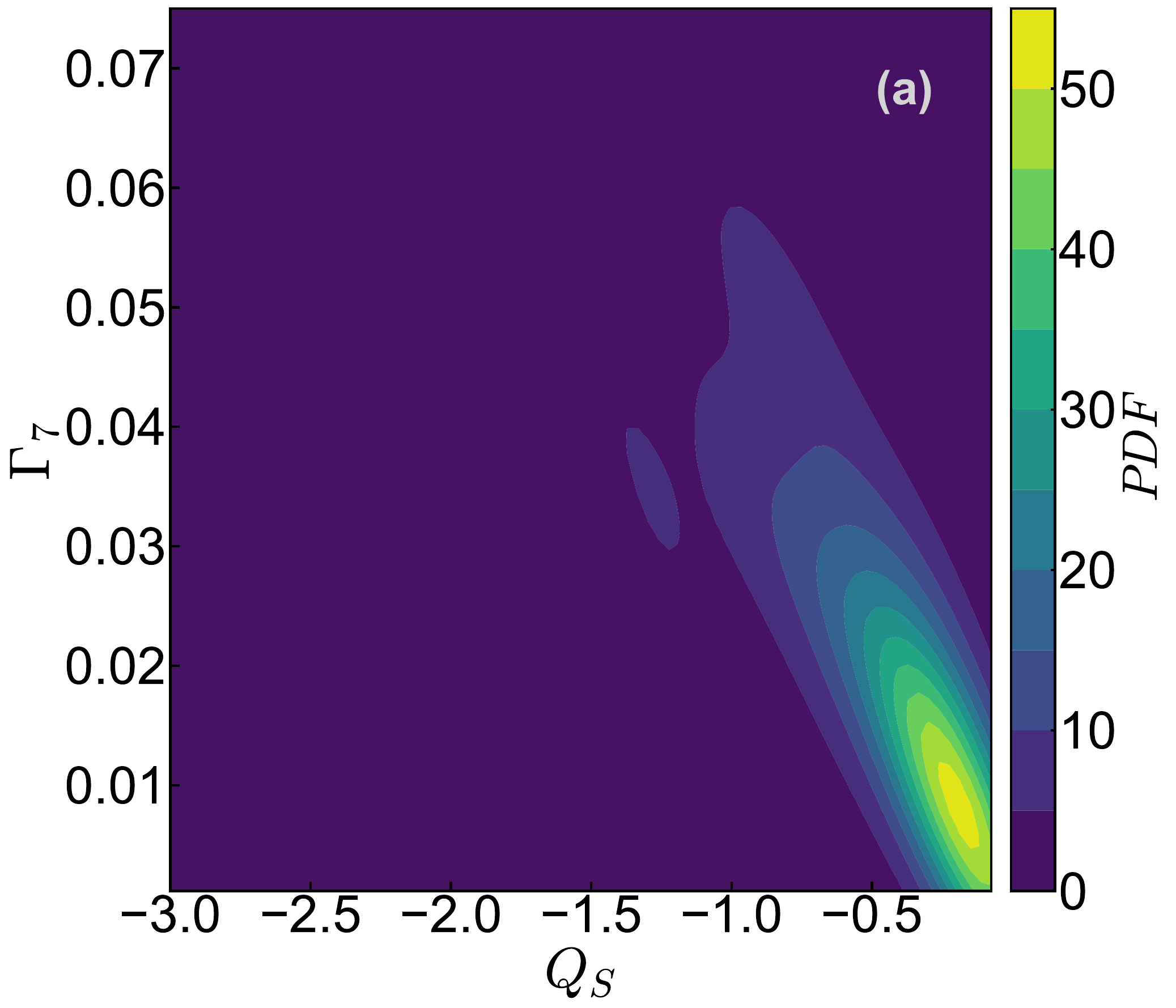}
    \end{subfigure}
    \hspace{0.02\textwidth}
    \begin{subfigure}[t]{0.48\textwidth}
        \includegraphics[width=\textwidth,height=0.965\textwidth]{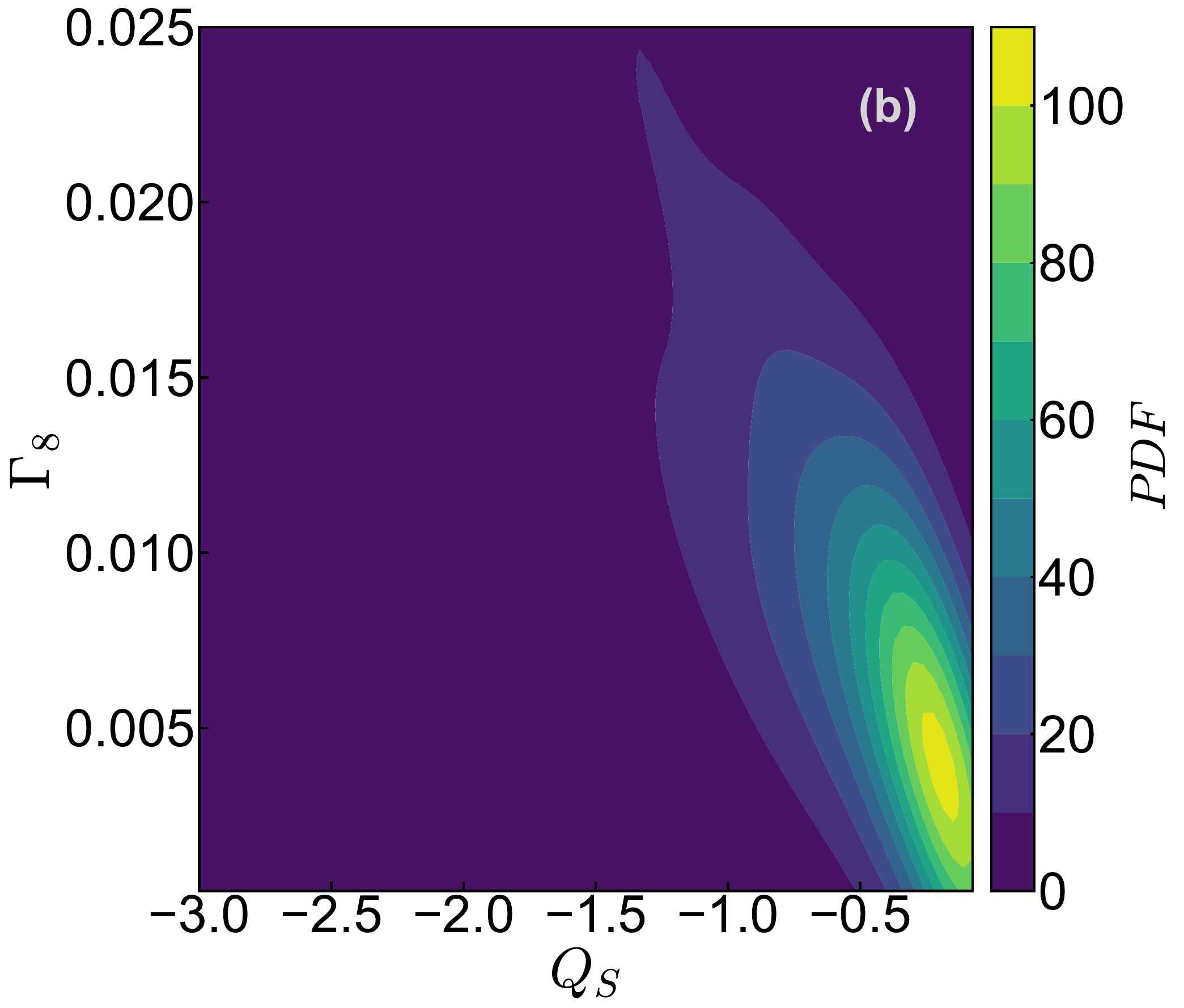}
    \end{subfigure}
    \vspace{1em}
    \begin{subfigure}[t]{0.48\textwidth}
        \includegraphics[width=\textwidth,height=0.95\textwidth]{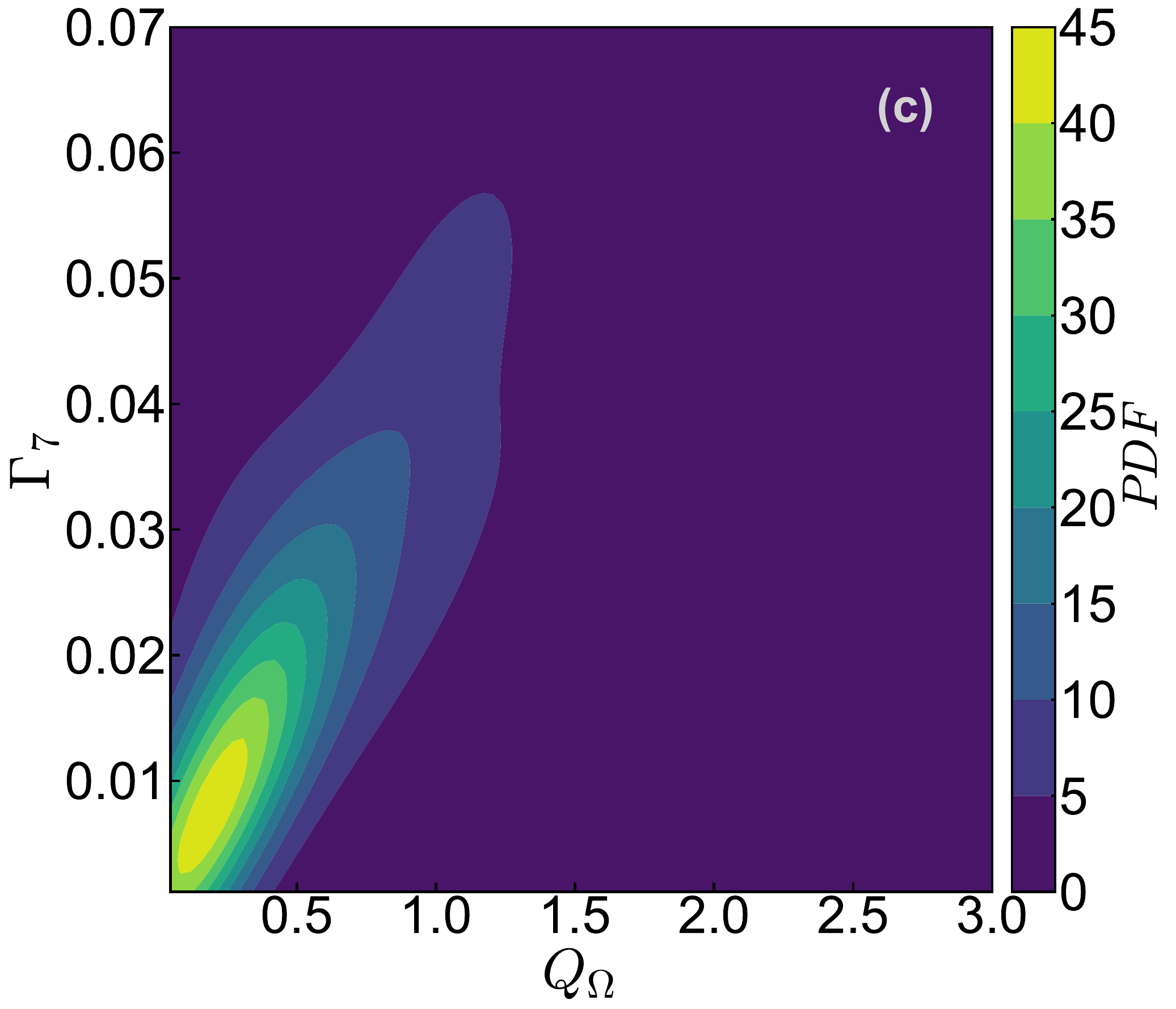}
    \end{subfigure}
    \hspace{0.02\textwidth}
    \begin{subfigure}[t]{0.48\textwidth}
        \includegraphics[width=\textwidth,height=0.965\textwidth]{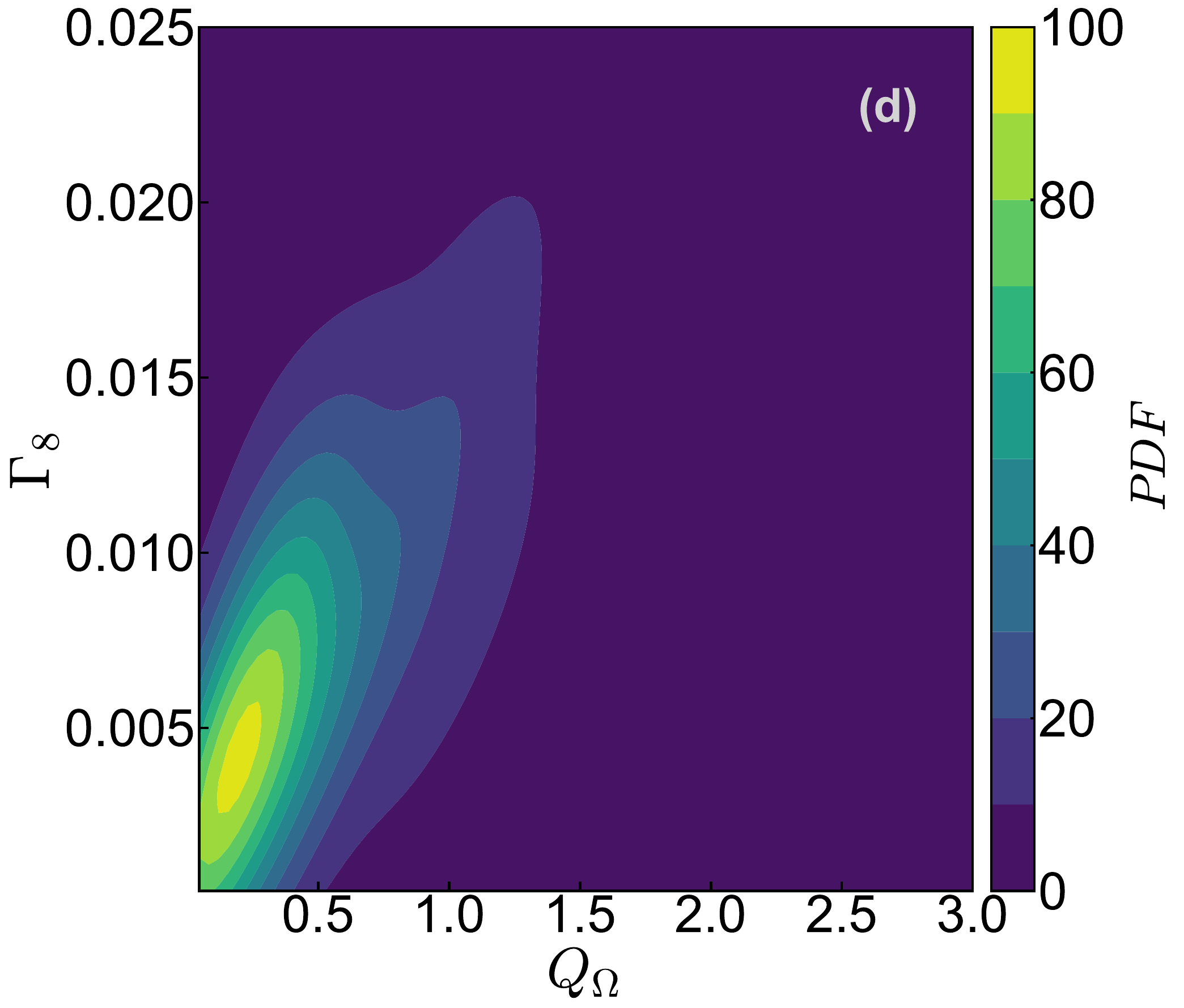}
    \end{subfigure}
    \caption{Joint PDFs of (a) $\Gamma_{7}$ vs.\ $Q_S$, (b) $\Gamma_{8}$ vs.\ $Q_S$, (c) $\Gamma_{7}$ vs.\ $Q_{\Omega}$, and (d) $\Gamma_{8}$ vs.\ $Q_{\Omega}$ for the $11\mathcal{P}4$ TGV case at $t=6$ ($\Gamma$'s corresponding to $x-$ component velocity).}
    \label{fig:elemenetlevelmodalassessmentRe200}
\end{figure}

\begin{figure}[htbp]
    \centering
    \begin{subfigure}[t]{0.48\textwidth}
        \includegraphics[width=\textwidth,height=0.95\textwidth]
        {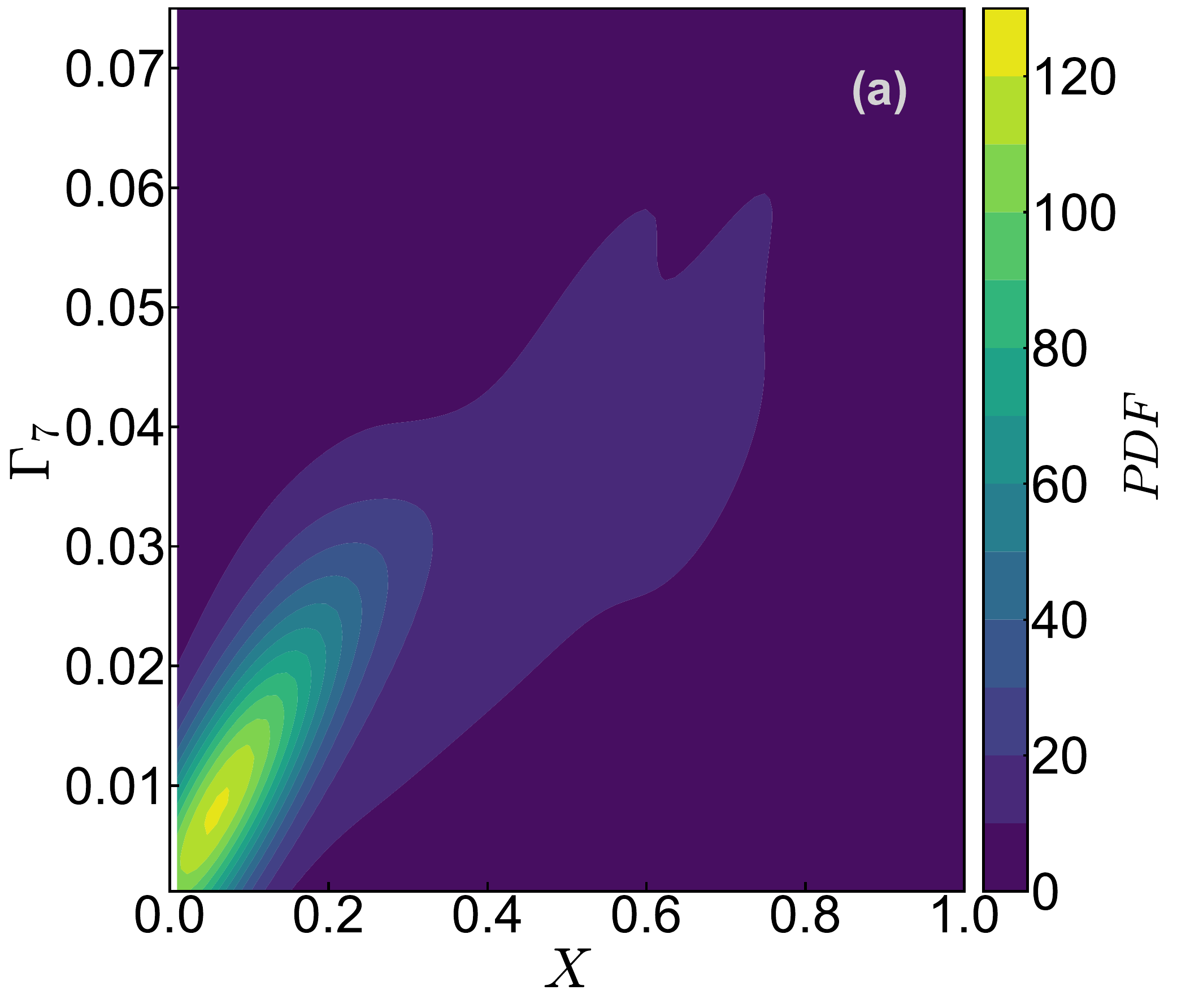}
    \end{subfigure}
    \hspace{0.02\textwidth}
    \begin{subfigure}[t]{0.48\textwidth}
        \includegraphics[width=\textwidth,height=0.965\textwidth]{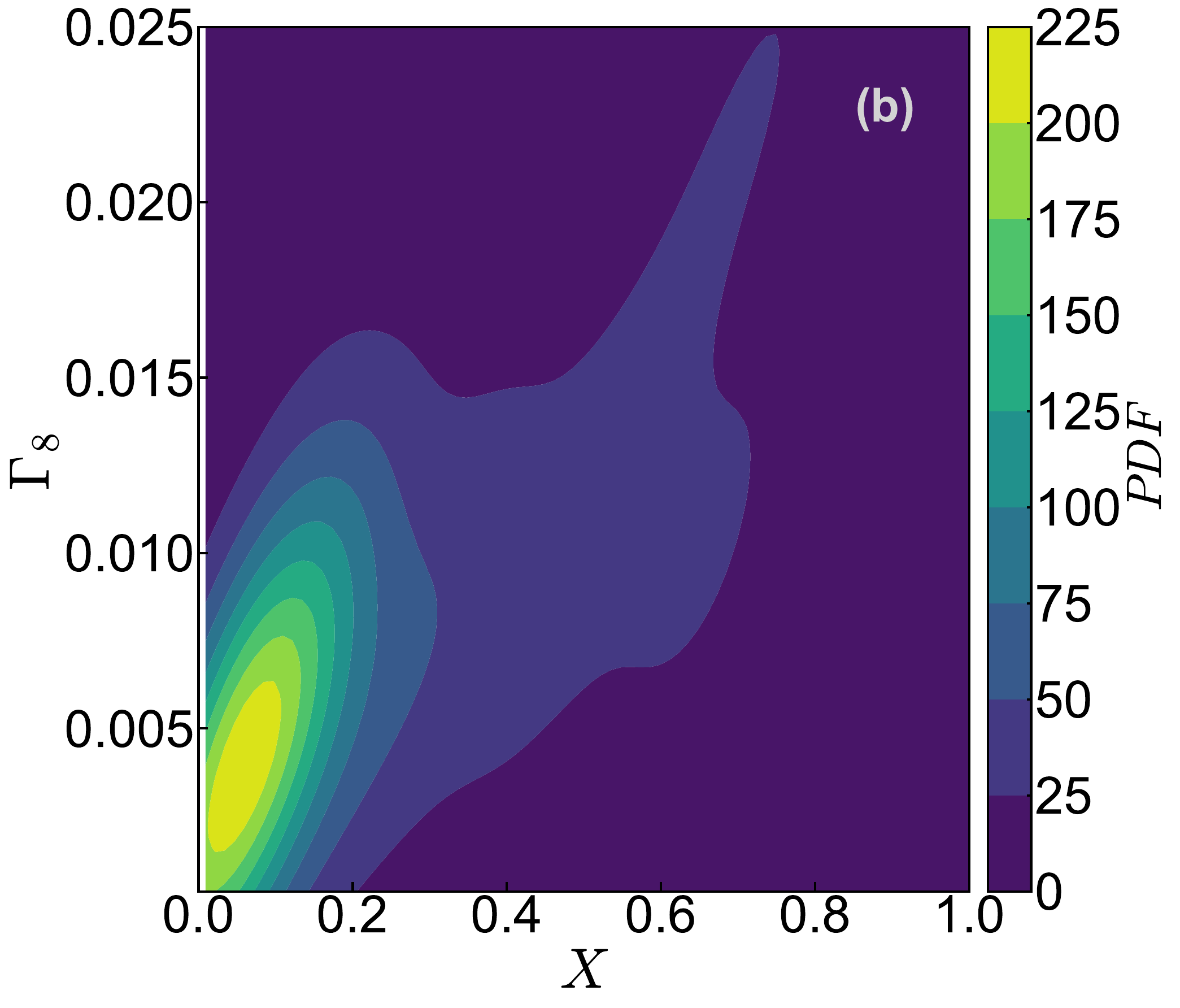}
    \end{subfigure}
    \caption{Joint PDFs of (a) $\Gamma_{7}$ vs.\ $X$ and (b) $\Gamma_{8}$ vs.\ $X$ for the $11\mathcal{P}4$ TGV case at $t=6$ ($\Gamma$'s corresponding to $x-$ component velocity).}
    \label{fig:GammavsXRe200}
\end{figure}

Figure \ref{fig:elemenetlevelmodalassessmentRe200} presents the joint PDFs of $\Gamma_7$ and $\Gamma_8$ of the $x-$component velocity versus $Q_S$ and $Q_{\Omega}$ across the entire domain for the $11\mathcal{P}4$ coarse-mesh TGV at the time of peak dissipation ($t=6$) for $\mathrm{Re_c}=200$. With a polynomial order of $4$, there are $8$ energy levels ($\Gamma$'s) in this case. A similar behavior is observed in all four subfigures, While the peaks of the PDFs are concentrated near low values of $Q$'s and $\Gamma$'s, larger magnitudes of $Q$'s are associated with larger magnitudes of $\Gamma$'s, indicating that regions of high strain coupled with strong rotation exhibit greater energy accumulation when the resolution is insufficient to capture small-scale motions. To further verify this, the variable $X$, which quantifies the simultaneous presence of shear and rotation, is computed from $Q_S$ and $Q_{\Omega}$, and its joint PDFs with $\Gamma_7$ and $\Gamma_8$ are shown in Fig. \ref{fig:GammavsXRe200}. As seen in this figure, larger magnitudes of $\Gamma_7$ and $\Gamma_8$ correspond to larger values of $X$. This indicates that a DEMF model should complement the Kolmogorov length scale by including strain and rotational invariants, enabling dissipation to respond to local strain and rotation to avoid under- or over-dissipation in elements with similar Kolmogorov scales but different shear–rotation dynamics.

\begin{figure}[htbp]
    \centering
    \begin{subfigure}[t]{0.48\textwidth}
        \includegraphics[width=\textwidth,,height=0.95\textwidth]{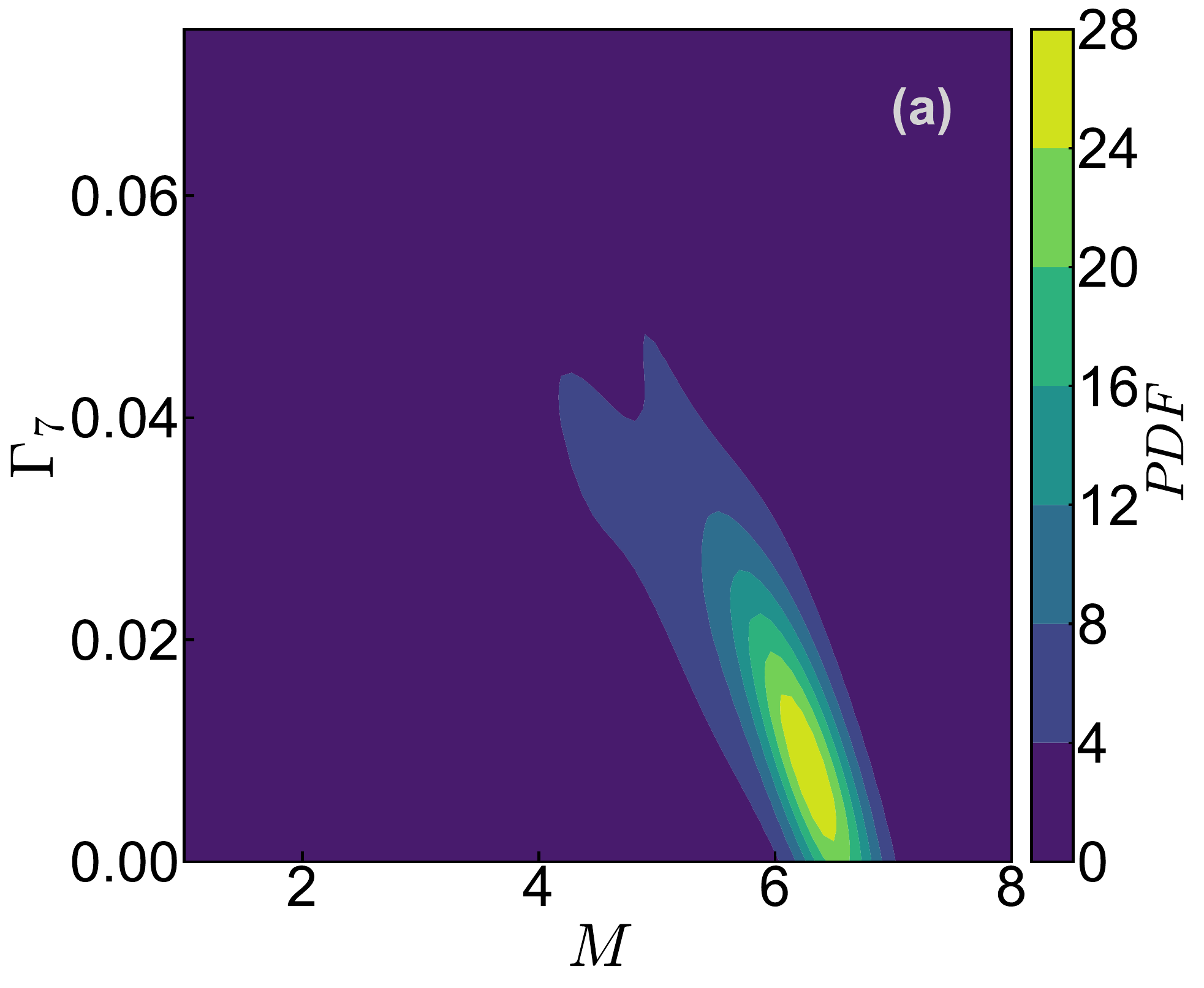}
    \end{subfigure}
    \hspace{0.02\textwidth}
    \begin{subfigure}[t]{0.48\textwidth}
        \includegraphics[width=\textwidth,,height=0.96\textwidth]{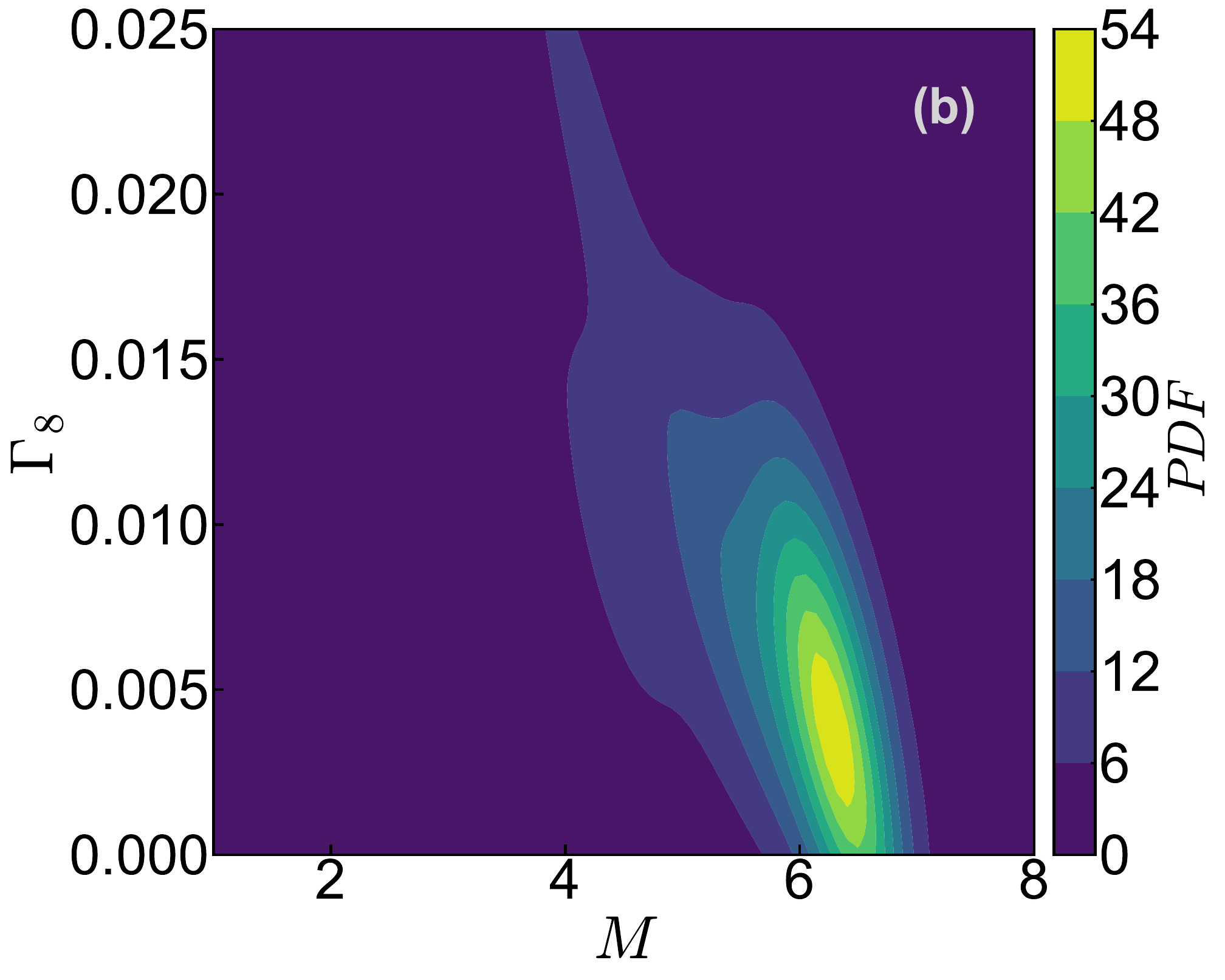}
    \end{subfigure}

    \vspace{1em}

    \begin{subfigure}[t]{0.48\textwidth}
        \includegraphics[width=\textwidth,,height=0.95\textwidth]{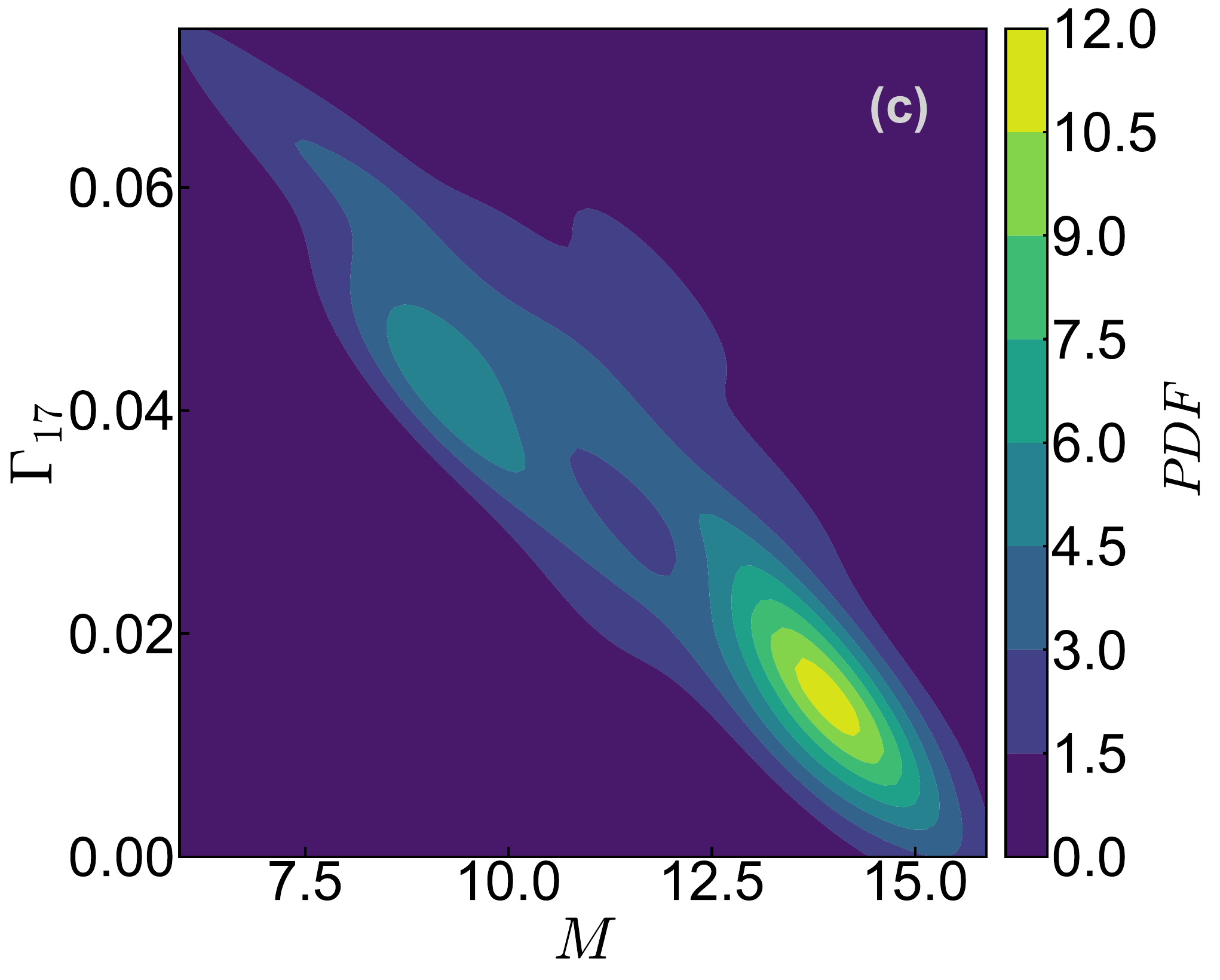}
    \end{subfigure}
    \hspace{0.02\textwidth}
    \begin{subfigure}[t]{0.48\textwidth}
        \includegraphics[width=\textwidth,,height=0.96\textwidth]{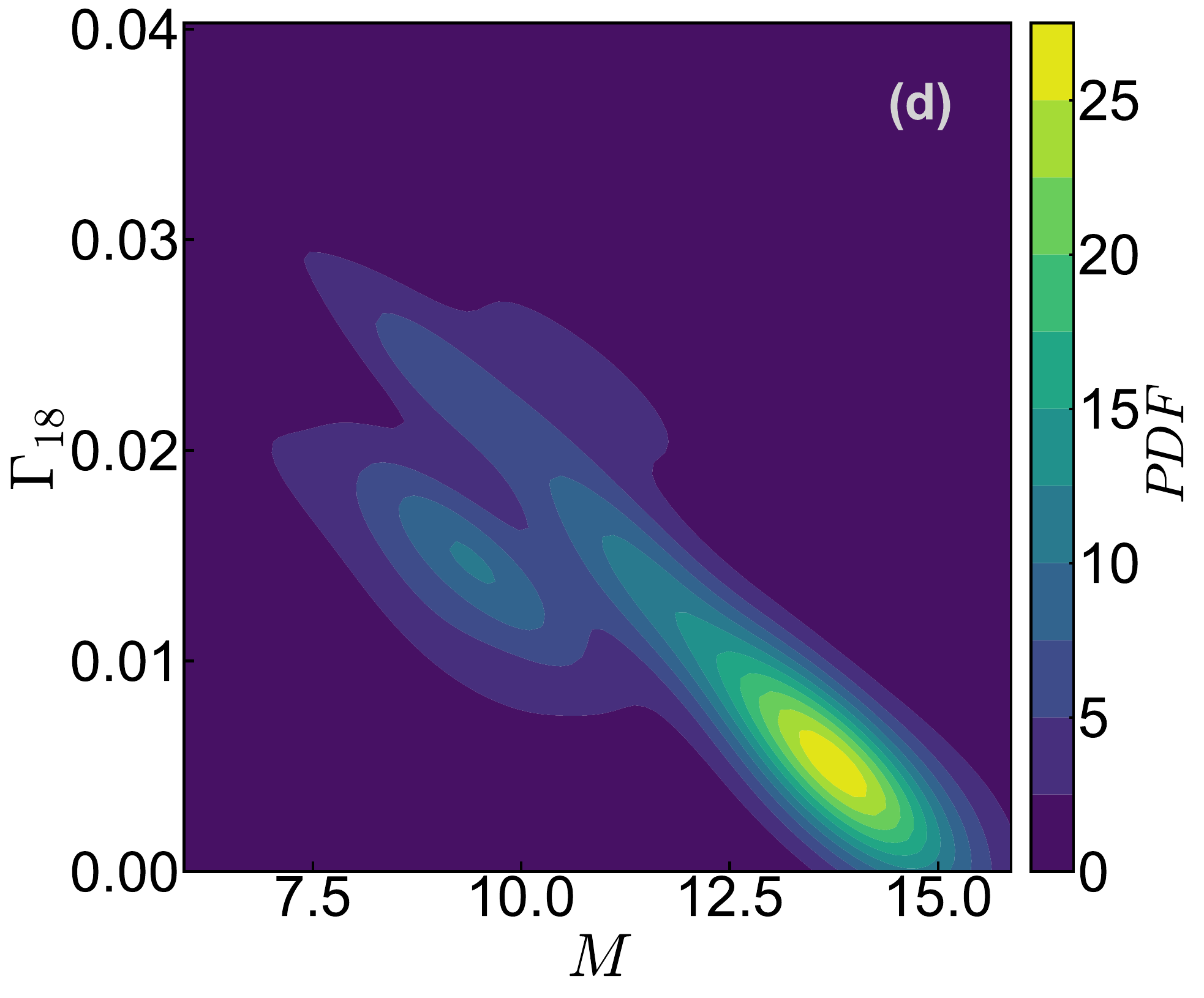}
    \end{subfigure}

    \caption{Joint PDFs of the cut-off mode $M$ vs (a) $\Gamma_7$ for the $11\mathcal{P}4$ TGV case at $t=6$, 
             (b) $\Gamma_8$ for the $11\mathcal{P}4$ TGV case at $t=6$, 
             (c) $\Gamma_{17}$ for the $7\mathcal{P}10$ TGV case at $t=9$, 
             and (d) $\Gamma_{18}$ for the $7\mathcal{P}10$ TGV case at $t=9$ ($\Gamma$'s corresponding to $x$-component velocity).}
    \label{fig:MvsGamma}
\end{figure}

\begin{figure}[htbp]
    \centering
    \begin{subfigure}[t]{0.85\textwidth} 
        \includegraphics[width=\textwidth]{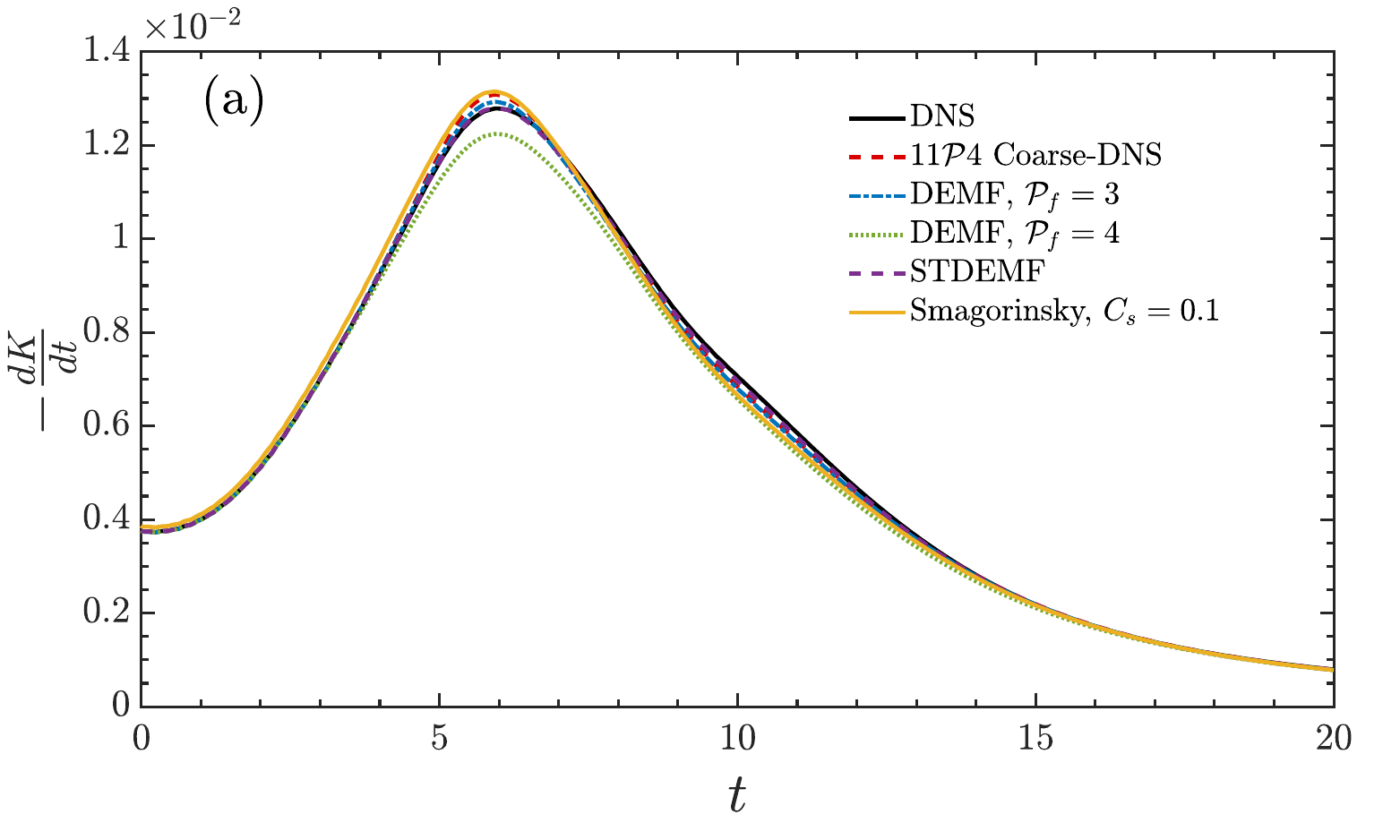}
    \end{subfigure}
    \vspace{1em}
    \begin{subfigure}[t]{0.85\textwidth}
        \includegraphics[width=\textwidth]{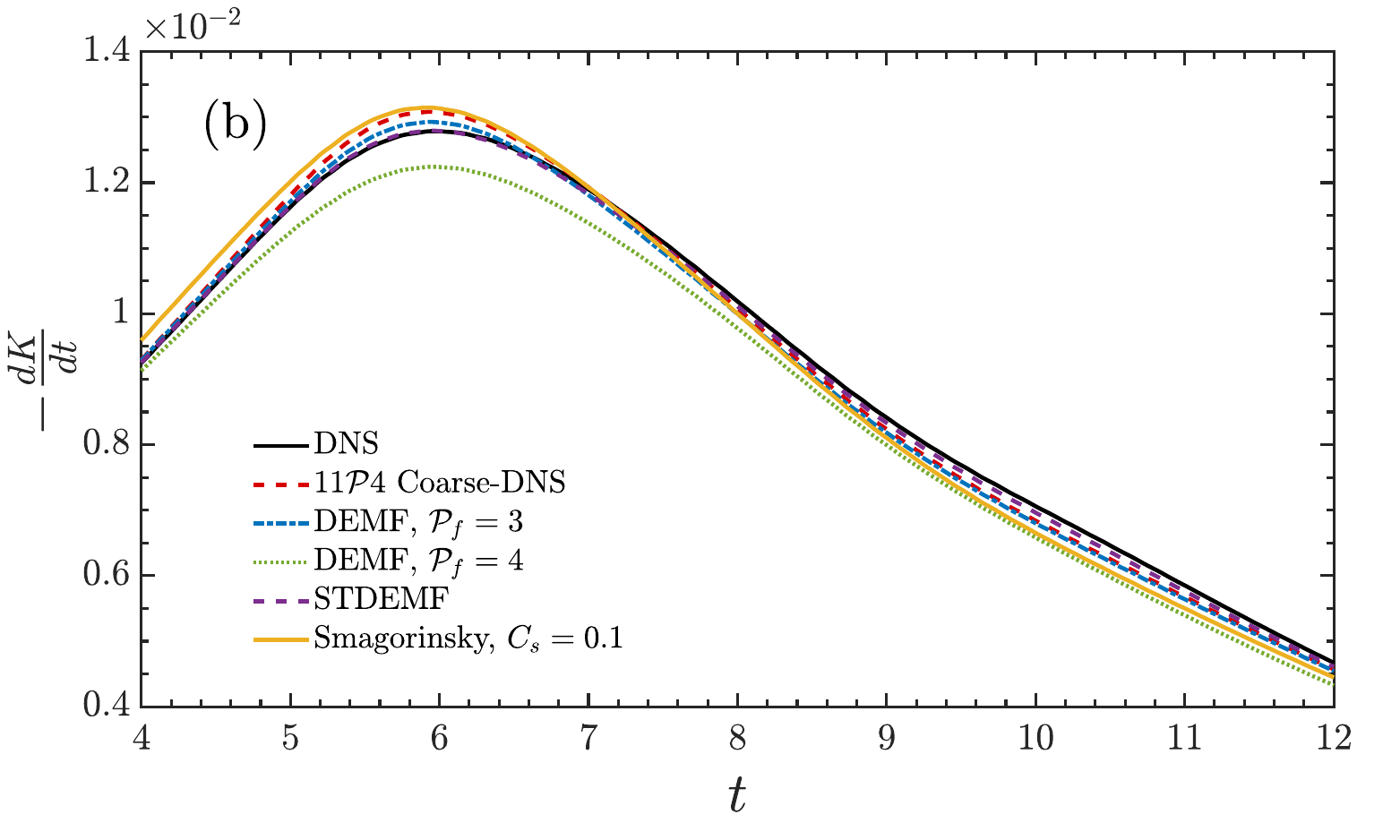}
    \end{subfigure}
    \caption{Comparison of the temporal evolution of the kinetic energy dissipation rate of the TGV flow for the $11\mathcal{P}4$ case at $\mathrm{Re}_c=200$ between the coarse DNS, DEMF model, STDEMF model, and the Smagorinsky eddy-viscosity model (a) over the entire simulation time and (b) zoomed in on $t=[4,12]$ for closer comparison.}
    \label{fig:Re200 model assessments}
\end{figure}

\begin{figure}[htbp]
    \centering
        \includegraphics[width=0.85\textwidth]{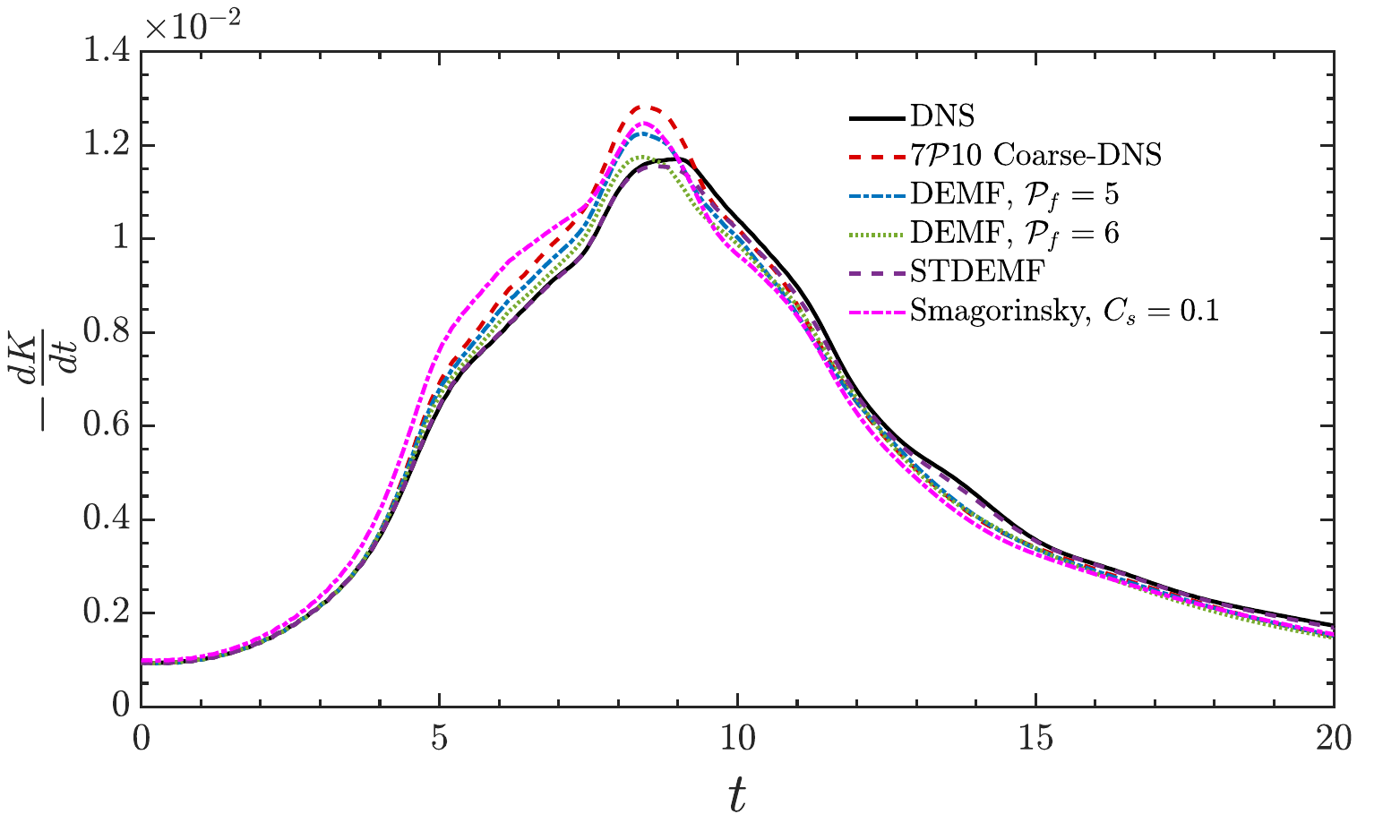}
    \caption{ Comparison of the temporal evolution of the kinetic energy dissipation rate between the coarse DNS, DEMF model, STDEMF model, and the Smagorinsky eddy-viscosity model for the $7\mathcal{P}10$ case of the TGV flow at $\mathrm{Re}_c=800$.}
    \label{fig:Re800 model assessments}
\end{figure}

Having examined the relationship between $X$ and the higher-index $\Gamma$'s, we now evaluate whether the cut-off mode in Eq. (\ref{eq:cutoff mode}) correctly captures the flow physics. Larger energy build-up, indicated by bigger magnitudes of higher-index $\Gamma$, is expected to correspond to more dissipation through lower cut-off modes. To verify this, Fig. \ref{fig:MvsGamma} presents the joint PDFs of higher-index $\Gamma$ versus the cut-off mode $M$ for the entire domain of the $11\mathcal{P}4$ case at $\mathrm{Re_c}=200$ and $t=6$, and the $7\mathcal{P}10$ case at $\mathrm{Re_c}=800$ and $t=9$. From all four subfigures, larger $\Gamma$ values are associated with lower cut-off modes. Although there are some differences between the PDF contours in the $7\mathcal{P}10$ and $11\mathcal{P}4$ cases, the same overall trend is observed, with high magnitudes of the higher-index $\Gamma$ corresponding to lower cut-off modes. This confirms that the modal filter dissipates more energy in elements with a higher degree of unresolved turbulence and greater energy accumulation. The higher cut-off modes for the $\mathrm{Re_c}=800$ case arise from the higher polynomial order used compared to the $\mathrm{Re_c}=200$ case. The constant $c=0.25$, which provides optimal agreement with DNS (Figs. \ref{fig:Re200 model assessments} and \ref{fig:Re800 model assessments}), is used for both cases in Eq. (\ref{eq:cutoff mode}).

\begin{figure}[htbp]
    \begin{subfigure}[t]{0.48\textwidth} 
        \includegraphics[width=\textwidth,height=1\textwidth,keepaspectratio]{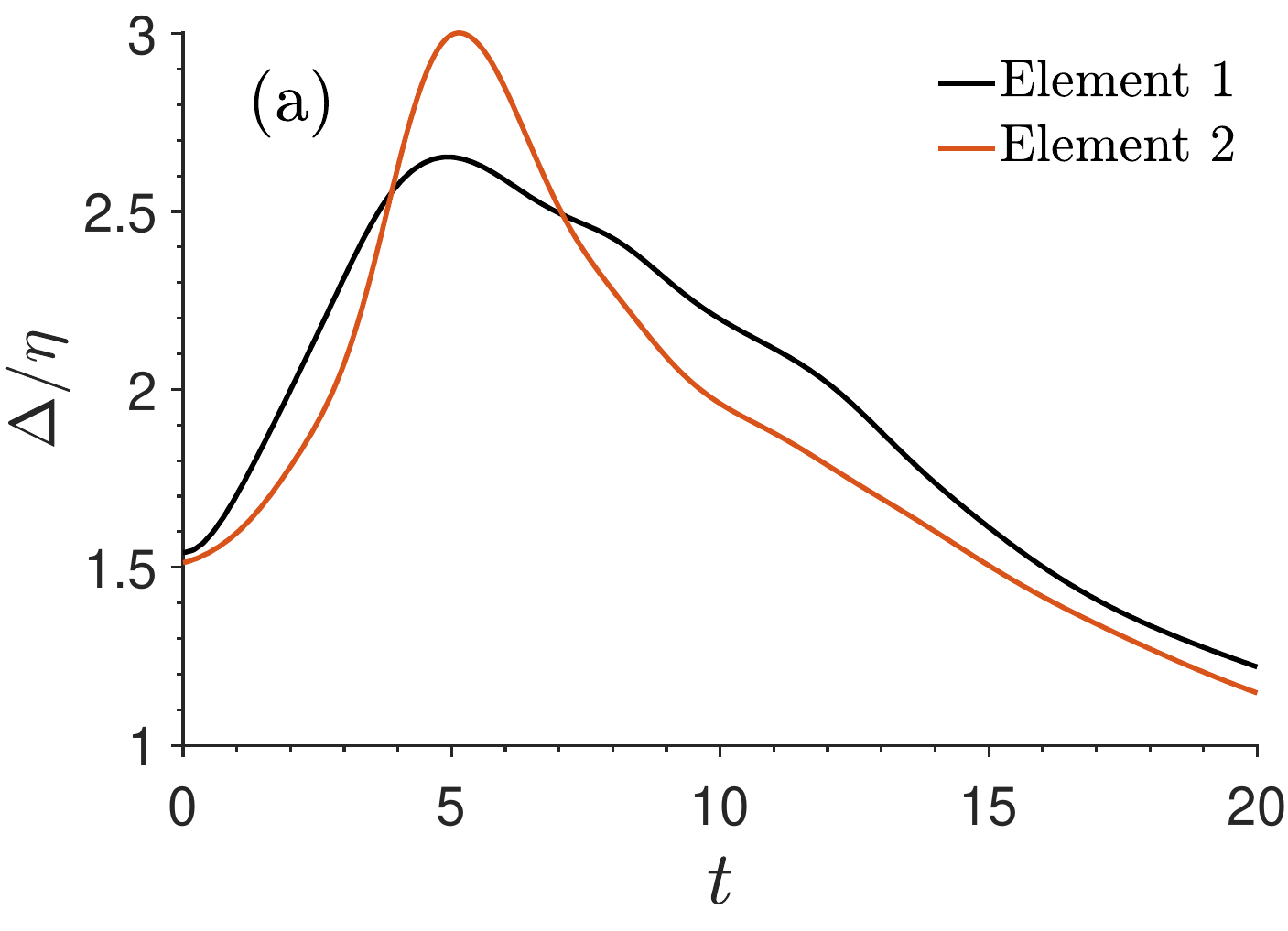}
    \end{subfigure}
    \hspace{0.02\textwidth}
    \begin{subfigure}[t]{0.48\textwidth}
        \includegraphics[width=\textwidth,height=1\textwidth,keepaspectratio]{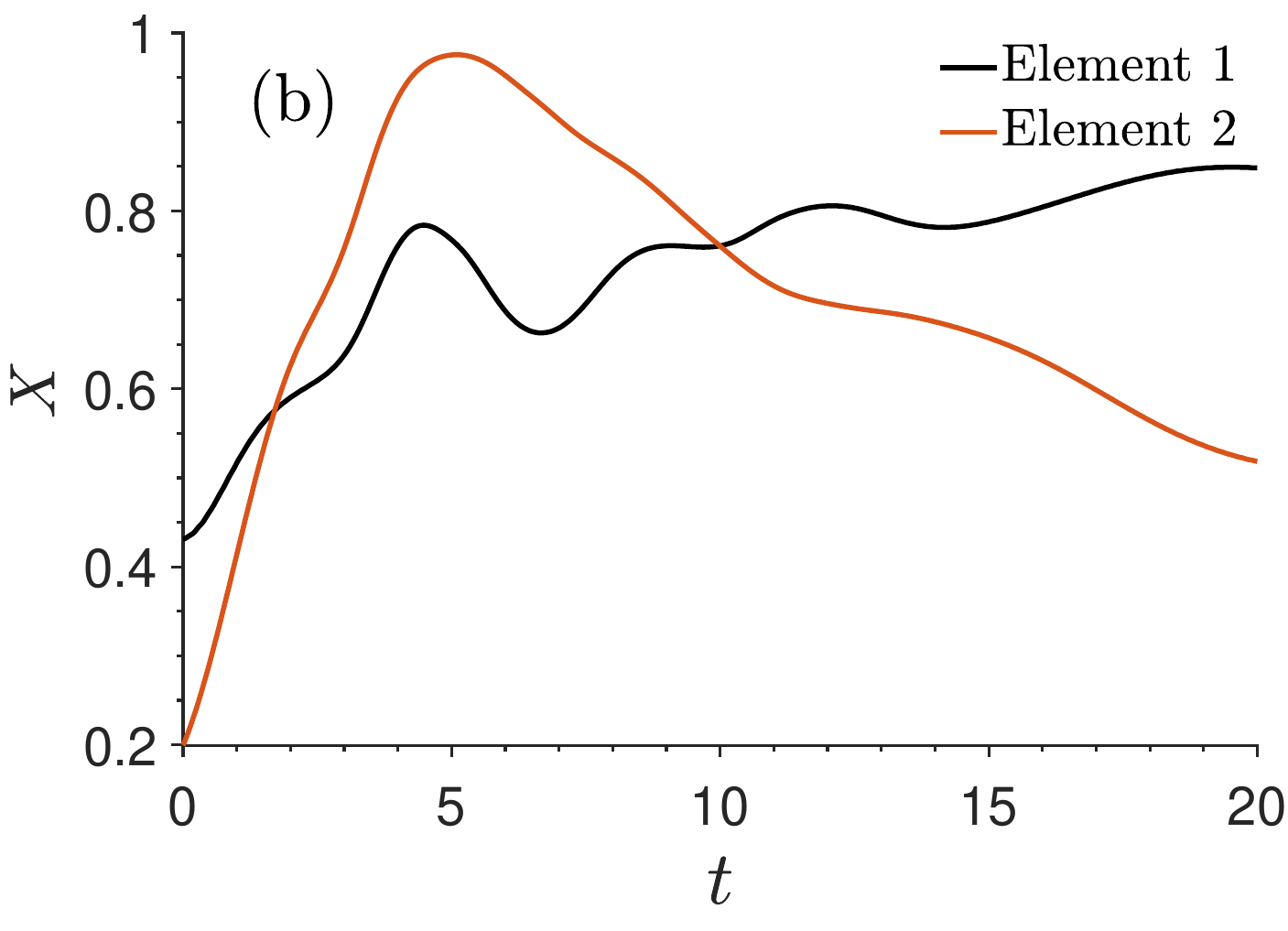}
    \end{subfigure}
    \vspace{1em}
    \begin{subfigure}[t]{0.48\textwidth}
        \centering
        \includegraphics[width=\textwidth,height=1\textwidth,keepaspectratio]{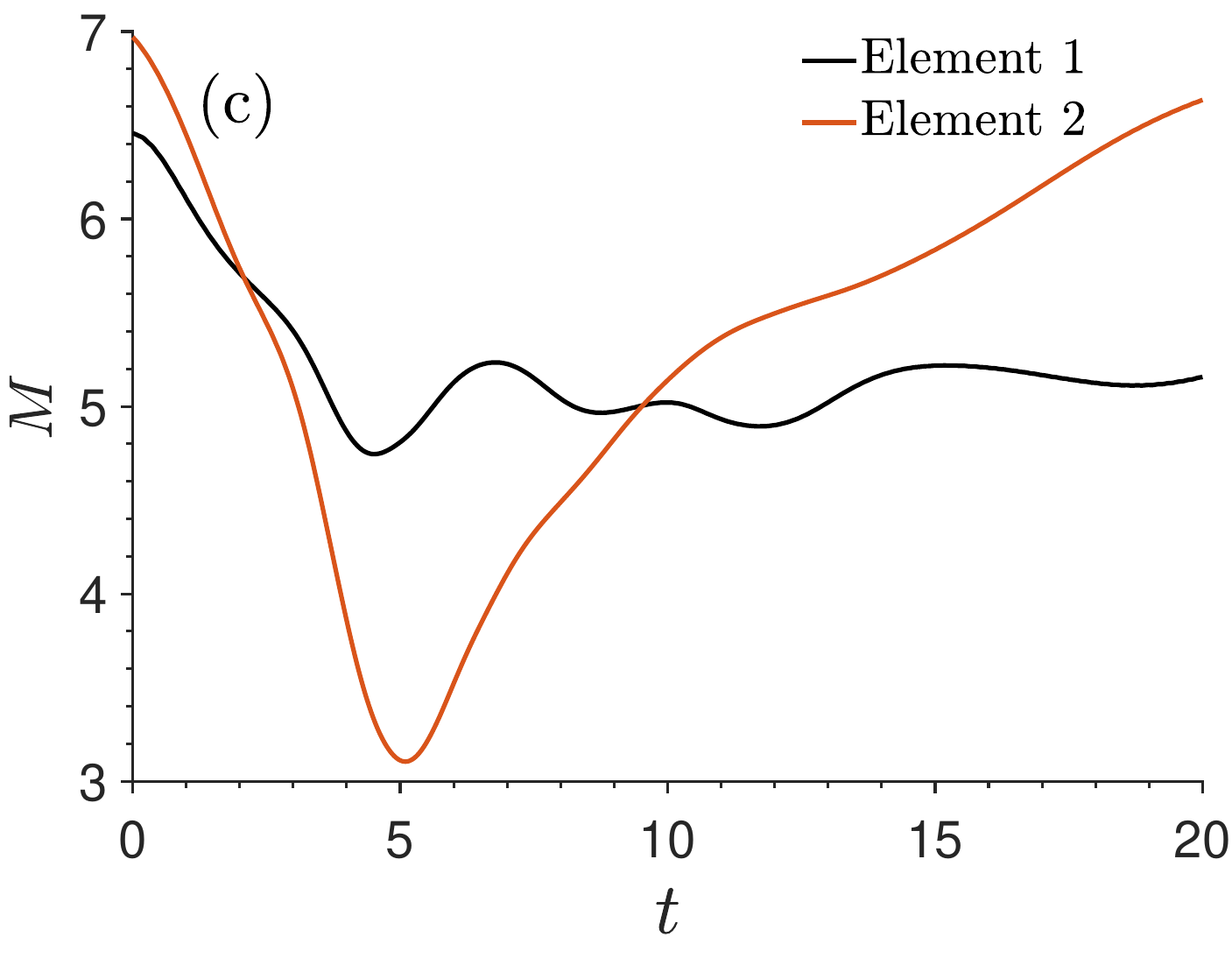}
    \end{subfigure}
    \caption{Temporal evolution of the STDEMF model parameters for two distinct elements in the $11\mathcal{P}4$ case of the TGV flow at $\mathrm{Re}_c=200$: (a) $\Delta/\eta$, (b) $X$, and (c) the cut-off mode $M$.}
    \label{fig:tgvmodelparameters}
\end{figure}

As another example illustrating the self-tuning feature of the STDEMF model and its associated parameters across elements, the temporal evolutions of $\Delta/\eta$, $X$, and the cut-off mode $M$ are analyzed for two representative elements in the $11\mathcal{P}4$ TGV case. As shown in Fig. \ref{fig:tgvmodelparameters}, in both elements, $\eta$ initially decreases, reflecting turbulence growth, and later increases, with the magnitude of these changes varying by location. The variable $X$ for both elements increases initially until $t=5$, but then evolves differently. For element $1$, $X$ decreases briefly after $t=5$ before increasing again until $t=20$, whereas for element $2$, $X$ decreases consistently after $t=5$. These variations in $X$ are related to the evolution of the flow in the TGV case. Initially, during the transition from laminar to turbulent flow, both rotation and shear increase. The magnitude of this increase varies across locations depending on the presence of the initial large structures. After some time, as these large structures stretch and break into smaller vortices, the subsequent evolution depends on whether the region retains the resulting eddies. Correspondingly, the cut-off mode in both elements drops sharply at early times, with element $2$ experiencing a steeper decrease. After $t=5$, the cut-off mode increases in element $2$, while in element $1$ it fluctuates around $M \approx 5$, indicating higher dissipation at later times for this location.

\begin{figure}[!ht]
    \centering
    \includegraphics[width=0.8\textwidth]{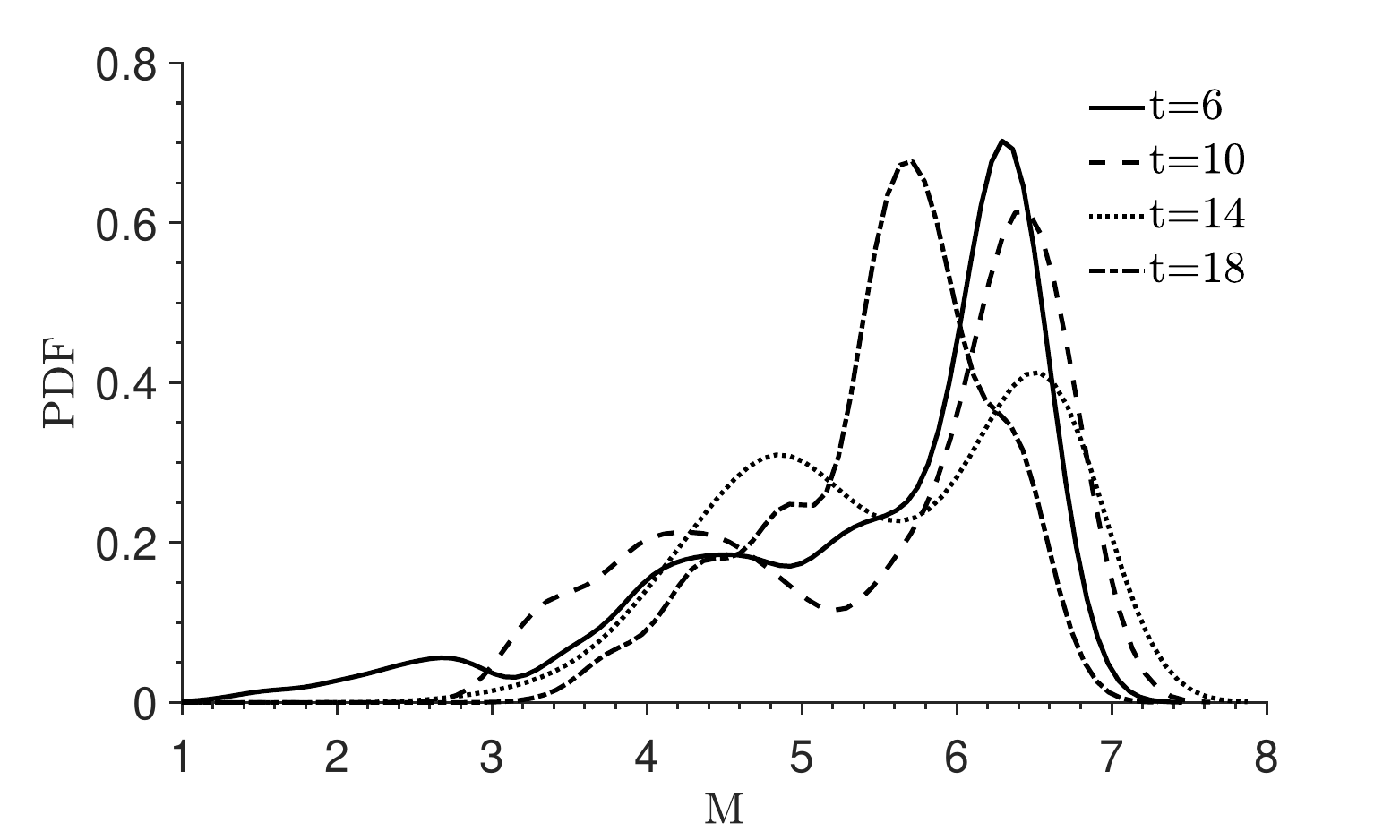}
    \caption{\label{fig:tgvMPDF} Probability density function of the cut-off mode for the $11\mathcal{P}4$ case of the TGV flow at different times.}
\end{figure}

Examining the entire computational domain, the PDF of the cut-off mode $M$ in the $11\mathcal{P}4$ TGV case ($\mathrm{Re}_c=200$) at $t=6$, $t=10$, $t=14$ and $t=18$ is shown in Fig. \ref{fig:tgvMPDF}. For the $11\mathcal{P}4$ case, the peak of dissipation occurs at $t=6$ (see Fig.~\ref{fig:tgvvalid}). Therefore, the selected times correspond to the peak dissipation and subsequent stages during which the turbulence decays. Unlike the HID case, the standard deviation here is high, reflecting spatial discrepancies in turbulent properties across the domain, which in turn affect the cut-off mode and the amount of energy removal inside each element. The TGV flow is initialized with large-scale structures that break into smaller eddies and spread across the domain. At earlier times, the flow structures are present in some parts of the domain, leading to stronger turbulence activity in those regions. This manifests as low cut-off modes in those elements. As time evolves, the large turbulent structures break down into smaller eddies that spread across the domain. This attenuates the turbulent activity in the regions initially containing those structures and causes the cut-off mode in those regions to shift toward higher values. Consequently, as time progresses, the PDF of the cut-off mode becomes narrower, and low cut-off mode numbers gradually disappear from the distribution. It can also be observed that as the flow evolves toward a more isotropic state, the cut-off mode distribution becomes narrower, and the two peaks gradually move to merge into a single one. The mean (standard deviation) of $M$ is $5.409 (1.174)$ at $t=6$, $5.515 (1.136)$ at $t=10$, $5.363 (1.408)$ at $t=14$ and $5.22 (1.38)$ at $t=18$.

Finally, the STDEMF model is applied on coarse meshes for both the $11\mathcal{P}4$ ($\mathrm{Re}_c=200$) and $7\mathcal{P}10$ ($\mathrm{Re}_c=800$) cases and compared with the DNS results, the Smagorinsky eddy-viscosity model, and the DEMF model. All TGV cases employ the split form with the Chandrasekar averaging function and the low-dissipation Roe Riemann solver for the convective fluxes. For the viscous flux, the BR1 method with a penalty parameter of zero is used. For both the $11\mathcal{P}4$ and $7\mathcal{P}10$ cases, $c=0.25$ is used in the STDEMF model. Figure \ref{fig:Re200 model assessments}(a) shows the kinetic energy dissipation rate for the $11\mathcal{P}4$ case, while Fig. \ref{fig:Re200 model assessments}(b) zooms in on $t=[4,12]$ for closer comparison. The coarse mesh without a model begins to over-predict the dissipation rate after $t=4$ and underpredicts it after $t=7$. The DEMF model with $\mathcal{P}_f=3$ and $\mathcal{P}_f=4$ brackets the DNS results. However, $\mathcal{P}_f=4$ removes too much energy, leading to underprediction, while $\mathcal{P}_f=3$ gets closer to DNS before $t=7$ but behaves like the coarse mesh for $t>7$ and underpredicts the DNS. The Smagorinsky model overpredicts both the DNS and coarse mesh for $t<6$ and underpredicts them after $t=7$, indicating that it is not a suitable option for this case. The STDEMF model shows the closest agreement with DNS, slightly underpredicting dissipation between $t=[8,12]$. Despite the resolution ratio between DNS and the $11\mathcal{P}4$ case being roughly $18$, the differences in energy dissipation rates are small.

Figure \ref{fig:Re800 model assessments} presents the results for the $7\mathcal{P}10$ case of the TGV flow at $\mathrm{Re_c}=800$. The coarse-DNS overpredicts the DNS after $t=4$ and underpredicts it after $t=9$. Applying the DEMF model improves the agreement with DNS before $t=9$. However, for $t>9$, the filtered results underpredict the DNS results more than the coarse-DNS until $t=11$, after which they coincide with the coarse-DNS results. $\mathcal{P}_f=6$ performs better than $\mathcal{P}_f=5$ until $t=9$, and at later times ($t>11$) both coincide with the coarse-mesh results. The Smagorinsky model results show deviations from the DNS results as early as $t=2$, while the coarse mesh and other models remain in good agreement. It overpredicts the coarse-mesh dissipation and moves further from the DNS results until $t=8$. This is because the flow is mostly laminar at initial times $t<4$, and adding viscosity through Smagorinsky is counterproductive. During $t>9$, the Smagorinsky model underpredicts the DNS results and also the coarse-mesh results, indicating that it is not a suitable option in this case. The STDEMF model shows good agreement with DNS, slightly underpredicting dissipation for $t=[8.5,12]$. A key advantage of the DEMF and STDEMF models is that, unlike the Smagorinsky model, they naturally vanish in the laminar limit through the implemented sensor. 

In the DEMF model, the higher $\mathcal{P}_f$ in the $7\mathcal{P}10$ case compared to the $11\mathcal{P}4$ case is due to its higher polynomial order. The STDEMF model incorporates polynomial order into the filter kernel and the cut-off mode formulation, allowing the same value of $c$ to be used consistently for cases with different polynomial orders. To further assess the performance of the developed model in wall-bounded turbulence, the study next considers the periodic channel flow.

\subsection{Periodic Channel Flow}

The periodic channel flow case at $\mathrm{Re_c} = 7,095$, corresponding to $\mathrm{Re}_{\tau} = 395$ of \cite{Moser1999}, is simulated in this study. The bulk velocity $\bar{U}$, bulk density $\bar{\rho}$, wall temperature $T_w$, and channel half-height $\delta$ are used as reference values. The reference Mach number, $\mathrm{Ma}_c$, based on the reference temperature and velocity, is $0.3$. The friction Reynolds number is defined as
\begin{equation}
\mathrm{Re}_{\tau} = \frac{u_{\tau}}{\bar{U}}\,\mathrm{Re_c}
\end{equation}
where the friction velocity is
\begin{equation}
u_{\tau} = \sqrt{\,\frac{\left.\dfrac{\partial v_1}{\partial y}\right|_{y=0}}{\mathrm{Re_c}}}.
\end{equation}

Periodic boundary conditions are applied in the streamwise, $x$, and spanwise, $z$, directions, while no-slip isothermal walls are imposed in the wall-normal, $y$, direction. The initial conditions set the density to $\rho = 1$, the velocity in the wall-normal and spanwise directions to zero, and the streamwise velocity to a parabolic profile with mean $\bar{U}$
\begin{equation}
 v_1(y) = -6 \left[ \left(\frac{y}{2}\right)^2-\left(\frac{y}{2}\right)\right].
\end{equation}
 The initial temperature is set to 
 \begin{equation}
 T\left(y\right)= T_w + \frac{3\left(\gamma-1\right)}{4}\mathrm{Pr}_c\mathrm{Ma}_c\left[1-\left(y-1\right)^4\right]
 \end{equation}
 with $T_w = 1$.
 
The transition to turbulence follows the process described by \cite{Jacobs_2003}, and the source term of \cite{Lenormand_2000} is applied to maintain a constant mass flow rate. The heat capacity ratio and Prandtl number are $\gamma = 1.4$ and $\mathrm{Pr}_c = 0.72$, respectively. Table \ref{tab:channel} summarizes the domain sizes and grid details of the coarse-DNS case. In this study, $\delta=1$, matching the domain size of \cite{Moser1999}, which ensures that the largest turbulent scales are captured. The mesh is refined near the walls to enhance the resolution in the near-wall region.
\begin{table}[h]
\centering
\begin{threeparttable}
\caption{Details of the coarse-DNS grid used for the periodic channel flow.}
\label{tab:channel}
\begin{tabular}{
    c@{\hspace{0.25cm}}
    c@{\hspace{0.25cm}}
    c@{\hspace{0.25cm}}
    c@{\hspace{0.25cm}}
    c@{\hspace{0.25cm}}
    c@{\hspace{0.25cm}}
    c@{\hspace{0.25cm}}
}
\toprule
Case & Case ID & $L_{x} \times L_{y} \times L_{z}$ & $N_x \times N_y \times N_z$& $\mathcal{P}$ & $N_\mathrm{total}^{\tnote{a}}$& $N_{y^+<10}^{\tnote{b}}$\\
\midrule
Channel-Coarse DNS & $14\mathcal{P}6$ & $2\pi\delta \times 2\delta \times \pi \delta$ & $14\times12\times16$ & 6 & $921,984$ & 6 \\
\bottomrule
\end{tabular}
\begin{tablenotes}
\item[a] Total number of solution points
\item[b] Number of points inside $y^{+}=10$
\end{tablenotes}
\end{threeparttable}
\end{table}

The superscript $(^+)$ indicates scaling with the friction velocity $u_{\tau}$, with $y^+ = y \, \mathrm{Re}_{\tau}$ representing the scaled wall-normal distance in wall units. The grid resolution lies within the suitable range for LES without wall modeling, as recommended by \cite{Choi}. In this case, the elements are clustered toward the wall in the wall-normal direction while maintaining uniform size in the streamwise and spanwise directions. 

The coarse-DNS simulations and the filtered coarse-DNS simulations employing the DEMF and STDEMF models were performed, and the results are compared with the DNS of \cite{Moser1999}. The root mean square (rms) of velocity fluctuation in the \(x_i\)-direction is defined as $v_{i,\mathrm{rms}} = \sqrt{\overline{v'_i v'_i}},$ where $v'_i = v_i - \overline{v_i}$ and \(\overline{(\cdot)}\) denotes time average. The scaled rms velocity fluctuation is given by $v^{+}_{i,\mathrm{rms}} = v_{i,\mathrm{rms}}/u_{\tau}$. All channel flow cases in this study employ the split form with the Chandrasekar averaging function and the low-dissipation Roe Riemann solver for the convective fluxes. For the viscous flux, the BR2 method with a penalty parameter of $1.5$ is used in all cases. Unlike the periodic box flow cases, HID and TGV, which employed BR1 with a penalty parameter of zero, the presence of walls in channel flow necessitates BR2 implementation. The penalty parameter of BR2 mitigates jumps in velocity gradients at element interfaces in the coarse-mesh cases. 
\begin{figure}[htbp]
    \centering
    \begin{subfigure}[t]{0.65\textwidth} 
        \includegraphics[width=\textwidth]{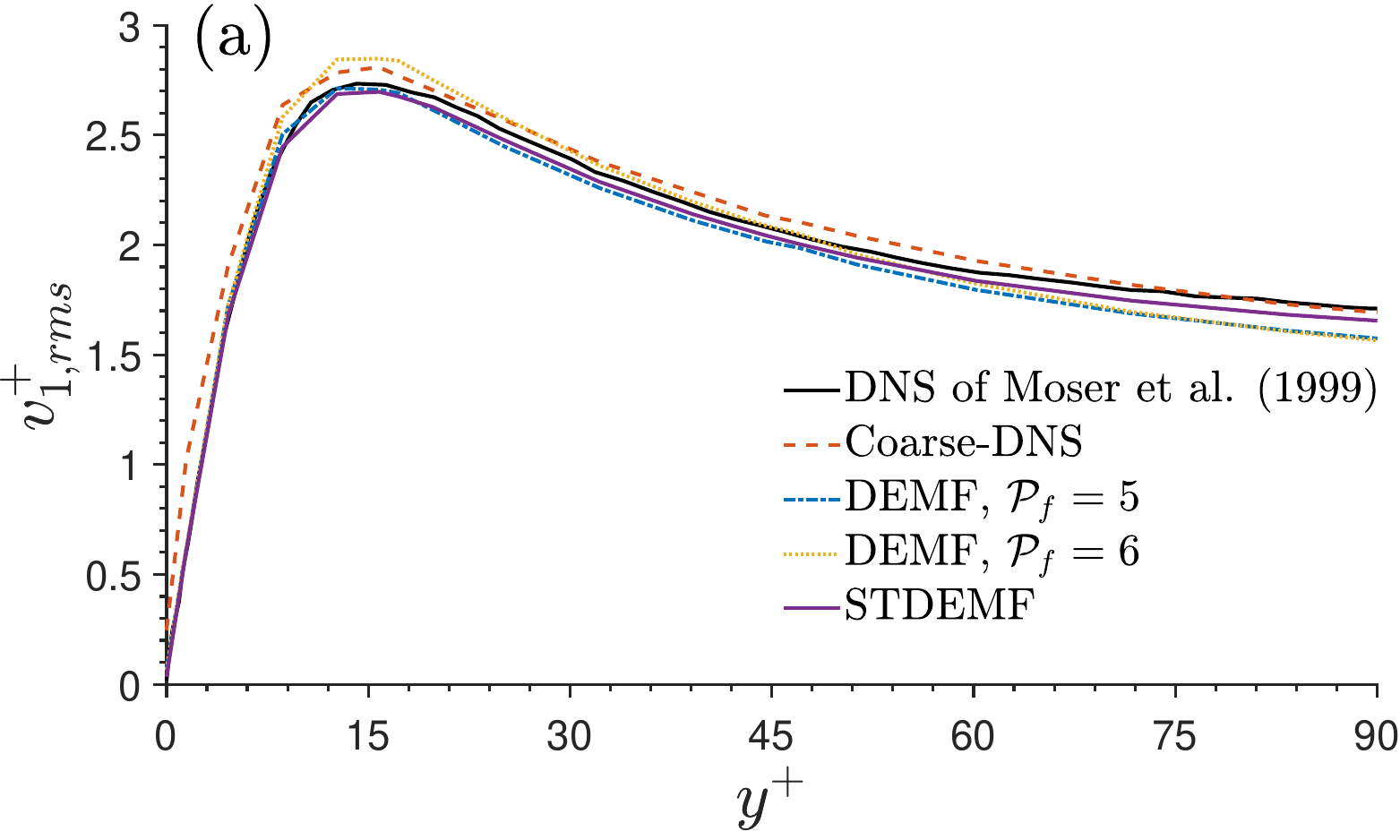}
    \end{subfigure}
    \vspace{1em}
    \begin{subfigure}[t]{0.65\textwidth}
        \includegraphics[width=\textwidth]{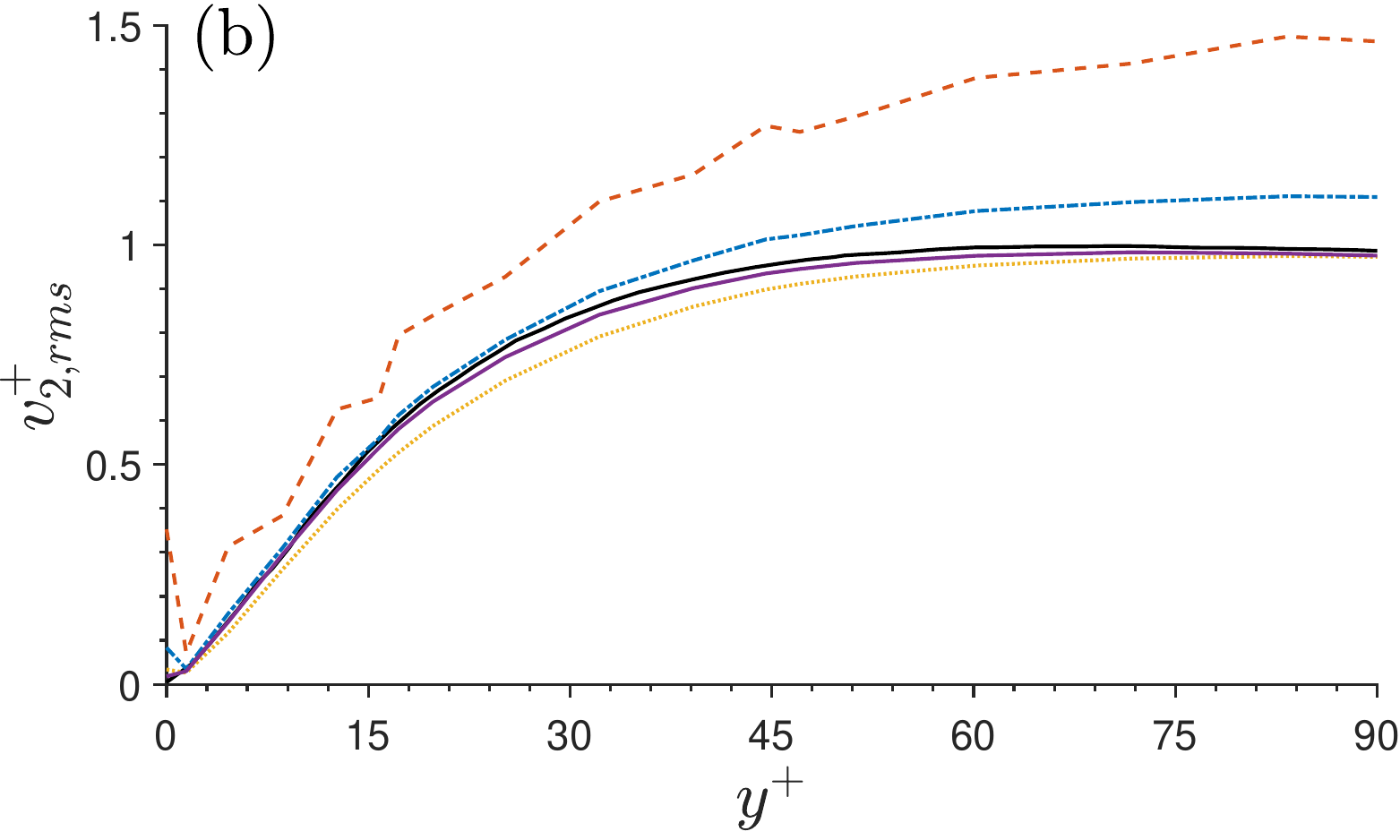}
    \end{subfigure}
    \vspace{1em}
    \begin{subfigure}[t]{0.65\textwidth} 
        \includegraphics[width=\textwidth]{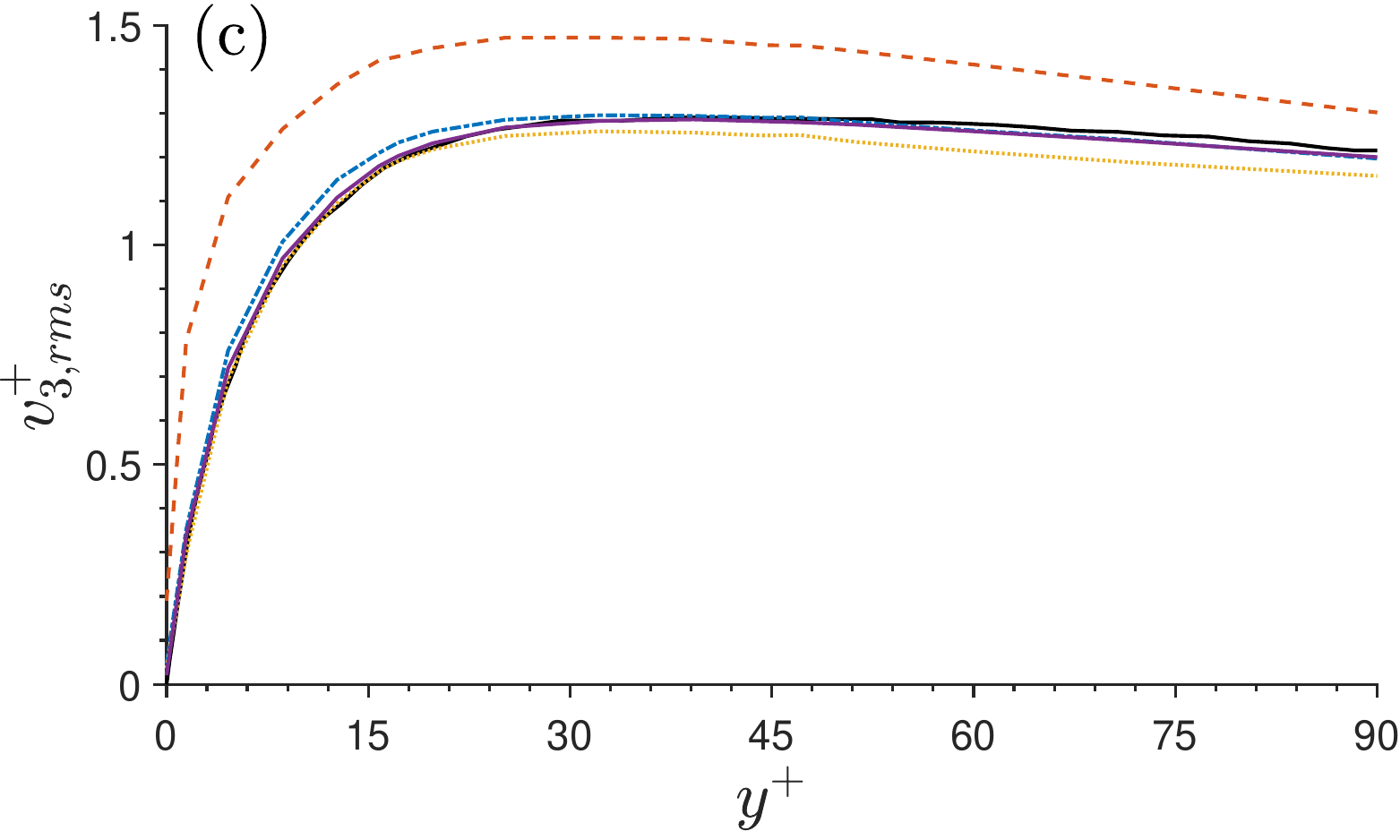}
    \end{subfigure}
    \caption{Comparison of scaled rms velocity fluctuations in the (a) streamwise, (b) wall-normal, and (c) spanwise directions between DNS of \cite{Moser1999}, the coarse DNS, the DEMF model of \cite{Ranjbar_2024POF}, and the STDEMF model.}
    \label{fig:channelrms}
\end{figure}

Figure \ref{fig:channelrms} shows that the rms velocity fluctuations are noticeably improved by applying the DEMF models. Larger rms values are obtained for all three velocity components in the coarse-mesh case, with more pronounced differences in the wall-normal and spanwise directions. This is because of the lack of small-scale mechanisms that dissipate energy. Additionally, in the coarse-mesh case, nonzero values are observed for velocity fluctuations at the wall, while applying the modal filter drives the rms values toward zero. The nonzero wall velocities arise from the weak enforcement of the boundary condition, as explained in \cite{collis2002}. Moreover, for the wall-normal component, jumps in the rms profiles are observed, which are smoothed out by the modal filter. For the DEMF model, filtering with $\mathcal{P}_f=5$ and $\mathcal{P}_f=6$ provides the best results. Specifically, $\mathcal{P}_f=6$ performs better in the wall-normal direction, whereas in the streamwise direction, $\mathcal{P}_f=5$ matches the DNS results more closely up to $y^{+}=15$, and $\mathcal{P}_f=6$ shows better agreement from $y^{+}=15$ to $y^{+}=50$, with both behaving similarly beyond that point. In the spanwise direction, $\mathcal{P}_f=6$ aligns more closely with the DNS results up to $y^{+}=18$, while $\mathcal{P}_f=5$ provides better agreement beyond that. Figure \ref{fig:channelrms} also shows that increasing the filter strength in the DEMF model of \cite{Ranjbar_2024POF} eliminates the near-wall jump, causing $v_{2,\mathrm{rms}}$ to approach zero at the wall.  The STDEMF model with $c=0.5$ outperforms the DEMF and yields the closest agreement with DNS by providing results that lie between the DEMF results obtained with $\mathcal{P}_f=5$ and $\mathcal{P}_f=6$. The STDEMF results are obtained by performing averaging over the homogeneous plane (i.e., elements located at the same distance from the wall). Alternative approaches, including temporal averaging and averaging across the spanwise direction only, were also performed, yielding results that are nearly identical. The Smagorinsky model did not produce good results in this flow. For $C_s < 0.15$, the simulations became unstable and crashed, while for $C_s = 0.2$, the results deviated significantly from DNS and other models. Consequently, the results of the Smagorinsky model are not reported here.

\begin{figure}[!ht]
    \centering
    \includegraphics[width=0.8\textwidth]{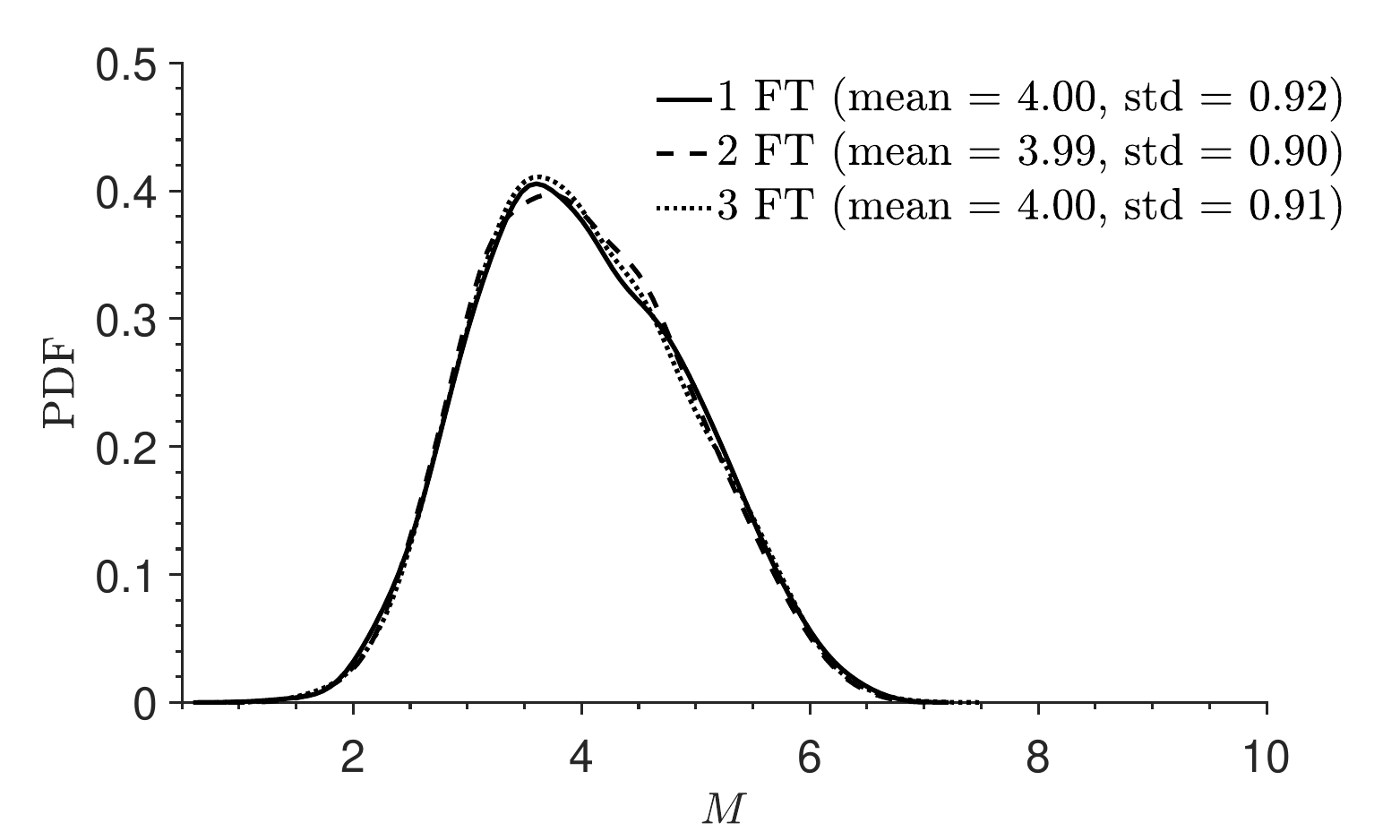}
    \caption{\label{fig:channelMPDF} Probability density function of the cut-off mode for the $14\mathcal{P}6$ case of periodic channel flow.}
\end{figure}

The PDF of the cut-off mode across the entire domain is presented at three different times. Figure \ref{fig:channelMPDF} shows the PDF after the flow reaches a statistically stationary state and for different flow-through times (FT, defined as the time required for the flow to traverse the domain). The PDFs exhibit similar behavior at all times, confirming statistical stationarity. The distribution has a mean of about $4.0$ and a standard deviation of about $0.91$. The lower cut-off mode yielding optimal results in STDEMF compared to DEMF is due to the use of the hyperbolic tangent filter kernel in STDEMF.

\subsection{Model Applicability and Parameter Sensitivity}

Having assessed the model on the three flow configurations, we now examine its range of applicability, sensitivity to the parameter $c$, computational cost, and behavior in the laminar limit.

\subsubsection{High Mach Number Limit}

Observations from the flow cases in this study indicate that regions with larger negative values of $Q_S$ and larger positive values of $Q_\Omega$ correspond to higher-index $\Gamma$ values. Accordingly, the cut-off mode formulation was developed to account for this behavior. Equation~(\ref{eq:xviscous}) was derived for incompressible and low-Mach-number compressible flows. For high-Mach-number cases, where dilatational effects may become significant, the correlation between more positive $Q_S$ values and higher-index $\Gamma$ should be further examined, and Eq.~(\ref{eq:xviscous}) may need to be revised.

\subsubsection{Sensitivity Analysis}
\begin{figure}[htbp]
    \begin{subfigure}[t]{0.48\textwidth} 
        \includegraphics[width=\textwidth,height=1\textwidth,keepaspectratio]{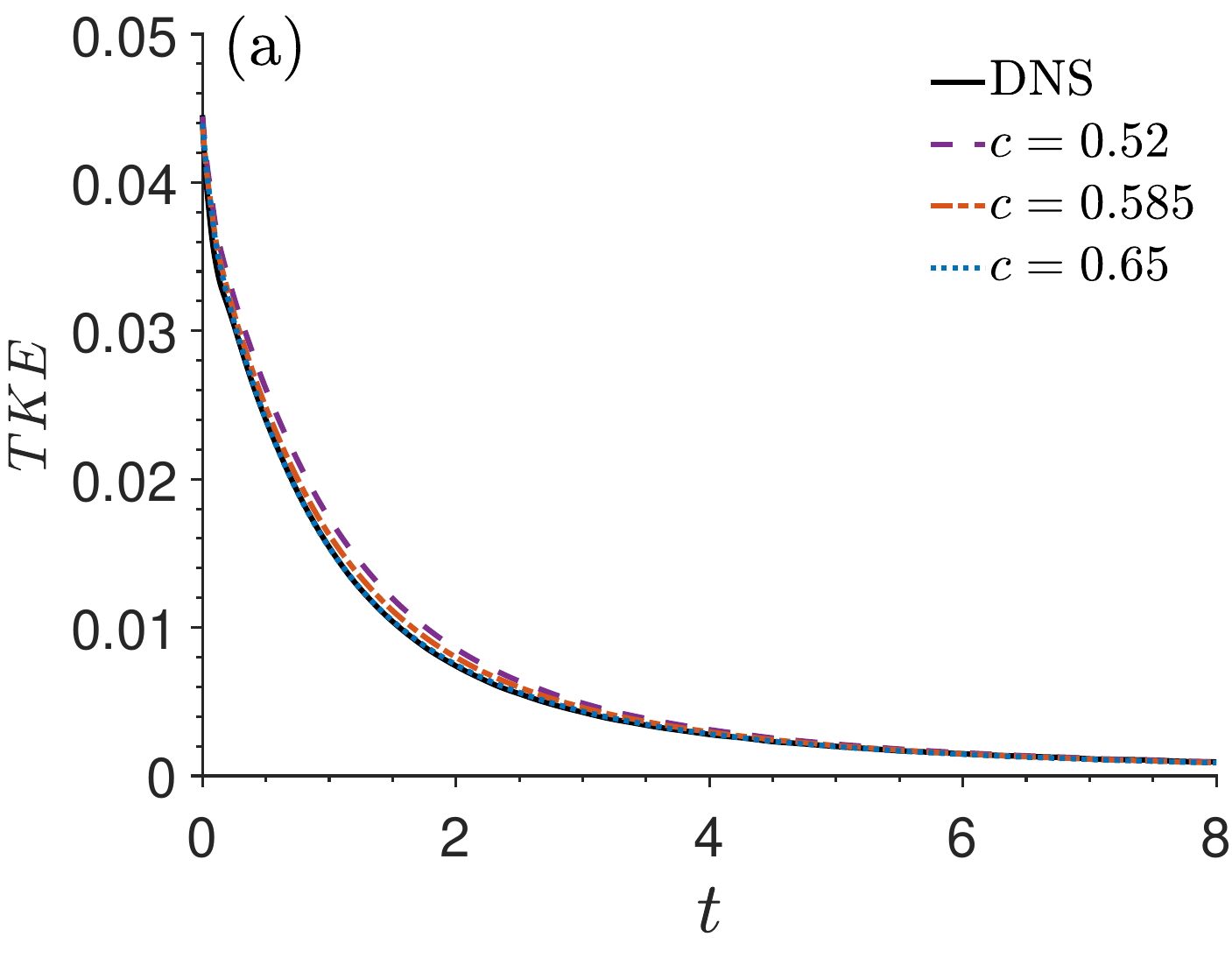}
    \end{subfigure}
    \hspace{0.02\textwidth}
    \begin{subfigure}[t]{0.48\textwidth}
        \includegraphics[width=\textwidth,height=1\textwidth,keepaspectratio]{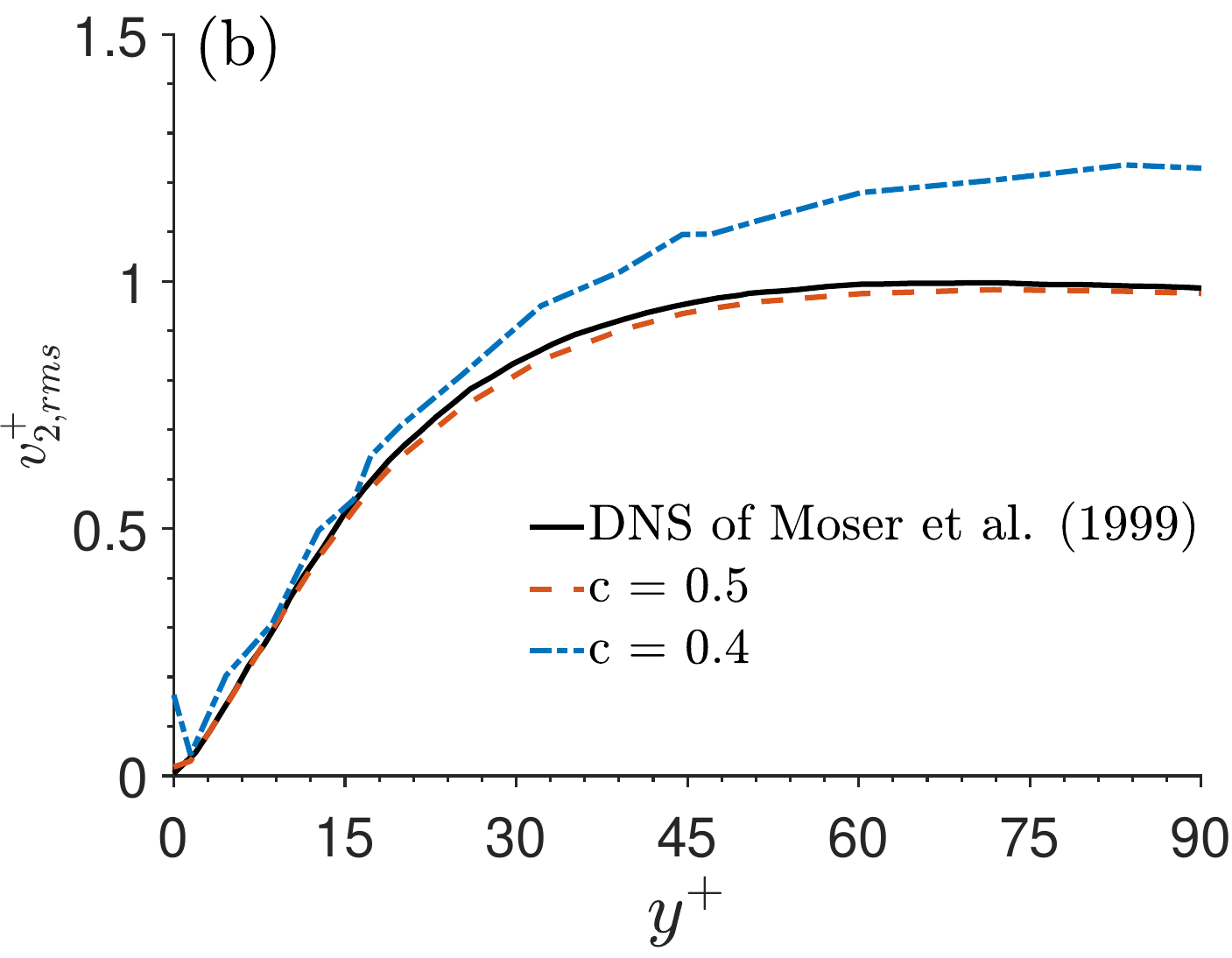}
    \end{subfigure}
    \vspace{1em}
    \begin{subfigure}[t]{0.48\textwidth}
        \centering
        \includegraphics[width=\textwidth,height=1\textwidth,keepaspectratio]{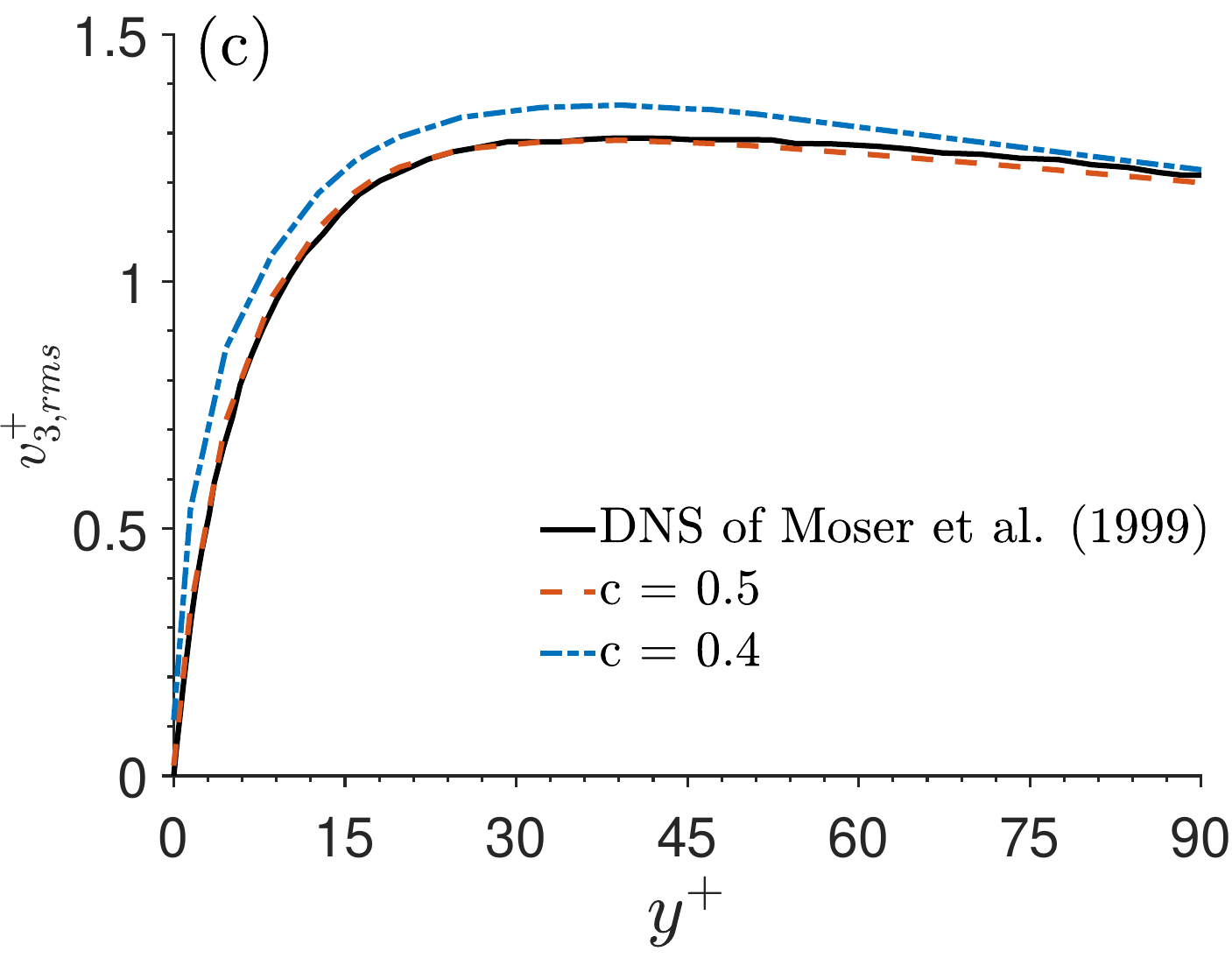}
    \end{subfigure}
    \caption{Sensitivity analysis of the STDEMF model with respect to the parameter $c$ in (a) HID flow and (b, c) channel flow.}
    \label{fig:c-Sensativity}
\end{figure}

The present model includes a tunable parameter, $c$. Tests conducted in this study show that optimal results are obtained for $c \in [0.2, 0.7]$. To assess the model’s sensitivity to this parameter, the HID and periodic channel flows are analyzed by varying $c$ and observing its effect on the results. For the HID flow, the optimal value yielding the best agreement was $c = 0.65$. The deviation in TKE with changes in $c$ is quantified by selecting TKE values at 1000 equally spaced times between $t = 0$ and $t = 8$, and defining the relative difference as
\begin{equation}
\label{eq:sensativity}
D = \sqrt{ \dfrac{1}{1000} 
\sum_{i=1}^{1000} 
\left( 
\dfrac{ TKE\big|^{c=c_1}_{t = i\Delta t} - TKE\big|^{c=c_2}_{t = i\Delta t} }{ TKE\big|^{c=c_1}_{t = i\Delta t} } 
\right)^2 }.
\end{equation}
Figure \ref{fig:c-Sensativity}(a) shows the results for the optimal case with $c = 0.65$, as well as cases with a 10\% lower value ($c = 0.585$) and a 20\% lower value ($c = 0.52$). A $10\%$ reduction in $c$ results in a $4.2\%$ change in TKE ($D = 0.042$), while a $20\%$ reduction leads to a $10.5\%$ change ($D = 0.105$). 

For the periodic channel flow, the unfiltered case is unstable, and stabilization requires an appropriate level of modal filtering. The optimal value is $c = 0.5$, and a $20\%$ variation in $c$ leads to a $15.1\%$ difference ($D = 0.151$) in the $y$-component and a $11.4\%$ difference ($D = 0.114$) in the $z$-component of velocity fluctuations. The results for both cases are shown in Fig.~\ref{fig:c-Sensativity}(b) and (c), where nonzero velocity fluctuations at the wall and jumps across the $y$-component profile are visible. The larger error in the $y$-component compared to the $z$-component velocity fluctuations arises from the flow’s instability. For the channel flow, $D$ is evaluated using rms values instead of TKE in Eq.~(\ref{eq:sensativity}), with the summation performed over all spatial points rather than time samples.

\subsubsection{Computational Cost}

To assess the computational cost of the model, the $7\mathcal{P}10$ case of the TGV flow and the $14\mathcal{P}6$ case of the channel flow are considered. The computational time was compared between the unfiltered and filtered coarse meshes using the same number of processors. For the TGV flow, the filtered case showed a $1.31\%$ increase in computational cost when the model was turned on. For the channel flow, three different averaging options were tested, resulting in slight variations. Averaging over time led to an increase of $0.96\%$, averaging over the homogeneous plane led to $1.59\%$, and averaging over the spanwise direction led to $2.5\%$. The low computational overhead of the method arises from evaluating the $\mathcal{N}$ and $\mathcal{M}$ matrices only once at the beginning and performing transformations between nodal and modal spaces using matrix-vector multiplications.

\subsubsection{Vanishing in the laminar limit}

\begin{figure}[!ht]
    \centering
    \includegraphics[width=0.6\textwidth]{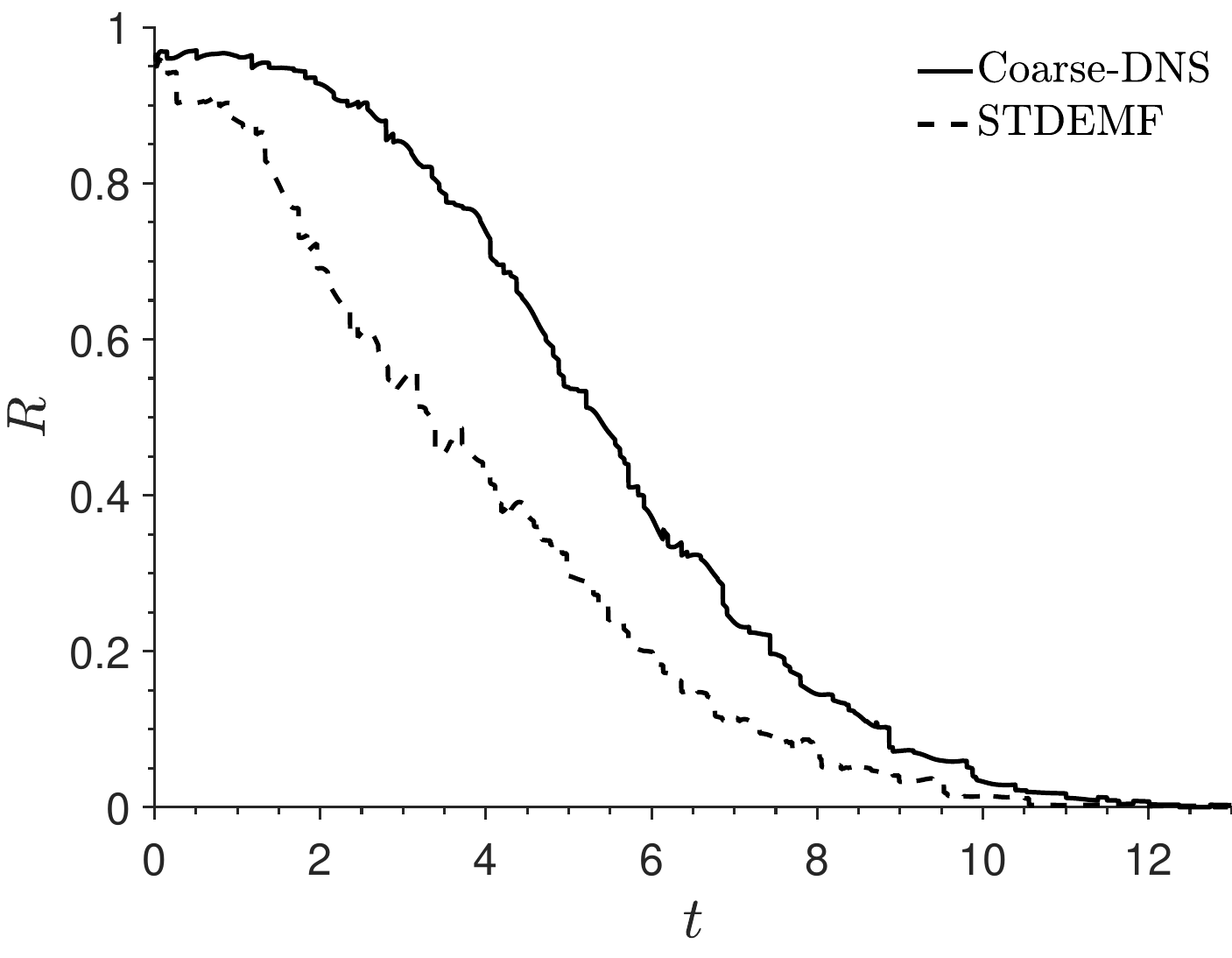}
    \caption{\label{fig:energy ratio} Temporal evolution of the ratio of the energy contained in modes above the cut-off mode to the energy of all modes for a representative element of the HID flow (element $1$ of Fig.~\ref{fig:isomodelparameters})}.
\end{figure}

As a final discussion, the behavior of the STDEMF model is examined as the flow transitions toward a laminar state. To this end, the simulation of the HID flow is continued until the flow reaches the laminar regime, and the ratio of the energy contained in the modes above the cut-off mode to the total energy of all modes is evaluated over time for element $1$ of the HID flow (shown in Fig.~\ref{fig:isomodelparameters}). It is expected that, as the flow becomes laminar, the energy in the modes above the cut-off mode approaches zero, and consequently, the model has no effect. Figure~\ref{fig:energy ratio} presents the temporal evolution of this energy ratio ($R$) for both the coarse-DNS and the filtered cases with the STDEMF model. In the coarse-DNS case, the cut-off mode is evaluated, but the filter is not applied, and the energy ratio is computed based on this cut-off mode. This allows us to assess how much energy is removed by the STDEMF compared to the unfiltered case. At early times, the ratio is high in both cases due to strong turbulence and low cut-off mode values. As the flow evolves, this ratio decreases. In the coarse-DNS case, the reduction results from physical viscous dissipation and the gradual shift of the cut-off mode to higher values over time, whereas in the STDEMF case, in addition to these two effects, numerical dissipation introduced by the STDEMF further attenuates the energy. As can be seen in Fig.~\ref {fig:energy ratio}, as the flow approaches the laminar regime, the energy in the modes above the cut-off mode becomes negligible, confirming that STDEMF has no effect under near-laminar conditions.

%% file: include/conclusion.tex
\section{Conclusions}
Explicit modal filtering can be used as a mechanism to add controlled dissipation to high-order DG methods, helping stabilize the scheme and making it more robust, particularly in problems involving shock waves or unresolved turbulent simulations. This work implements modal filtering for large-eddy simulation by removing the built-up energy due to the missing sub-grid scales in unresolved turbulent flow simulations. The presented method improves the dynamic explicit modal filter (DEMF) of \cite{Ranjbar_2024POF} by self-tuning the model parameters and dynamically adjusting the amount of energy removal. It also extends the methodology for computing the modes from the physical nodal values via orthogonality, allowing its use on any grid with any orthogonal polynomial.

The new self-tuning dynamic explicit modal filter (STDEMF) is coupled with the sensor developed by \cite{Ranjbar_2024POF} and implements a hyperbolic tangent filter kernel, which automatically adapts the modal filtering to different polynomial orders within each element. The filter kernel has a parameter, the cut-off mode, within each element, which adapts the model to local flow characteristics, including the Kolmogorov length scale, $\eta$, the second invariant of the strain-rate tensor, and the rotation tensor, and allows dynamic adjustment for energy removal. Turbulence unresolvedness, revealed through increased magnitudes of higher-index energy levels, is intensified as $\eta$ decreases \citep{Ranjbar_2024POF} and, as shown in this study, as $X$ (representing shear and rotation) increases. The amount of the added dissipation through STDEMF is controlled through the cut-off mode and increases as $\eta$ decreases and $X$ increases.

The proposed STDEMF model has been evaluated on three benchmark problems, the HID, the TGV, and the periodic channel flow, each with distinct turbulent characteristics. Comparisons with the DEMF model and the Smagorinsky eddy viscosity model show that the STDEMF outperforms both by dynamically adjusting the cut-off mode and, consequently, the energy removal in unresolved turbulent regions.

Future work will assess the model in more complex flows, including spatially developing turbulent boundary layers, such as the flow over a flat plate, and flows involving separation followed by reattachment, such as the flow over a backward-facing step. Additionally, supervised machine learning algorithms can be explored as an alternative approach for determining the relationship between the cut-off mode and the introduced parameters.

%% file: include/acknowledgements.tex
\section{Acknowledgements}
Simulations were performed using the High-Performance Computing (HPC) resources supported by the University of Arizona TRIF, UITS, and Office of Research and Partnership (ORP) and maintained by the Arizona Research Technologies team.

%% file: include/authordeclaration.tex
\section{Author Declaration}
\subsection{Conflict of Interest}
The authors have no conflicts to disclose.
\subsection{Author Contributions}
\textbf{Mohammadmahdi Ranjbar}: Conceptualization; Methodology; Visualization; Formal analysis; Validation; Writing - original draft. \textbf{Ali Mostafavi}: Conceptualization; Formal analysis; Writing - review and editing. \textbf{Farzad Mashayek}: Conceptualization; Methodology; Supervision; Writing-review and editing.

%% file: include/dataavailability.tex
\section{Data Availability}
Data will be made available upon request.

%% file: main.bbl
\begin{thebibliography}{87}
\providecommand{\natexlab}[1]{#1}
\providecommand{\url}[1]{\texttt{#1}}
\expandafter\ifx\csname urlstyle\endcsname\relax
  \providecommand{\doi}[1]{doi: #1}\else
  \providecommand{\doi}{doi: \begingroup \urlstyle{rm}\Url}\fi

\bibitem[Andreassen et~al.(1994)Andreassen, Lie, and Wasberg]{Andreassen_1994}
{\O}.~Andreassen, I.~Lie, and C.~E. Wasberg.
\newblock The spectral viscosity method applied to simulation of waves in a stratified atmosphere.
\newblock \emph{Journal of Computational Physics}, 110\penalty0 (2):\penalty0 257–273, 1994.
\newblock ISSN 0021-9991.
\newblock \doi{10.1006/jcph.1994.1023}.
\newblock URL \url{http://dx.doi.org/10.1006/jcph.1994.1023}.

\bibitem[Arnold et~al.(2002)Arnold, Brezzi, Cockburn, and Marini]{Arnold_2002}
D.~N. Arnold, F.~Brezzi, B.~Cockburn, and L.~D. Marini.
\newblock Unified analysis of discontinuous galerkin methods for elliptic problems.
\newblock \emph{SIAM Journal on Numerical Analysis}, 39\penalty0 (5):\penalty0 1749–1779, January 2002.
\newblock ISSN 1095-7170.
\newblock \doi{10.1137/s0036142901384162}.
\newblock URL \url{http://dx.doi.org/10.1137/s0036142901384162}.

\bibitem[Bassi and Rebay(1996)]{Bassi_1996}
F.~Bassi and S.~Rebay.
\newblock \emph{Discontinuous finite element high order accurate numerical solution of the compressible Navier-Stokes equations}, page 295–302.
\newblock Oxford University PressOxford, February 1996.
\newblock ISBN 9781383022650.
\newblock \doi{10.1093/oso/9780198514800.003.0022}.
\newblock URL \url{http://dx.doi.org/10.1093/oso/9780198514800.003.0022}.

\bibitem[Bassi et~al.(2015)Bassi, Botti, Colombo, Ghidoni, and Massa]{Bassi_2015}
F.~Bassi, L.~Botti, A.~Colombo, A.~Ghidoni, and F.~Massa.
\newblock Linearly implicit rosenbrock-type runge–kutta schemes applied to the discontinuous galerkin solution of compressible and incompressible unsteady flows.
\newblock \emph{Computers \& Fluids}, 118:\penalty0 305–320, 2015.
\newblock ISSN 0045-7930.
\newblock \doi{10.1016/j.compfluid.2015.06.007}.
\newblock URL \url{http://dx.doi.org/10.1016/j.compfluid.2015.06.007}.

\bibitem[Batchelor and Townsend(1948)]{batchelor1948}
G.~K. Batchelor and A.~A. Townsend.
\newblock Decay of isotropic turbulence in the initial period.
\newblock \emph{Proceedings of the Royal Society of London. Series A. Mathematical and Physical Sciences}, 193\penalty0 (1035):\penalty0 539--558, 1948.

\bibitem[Beck et~al.(2014)Beck, Bolemann, Flad, Frank, Gassner, Hindenlang, and Munz]{Beck_2014}
A.~D. Beck, T.~Bolemann, D.~Flad, H.~Frank, G.~J. Gassner, F.~Hindenlang, and C.‐D. Munz.
\newblock High‐order discontinuous galerkin spectral element methods for transitional and turbulent flow simulations.
\newblock \emph{International Journal for Numerical Methods in Fluids}, 76\penalty0 (8):\penalty0 522–548, 2014.
\newblock ISSN 1097-0363.
\newblock \doi{10.1002/fld.3943}.
\newblock URL \url{http://dx.doi.org/10.1002/fld.3943}.

\bibitem[Blaisdell et~al.(1993)Blaisdell, Mansour, and Reynolds]{Blaisdell_1993}
G.~A. Blaisdell, N.~N. Mansour, and W.~C. Reynolds.
\newblock Compressibility effects on the growth and structure of homogeneous turbulent shear flow.
\newblock \emph{Journal of Fluid Mechanics}, 256:\penalty0 443–485, 1993.
\newblock ISSN 1469-7645.
\newblock \doi{10.1017/s0022112093002848}.
\newblock URL \url{http://dx.doi.org/10.1017/s0022112093002848}.

\bibitem[Bozorgpour(2024)]{Bozorgpour2024}
R.~Bozorgpour.
\newblock A comprehensive review on turbulence modeling of ejectors.
\newblock \emph{OSF Preprints}, September 2024.
\newblock \doi{https://osf.io/n6wgf_v1}.

\bibitem[Carpenter et~al.(2014)Carpenter, Fisher, Nielsen, and Frankel]{Carpenter_2014}
M.~H. Carpenter, T.~C. Fisher, E.~J. Nielsen, and S.~H. Frankel.
\newblock Entropy stable spectral collocation schemes for the navier--stokes equations: Discontinuous interfaces.
\newblock \emph{SIAM Journal on Scientific Computing}, 36\penalty0 (5):\penalty0 B835–B867, 2014.
\newblock ISSN 1095-7197.
\newblock \doi{10.1137/130932193}.
\newblock URL \url{http://dx.doi.org/10.1137/130932193}.

\bibitem[Chandrashekar(2013)]{Chandrashekar_2013}
P.~Chandrashekar.
\newblock Discontinuous galerkin method for navier–stokes equations using kinetic flux vector splitting.
\newblock \emph{Journal of Computational Physics}, 233:\penalty0 527–551, 2013.
\newblock ISSN 0021-9991.
\newblock \doi{10.1016/j.jcp.2012.09.017}.
\newblock URL \url{http://dx.doi.org/10.1016/j.jcp.2012.09.017}.

\bibitem[Chen et~al.(1993)Chen, Du, and Tadmor]{Chen_1993}
G.-Q. Chen, Q.~Du, and E.~Tadmor.
\newblock Spectral viscosity approximations to multidimensional scalar conservation laws.
\newblock \emph{Mathematics of Computation}, 61\penalty0 (204):\penalty0 629, 1993.
\newblock ISSN 0025-5718.
\newblock \doi{10.2307/2153244}.
\newblock URL \url{http://dx.doi.org/10.2307/2153244}.

\bibitem[Chen et~al.(2022)Chen, Tang, and Xu]{Chen_2022}
L.~Chen, T.~Tang, and C.~Xu.
\newblock Efficient svv stabilized triangular spectral element methods for incompressible flows of high reynolds numbers.
\newblock \emph{Advances in Aerodynamics}, 4\penalty0 (1), 2022.
\newblock ISSN 2524-6992.
\newblock \doi{10.1186/s42774-021-00090-x}.
\newblock URL \url{http://dx.doi.org/10.1186/s42774-021-00090-x}.

\bibitem[Chen and Shu(2017)]{Chen_2017}
T.~Chen and C.-W. Shu.
\newblock Entropy stable high order discontinuous galerkin methods with suitable quadrature rules for hyperbolic conservation laws.
\newblock \emph{Journal of Computational Physics}, 345:\penalty0 427–461, 2017.
\newblock ISSN 0021-9991.
\newblock \doi{10.1016/j.jcp.2017.05.025}.
\newblock URL \url{http://dx.doi.org/10.1016/j.jcp.2017.05.025}.

\bibitem[Choi and Moin(2012)]{Choi}
H.~Choi and P.~Moin.
\newblock Grid-point requirements for large eddy simulation: Chapman’s estimates revisited.
\newblock \emph{Physics of Fluids}, 24\penalty0 (1):\penalty0 011702, 01 2012.
\newblock ISSN 1070-6631.
\newblock \doi{10.1063/1.3676783}.
\newblock URL \url{https://doi.org/10.1063/1.3676783}.

\bibitem[Collis(2002)]{collis2002}
S.~S. Collis.
\newblock Discontinuous galerkin methods for turbulence simulation.
\newblock \emph{Studying Turbulence Using Numerical Simulation Databases-IX: Proceedings of the 2002 Summer Program}, 2002.

\bibitem[de~la Llave~Plata et~al.(2018)de~la Llave~Plata, Couaillier, and le~Pape]{de_la_Llave_Plata_2018}
M.~de~la Llave~Plata, V.~Couaillier, and M.-C. le~Pape.
\newblock On the use of a high-order discontinuous galerkin method for dns and les of wall-bounded turbulence.
\newblock \emph{Computers \& Fluids}, 176:\penalty0 320–337, 2018.
\newblock ISSN 0045-7930.
\newblock \doi{10.1016/j.compfluid.2017.05.013}.
\newblock URL \url{http://dx.doi.org/10.1016/j.compfluid.2017.05.013}.

\bibitem[Fehn et~al.(2021)Fehn, Kronbichler, Munch, and Wall]{Fehn_2021}
N.~Fehn, M.~Kronbichler, P.~Munch, and W.~A. Wall.
\newblock Numerical evidence of anomalous energy dissipation in incompressible euler flows: towards grid-converged results for the inviscid taylor–green problem.
\newblock \emph{Journal of Fluid Mechanics}, 932, 2021.
\newblock ISSN 1469-7645.
\newblock \doi{10.1017/jfm.2021.1003}.
\newblock URL \url{http://dx.doi.org/10.1017/jfm.2021.1003}.

\bibitem[Ferrer and Willden(2011{\natexlab{a}})]{FERRER2011}
E.~Ferrer and R.~H.~J. Willden.
\newblock A high order discontinuous galerkin finite element solver for the incompressible navier–stokes equations.
\newblock \emph{Computers \& Fluids}, 46\penalty0 (1):\penalty0 224--230, 2011{\natexlab{a}}.
\newblock ISSN 0045-7930.
\newblock \doi{https://doi.org/10.1016/j.compfluid.2010.10.018}.
\newblock URL \url{https://www.sciencedirect.com/science/article/pii/S0045793010002860}.

\bibitem[Ferrer and Willden(2011{\natexlab{b}})]{Ferrer_2011}
E.~Ferrer and R.~H.~J. Willden.
\newblock A high order discontinuous galerkin finite element solver for the incompressible navier–stokes equations.
\newblock \emph{Computers \& Fluids}, 46\penalty0 (1):\penalty0 224–230, 2011{\natexlab{b}}.
\newblock ISSN 0045-7930.
\newblock \doi{10.1016/j.compfluid.2010.10.018}.
\newblock URL \url{http://dx.doi.org/10.1016/j.compfluid.2010.10.018}.

\bibitem[Ferrer et~al.(2023)Ferrer, Rubio, Ntoukas, Laskowski, Mariño, Colombo, Mateo-Gabín, Marbona, Manrique~de Lara, Huergo, Manzanero, Rueda-Ramírez, Kopriva, and Valero]{Ferrer_2023}
E.~Ferrer, G.~Rubio, G.~Ntoukas, W.~Laskowski, O.~A. Mariño, S.~Colombo, A.~Mateo-Gabín, H.~Marbona, F.~Manrique~de Lara, D.~Huergo, J.~Manzanero, A.~M. Rueda-Ramírez, D.~A. Kopriva, and E.~Valero.
\newblock : A high-order discontinuous galerkin solver for flow simulations and multi-physics applications.
\newblock \emph{Computer Physics Communications}, 287:\penalty0 108700, 2023.
\newblock ISSN 0010-4655.
\newblock \doi{10.1016/j.cpc.2023.108700}.
\newblock URL \url{http://dx.doi.org/10.1016/j.cpc.2023.108700}.

\bibitem[Fisher and Carpenter(2013)]{Fisher_2013}
T.~C. Fisher and M.~H. Carpenter.
\newblock High-order entropy stable finite difference schemes for nonlinear conservation laws: Finite domains.
\newblock \emph{Journal of Computational Physics}, 252:\penalty0 518–557, November 2013.
\newblock ISSN 0021-9991.
\newblock \doi{10.1016/j.jcp.2013.06.014}.
\newblock URL \url{http://dx.doi.org/10.1016/j.jcp.2013.06.014}.

\bibitem[Frigo and Johnson(2005)]{Frigo_2005}
M.~Frigo and S.~G. Johnson.
\newblock The design and implementation of fftw3.
\newblock \emph{Proceedings of the IEEE}, 93\penalty0 (2):\penalty0 216–231, February 2005.
\newblock ISSN 0018-9219.
\newblock \doi{10.1109/jproc.2004.840301}.
\newblock URL \url{http://dx.doi.org/10.1109/jproc.2004.840301}.

\bibitem[Gao et~al.(2018)Gao, Ouyang, and Zhou]{Gao_2018}
P.~Gao, J.~Ouyang, and W.~Zhou.
\newblock Development of a finite element/discontinuous galerkin/level set approach for the simulation of incompressible two phase flow.
\newblock \emph{Advances in Engineering Software}, 118:\penalty0 45–59, 2018.
\newblock ISSN 0965-9978.
\newblock \doi{10.1016/j.advengsoft.2018.01.006}.
\newblock URL \url{http://dx.doi.org/10.1016/j.advengsoft.2018.01.006}.

\bibitem[Gassner et~al.(2016)Gassner, Winters, and Kopriva]{Gassner_2016}
G.~J. Gassner, A.~R. Winters, and D.~A. Kopriva.
\newblock Split form nodal discontinuous galerkin schemes with summation-by-parts property for the compressible euler equations.
\newblock \emph{Journal of Computational Physics}, 327:\penalty0 39–66, 2016.
\newblock ISSN 0021-9991.
\newblock \doi{10.1016/j.jcp.2016.09.013}.
\newblock URL \url{http://dx.doi.org/10.1016/j.jcp.2016.09.013}.

\bibitem[George(1992)]{george1992}
W.~K. George.
\newblock The decay of homogeneous isotropic turbulence.
\newblock \emph{Physics of Fluids A: Fluid Dynamics}, 4\penalty0 (7):\penalty0 1492--1509, 1992.

\bibitem[Ghiasi et~al.(2018)Ghiasi, Li, Komperda, and Mashayek]{Ghiasi_2018}
Z.~Ghiasi, D.~Li, J.~Komperda, and F.~Mashayek.
\newblock Near-wall resolution requirement for direct numerical simulation of turbulent flow using multidomain chebyshev grid.
\newblock \emph{International Journal of Heat and Mass Transfer}, 126:\penalty0 746–760, November 2018.
\newblock ISSN 0017-9310.
\newblock \doi{10.1016/j.ijheatmasstransfer.2018.05.114}.
\newblock URL \url{http://dx.doi.org/10.1016/j.ijheatmasstransfer.2018.05.114}.

\bibitem[Ghiasi et~al.(2019)Ghiasi, Komperda, Li, Peyvan, Nicholls, and Mashayek]{Ghiasi_2019}
Z.~Ghiasi, J.~Komperda, D.~Li, A.~Peyvan, D.~Nicholls, and F.~Mashayek.
\newblock Modal explicit filtering for large eddy simulation in discontinuous spectral element method.
\newblock \emph{Journal of Computational Physics: X}, 3:\penalty0 100024, 2019.
\newblock ISSN 2590-0552.
\newblock \doi{10.1016/j.jcpx.2019.100024}.
\newblock URL \url{http://dx.doi.org/10.1016/j.jcpx.2019.100024}.

\bibitem[Guermond and Prudhomme(2003)]{Guermond_2003}
J.-L Guermond and S.~Prudhomme.
\newblock Mathematical analysis of a spectral hyperviscosity les model for the simulation of turbulent flows.
\newblock \emph{ESAIM: Mathematical Modelling and Numerical Analysis}, 37\penalty0 (6):\penalty0 893–908, 2003.
\newblock ISSN 1290-3841.
\newblock \doi{10.1051/m2an:2003060}.
\newblock URL \url{http://dx.doi.org/10.1051/m2an:2003060}.

\bibitem[Hong et~al.(2025)Hong, Zhang, Zhao, and Zheng]{HONG2025}
Y.~Hong, W.~Zhang, L.~Zhao, and H.~Zheng.
\newblock Coupling finite element and multiscale finite element methods for the non-stationary stokes-darcy model.
\newblock \emph{Journal of Computational Physics}, 530:\penalty0 113899, 2025.
\newblock ISSN 0021-9991.
\newblock \doi{https://doi.org/10.1016/j.jcp.2025.113899}.
\newblock URL \url{https://www.sciencedirect.com/science/article/pii/S0021999125001822}.

\bibitem[Jacobs(2003)]{Jacobs_2003}
G.~B. Jacobs.
\newblock \emph{Numerical simulation of two-phase turbulent compressible flows with a multidomain spectral method.}
\newblock {Ph.D.} {T}hesis, University of Illinois at Chicago, Chicago, IL, 2003.

\bibitem[Kaber(1996)]{Kaber_1996}
S.~M.~O. Kaber.
\newblock A legendre pseudospectral viscosity method.
\newblock \emph{Journal of Computational Physics}, 128\penalty0 (1):\penalty0 165–180, 1996.
\newblock ISSN 0021-9991.
\newblock \doi{10.1006/jcph.1996.0201}.
\newblock URL \url{http://dx.doi.org/10.1006/jcph.1996.0201}.

\bibitem[Karamanos and Karniadakis(2000)]{Karamanos_2000}
G.-S. Karamanos and G.~E. Karniadakis.
\newblock A spectral vanishing viscosity method for large-eddy simulations.
\newblock \emph{Journal of Computational Physics}, 163\penalty0 (1):\penalty0 22–50, 2000.
\newblock ISSN 0021-9991.
\newblock \doi{10.1006/jcph.2000.6552}.
\newblock URL \url{http://dx.doi.org/10.1006/jcph.2000.6552}.

\bibitem[Kirby and Karniadakis(2002)]{Kirby_2002}
R.~M. Kirby and G.~E. Karniadakis.
\newblock Coarse resolution turbulence simulations with spectral vanishing viscosity—large-eddy simulations (svv-les).
\newblock \emph{Journal of Fluids Engineering}, 124\penalty0 (4):\penalty0 886–891, 2002.
\newblock ISSN 1528-901X.
\newblock \doi{10.1115/1.1511321}.
\newblock URL \url{http://dx.doi.org/10.1115/1.1511321}.

\bibitem[Kirby and Karniadakis(2003)]{Kirby_2003}
R.~M. Kirby and G.~E. Karniadakis.
\newblock De-aliasing on non-uniform grids: algorithms and applications.
\newblock \emph{Journal of Computational Physics}, 191\penalty0 (1):\penalty0 249–264, 2003.
\newblock ISSN 0021-9991.
\newblock \doi{10.1016/s0021-9991(03)00314-0}.
\newblock URL \url{http://dx.doi.org/10.1016/s0021-9991(03)00314-0}.

\bibitem[Kirby and Sherwin(2006)]{Kirby_2006}
R.~M. Kirby and S.~J. Sherwin.
\newblock Stabilisation of spectral/hp element methods through spectral vanishing viscosity: Application to fluid mechanics modelling.
\newblock \emph{Computer Methods in Applied Mechanics and Engineering}, 195\penalty0 (23–24):\penalty0 3128–3144, 2006.
\newblock ISSN 0045-7825.
\newblock \doi{10.1016/j.cma.2004.09.019}.
\newblock URL \url{http://dx.doi.org/10.1016/j.cma.2004.09.019}.

\bibitem[Koal et~al.(2012)Koal, Stiller, and Blackburn]{Koal_2012}
K.~Koal, J.~Stiller, and H.~M. Blackburn.
\newblock Adapting the spectral vanishing viscosity method for large-eddy simulations in cylindrical configurations.
\newblock \emph{Journal of Computational Physics}, 231\penalty0 (8):\penalty0 3389–3405, 2012.
\newblock ISSN 0021-9991.
\newblock \doi{10.1016/j.jcp.2012.01.014}.
\newblock URL \url{http://dx.doi.org/10.1016/j.jcp.2012.01.014}.

\bibitem[Komperda et~al.(2020)Komperda, Ghiasi, Li, Peyvan, Jaberi, and Mashayek]{Komperda_2020}
J.~Komperda, Z.~Ghiasi, D.~Li, A.~Peyvan, F.~Jaberi, and F.~Mashayek.
\newblock A hybrid discontinuous spectral element method and filtered mass density function solver for turbulent reacting flows.
\newblock \emph{Numerical Heat Transfer, Part B: Fundamentals}, 78\penalty0 (1):\penalty0 1–29, 2020.
\newblock ISSN 1521-0626.
\newblock \doi{10.1080/10407790.2020.1746608}.
\newblock URL \url{http://dx.doi.org/10.1080/10407790.2020.1746608}.

\bibitem[Kopriva(1998)]{Kopriva_1998}
D.~A. Kopriva.
\newblock A staggered-grid multidomain spectral method for the compressible navier–stokes equations.
\newblock \emph{Journal of Computational Physics}, 143\penalty0 (1):\penalty0 125–158, 1998.
\newblock ISSN 0021-9991.
\newblock \doi{10.1006/jcph.1998.5956}.
\newblock URL \url{http://dx.doi.org/10.1006/jcph.1998.5956}.

\bibitem[Kopriva(2009)]{Kopriva_2009}
D.~A. Kopriva.
\newblock \emph{Implementing spectral methods for partial differential equations: algorithms for scientists and engineers}.
\newblock Springer Netherlands, 2009.
\newblock ISBN 9789048122615.
\newblock \doi{10.1007/978-90-481-2261-5}.
\newblock URL \url{http://dx.doi.org/10.1007/978-90-481-2261-5}.

\bibitem[Lenormand et~al.(2000)Lenormand, Sagaut, and Ta~Phuoc]{Lenormand_2000}
E.~Lenormand, P.~Sagaut, and L.~Ta~Phuoc.
\newblock Large eddy simulation of subsonic and supersonic channel flow at moderate reynolds number.
\newblock \emph{International Journal for Numerical Methods in Fluids}, 32\penalty0 (4):\penalty0 369–406, February 2000.
\newblock ISSN 1097-0363.
\newblock \doi{10.1002/(sici)1097-0363(20000229)32:4<369::aid-fld943>3.0.co;2-6}.
\newblock URL \url{http://dx.doi.org/10.1002/(sici)1097-0363(20000229)32:4<369::aid-fld943>3.0.co;2-6}.

\bibitem[Li et~al.(2021{\natexlab{a}})Li, Komperda, Peyvan, Ghiasi, and Mashayek]{Li_2021}
D.~Li, J.~Komperda, A.~Peyvan, Z.~Ghiasi, and F.~Mashayek.
\newblock Assessment of turbulence models using dns data of compressible plane free shear layer flow.
\newblock \emph{Journal of Fluid Mechanics}, 931, November 2021{\natexlab{a}}.
\newblock ISSN 1469-7645.
\newblock \doi{10.1017/jfm.2021.919}.
\newblock URL \url{http://dx.doi.org/10.1017/jfm.2021.919}.

\bibitem[Li et~al.(2021{\natexlab{b}})Li, Peyvan, Ghiasi, Komperda, and Mashayek]{Li_2021b}
D.~Li, A.~Peyvan, Z.~Ghiasi, J.~Komperda, and F.~Mashayek.
\newblock Compressibility effects on energy exchange mechanisms in a spatially developing plane free shear layer.
\newblock \emph{Journal of Fluid Mechanics}, 910, January 2021{\natexlab{b}}.
\newblock ISSN 1469-7645.
\newblock \doi{10.1017/jfm.2020.932}.
\newblock URL \url{http://dx.doi.org/10.1017/jfm.2020.932}.

\bibitem[Liu et~al.(2024)Liu, Liu, and Zhao]{Liu_2024}
J.~Liu, Y.~Liu, and L.~Zhao.
\newblock Analysis of the staggered dg method for the quasi-newtonian stokes flows.
\newblock \emph{Journal of Scientific Computing}, 102\penalty0 (1), November 2024.
\newblock ISSN 1573-7691.
\newblock \doi{10.1007/s10915-024-02741-9}.
\newblock URL \url{http://dx.doi.org/10.1007/s10915-024-02741-9}.

\bibitem[Ma(1998)]{Ma_1998}
H.~Ma.
\newblock Chebyshev--legendre super spectral viscosity method for nonlinear conservation laws.
\newblock \emph{SIAM Journal on Numerical Analysis}, 35\penalty0 (3):\penalty0 893–908, 1998.
\newblock ISSN 1095-7170.
\newblock \doi{10.1137/s0036142995293912}.
\newblock URL \url{http://dx.doi.org/10.1137/s0036142995293912}.

\bibitem[Maday et~al.(1993)Maday, Kaber, and Tadmor]{Maday_1993}
Y.~Maday, S.~M.~Ould Kaber, and E.~Tadmor.
\newblock Legendre pseudospectral viscosity method for nonlinear conservation laws.
\newblock \emph{SIAM Journal on Numerical Analysis}, 30\penalty0 (2):\penalty0 321–342, 1993.
\newblock ISSN 1095-7170.
\newblock \doi{10.1137/0730016}.
\newblock URL \url{http://dx.doi.org/10.1137/0730016}.

\bibitem[Manrique~de Lara and Ferrer(2023)]{Manrique_de_Lara_2023}
F.~Manrique~de Lara and E.~Ferrer.
\newblock Accelerating high order discontinuous galerkin solvers using neural networks: 3d compressible navier-stokes equations.
\newblock \emph{Journal of Computational Physics}, 489:\penalty0 112253, 2023.
\newblock ISSN 0021-9991.
\newblock \doi{10.1016/j.jcp.2023.112253}.
\newblock URL \url{http://dx.doi.org/10.1016/j.jcp.2023.112253}.

\bibitem[Manzanero et~al.(2017)Manzanero, Rubio, Ferrer, Valero, and Kopriva]{Manzanero_2017}
J.~Manzanero, G.~Rubio, E.~Ferrer, E.~Valero, and D.~A. Kopriva.
\newblock Insights on aliasing driven instabilities for advection equations with application to gauss–lobatto discontinuous galerkin methods.
\newblock \emph{Journal of Scientific Computing}, 75\penalty0 (3):\penalty0 1262–1281, 2017.
\newblock ISSN 1573-7691.
\newblock \doi{10.1007/s10915-017-0585-6}.
\newblock URL \url{http://dx.doi.org/10.1007/s10915-017-0585-6}.

\bibitem[Manzanero et~al.(2020{\natexlab{a}})Manzanero, Ferrer, Rubio, and Valero]{Manzanero_2020}
J.~Manzanero, E.~Ferrer, G.~Rubio, and E.~Valero.
\newblock Design of a smagorinsky spectral vanishing viscosity turbulence model for discontinuous galerkin methods.
\newblock \emph{Computers \& Fluids}, 200:\penalty0 104440, 2020{\natexlab{a}}.
\newblock ISSN 0045-7930.
\newblock \doi{10.1016/j.compfluid.2020.104440}.
\newblock URL \url{http://dx.doi.org/10.1016/j.compfluid.2020.104440}.

\bibitem[Manzanero et~al.(2020{\natexlab{b}})Manzanero, Rubio, Kopriva, Ferrer, and Valero]{Manzanero_2020_b}
J.~Manzanero, G.~Rubio, D.~A. Kopriva, E.~Ferrer, and E.~Valero.
\newblock Entropy–stable discontinuous galerkin approximation with summation–by–parts property for the incompressible navier–stokes/cahn–hilliard system.
\newblock \emph{Journal of Computational Physics}, 408:\penalty0 109363, 2020{\natexlab{b}}.
\newblock ISSN 0021-9991.
\newblock \doi{10.1016/j.jcp.2020.109363}.
\newblock URL \url{http://dx.doi.org/10.1016/j.jcp.2020.109363}.

\bibitem[Mateo-Gabín et~al.(2022)Mateo-Gabín, Manzanero, and Valero]{Mateo_Gab_n_2022}
A.~Mateo-Gabín, J.~Manzanero, and E.~Valero.
\newblock An entropy stable spectral vanishing viscosity for discontinuous galerkin schemes: Application to shock capturing and les models.
\newblock \emph{Journal of Computational Physics}, 471:\penalty0 111618, December 2022.
\newblock ISSN 0021-9991.
\newblock \doi{10.1016/j.jcp.2022.111618}.
\newblock URL \url{http://dx.doi.org/10.1016/j.jcp.2022.111618}.

\bibitem[Moser et~al.(1999)Moser, Kim, and Mansour]{Moser1999}
R.~D. Moser, J.~Kim, and N.~N. Mansour.
\newblock Direct numerical simulation of turbulent channel flow up to $re_\tau = 590$.
\newblock \emph{Physics of Fluids}, 11\penalty0 (4):\penalty0 943--945, 04 1999.
\newblock \doi{10.1063/1.869966}.
\newblock URL \url{https://doi.org/10.1063/1.869966}.

\bibitem[Mostafavi et~al.(2024{\natexlab{a}})Mostafavi, Ranjbar, Yurkiv, Yarin, and Mashayek]{mostafavi2024mass}
A.~Mostafavi, M.~Ranjbar, V.~Yurkiv, A.~Yarin, and F.~Mashayek.
\newblock Mass-conserving phase-field numerical simulation of surfactant-induced two-phase flows.
\newblock \emph{Bulletin of the American Physical Society}, 2024{\natexlab{a}}.
\newblock URL \url{https://ui.adsabs.harvard.edu/abs/2024APS..DFDA10002M}.

\bibitem[Mostafavi et~al.(2024{\natexlab{b}})Mostafavi, Ranjbar, Yurkiv, Yarin, and Mashayek]{Mostafavi_2024}
A.~Mostafavi, M.~Ranjbar, V.~R. Yurkiv, A.~L. Yarin, and F.~Mashayek.
\newblock Moose-based finite element framework for mass-conserving two-phase flow simulations on adaptive grids using the diffuse interface approach and a lagrange multiplier.
\newblock \emph{Journal of Computational Physics}, 2024{\natexlab{b}}.
\newblock \doi{https://doi.org/10.1016/j.jcp.2025.113755}.
\newblock URL \url{https://www.sciencedirect.com/science/article/pii/S0021999125000385}.

\bibitem[Mostafavi et~al.(2025{\natexlab{a}})Mostafavi, Ranjbar, Yurkiv, Yarin, and Mashayek]{Mostafavi_2025}
A.~Mostafavi, M.~Ranjbar, V.~R. Yurkiv, A.~L. Yarin, and F.~Mashayek.
\newblock On the energy analysis of two-phase flows simulated with the diffuse interface method.
\newblock \emph{Physics of Fluids}, 37\penalty0 (7), July 2025{\natexlab{a}}.
\newblock ISSN 1089-7666.
\newblock \doi{10.1063/5.0276045}.
\newblock URL \url{http://dx.doi.org/10.1063/5.0276045}.

\bibitem[Mostafavi et~al.(2025{\natexlab{b}})Mostafavi, Saidi, and Moghtaderi]{MOSTAFAVI2025sh}
A.~Mostafavi, M.~S. Saidi, and M.~Moghtaderi.
\newblock Three-dimensional simulation of circulating tumor cells magnetic isolation using a viscoelastic-based ferrofluid solution.
\newblock \emph{Journal of Magnetism and Magnetic Materials}, 619:\penalty0 172847, 2025{\natexlab{b}}.
\newblock ISSN 0304-8853.
\newblock URL \url{https://www.sciencedirect.com/science/article/pii/S0304885325000782}.

\bibitem[Moura et~al.(2015)Moura, Sherwin, and Peiró]{Moura_2015}
R.~C. Moura, S.~J. Sherwin, and J.~Peiró.
\newblock Linear dispersion–diffusion analysis and its application to under-resolved turbulence simulations using discontinuous galerkin spectral/hp methods.
\newblock \emph{Journal of Computational Physics}, 298:\penalty0 695–710, 2015.
\newblock ISSN 0021-9991.
\newblock \doi{10.1016/j.jcp.2015.06.020}.
\newblock URL \url{http://dx.doi.org/10.1016/j.jcp.2015.06.020}.

\bibitem[Moura et~al.(2016)Moura, Sherwin, and Peiró]{Moura_2016}
R.~C. Moura, S.~J. Sherwin, and J.~Peiró.
\newblock Eigensolution analysis of spectral/hp continuous galerkin approximations to advection–diffusion problems: Insights into spectral vanishing viscosity.
\newblock \emph{Journal of Computational Physics}, 307:\penalty0 401–422, 2016.
\newblock ISSN 0021-9991.
\newblock \doi{10.1016/j.jcp.2015.12.009}.
\newblock URL \url{http://dx.doi.org/10.1016/j.jcp.2015.12.009}.

\bibitem[Okamoto et~al.(2025)Okamoto, Ishihara, Yokokawa, and Kaneda]{Okamoto_2025}
N.~Okamoto, T.~Ishihara, M.~Yokokawa, and Y.~Kaneda.
\newblock Effects of finite arithmetic precision on large-scale direct numerical simulation of box turbulence by spectral method.
\newblock \emph{Phys. Rev. Fluids}, 10:\penalty0 064603, Jun 2025.
\newblock \doi{10.1103/PhysRevFluids.10.064603}.
\newblock URL \url{https://link.aps.org/doi/10.1103/PhysRevFluids.10.064603}.

\bibitem[Orlando(2024)]{Orlando_2024}
G.~Orlando.
\newblock An implicit dg solver for incompressible two‐phase flows with an artificial compressibility formulation.
\newblock \emph{International Journal for Numerical Methods in Fluids}, 96\penalty0 (12):\penalty0 1932–1959, 2024.
\newblock ISSN 1097-0363.
\newblock \doi{10.1002/fld.5328}.
\newblock URL \url{http://dx.doi.org/10.1002/fld.5328}.

\bibitem[Pang et~al.(2021)Pang, Yang, Gao, and Chen]{PANG2021}
C.~Pang, H.~Yang, Z.~Gao, and S.~Chen.
\newblock Enhanced adaptive mesh refinement method using advanced vortex identification sensors in wake flow.
\newblock \emph{Aerospace Science and Technology}, 115:\penalty0 106796, 2021.
\newblock ISSN 1270-9638.
\newblock \doi{https://doi.org/10.1016/j.ast.2021.106796}.
\newblock URL \url{https://www.sciencedirect.com/science/article/pii/S1270963821003060}.

\bibitem[Pasquetti(2005)]{PASQUETTI_2005}
R.~Pasquetti.
\newblock Spectral vanishing viscosity method for les: sensitivity to the svv control parameters.
\newblock \emph{Journal of Turbulence}, 6:\penalty0 N12, 2005.
\newblock ISSN 1468-5248.
\newblock \doi{10.1080/14685240500125476}.
\newblock URL \url{http://dx.doi.org/10.1080/14685240500125476}.

\bibitem[Pasquetti(2006)]{Pasquetti_2006}
R.~Pasquetti.
\newblock Spectral vanishing viscosity method for large-eddy simulation of turbulent flows.
\newblock \emph{Journal of Scientific Computing}, 27\penalty0 (1–3):\penalty0 365–375, 2006.
\newblock ISSN 1573-7691.
\newblock \doi{10.1007/s10915-005-9029-9}.
\newblock URL \url{http://dx.doi.org/10.1007/s10915-005-9029-9}.

\bibitem[Pasquetti et~al.(2007)Pasquetti, Séverac, Serre, Bontoux, and Schäfer]{Pasquetti_2007}
R.~Pasquetti, E.~Séverac, E.~Serre, P.~Bontoux, and M.~Schäfer.
\newblock From stratified wakes to rotor–stator flows by an svv–les method.
\newblock \emph{Theoretical and Computational Fluid Dynamics}, 22\penalty0 (3–4):\penalty0 261–273, 2007.
\newblock ISSN 1432-2250.
\newblock \doi{10.1007/s00162-007-0070-1}.
\newblock URL \url{http://dx.doi.org/10.1007/s00162-007-0070-1}.

\bibitem[Peyvan et~al.(2021)Peyvan, Komperda, Li, Ghiasi, and Mashayek]{PEYVAN2021}
A.~Peyvan, J.~Komperda, D.~Li, Z.~Ghiasi, and F.~Mashayek.
\newblock Flux reconstruction using jacobi correction functions in discontinuous spectral element method.
\newblock \emph{Journal of Computational Physics}, 435:\penalty0 110261, 2021.
\newblock ISSN 0021-9991.
\newblock \doi{https://doi.org/10.1016/j.jcp.2021.110261}.
\newblock URL \url{https://www.sciencedirect.com/science/article/pii/S002199912100156X}.

\bibitem[Pochet et~al.(2013)Pochet, Hillewaert, Geuzaine, Remacle, and Marchandise]{Pochet_2013}
F.~Pochet, K.~Hillewaert, P.~Geuzaine, J.-F. Remacle, and É. Marchandise.
\newblock A 3d strongly coupled implicit discontinuous galerkin level set-based method for modeling two-phase flows.
\newblock \emph{Computers \& Fluids}, 87:\penalty0 144–155, 2013.
\newblock ISSN 0045-7930.
\newblock \doi{10.1016/j.compfluid.2013.04.010}.
\newblock URL \url{http://dx.doi.org/10.1016/j.compfluid.2013.04.010}.

\bibitem[Qi et~al.(2024)Qi, Wang, Zhu, Tian, and Zhao]{Qi_2024}
X.~Qi, Z.~Wang, J.~Zhu, L.~Tian, and N.~Zhao.
\newblock High-order discontinuous galerkin method with immersed boundary treatment for compressible flows on parallel adaptive cartesian grids.
\newblock \emph{Physics of Fluids}, 36\penalty0 (11), November 2024.
\newblock ISSN 1089-7666.
\newblock \doi{10.1063/5.0238605}.
\newblock URL \url{http://dx.doi.org/10.1063/5.0238605}.

\bibitem[Quaegebeur and Nadarajah(2019)]{Quaegebeur_2019}
S.~Quaegebeur and S.~Nadarajah.
\newblock Stability of energy stable flux reconstruction for the diffusion problem using the interior penalty and bassi and rebay ii numerical fluxes for linear triangular elements.
\newblock \emph{Journal of Computational Physics}, 380:\penalty0 88–118, March 2019.
\newblock ISSN 0021-9991.
\newblock \doi{10.1016/j.jcp.2018.12.017}.
\newblock URL \url{http://dx.doi.org/10.1016/j.jcp.2018.12.017}.

\bibitem[Ranjbar et~al.(2023)Ranjbar, Belknap~Fernandez, Mashayek, and Komperda]{Ranjbar_2023}
M.~Ranjbar, D.~A. Belknap~Fernandez, F.~Mashayek, and J.~Komperda.
\newblock Application of modal representation for instantaneous statistical analysis of flows simulated with spectral element method.
\newblock In \emph{AIAA AVIATION 2023 Forum}. American Institute of Aeronautics and Astronautics, 2023.
\newblock \doi{10.2514/6.2023-3688}.
\newblock URL \url{http://dx.doi.org/10.2514/6.2023-3688}.

\bibitem[Ranjbar et~al.(2024{\natexlab{a}})Ranjbar, Komperda, and Mashayek]{Ranjbar_2024AIAA}
M.~Ranjbar, J.~Komperda, and F.~Mashayek.
\newblock Dynamic explicit modal filtering for large-eddy simulation of turbulent flows with spectral element method.
\newblock In \emph{AIAA SCITECH 2024 Forum}. American Institute of Aeronautics and Astronautics, 2024{\natexlab{a}}.
\newblock \doi{10.2514/6.2024-0295}.
\newblock URL \url{http://dx.doi.org/10.2514/6.2024-0295}.

\bibitem[Ranjbar et~al.(2024{\natexlab{b}})Ranjbar, Mostafavi, Komperda, and Mashayek]{Ranjbar2024APS}
M.~Ranjbar, A.~Mostafavi, J.~Komperda, and F.~Mashayek.
\newblock Dynamic modal filtering for turbulence modeling in spectral element method.
\newblock \emph{Bulletin of the American Physical Society}, 2024{\natexlab{b}}.
\newblock URL \url{https://meetings.aps.org/Meeting/ DFD24/Session/T39.7}.

\bibitem[Ranjbar et~al.(2024{\natexlab{c}})Ranjbar, Mostafavi, Rajendran, Komperda, and Mashayek]{Ranjbar_2024POF}
M.~Ranjbar, A.~Mostafavi, P.~T. Rajendran, J.~Komperda, and F.~Mashayek.
\newblock Modal analysis of turbulent flows simulated with spectral element method.
\newblock \emph{Physics of Fluids}, 36\penalty0 (12), 2024{\natexlab{c}}.
\newblock ISSN 1089-7666.
\newblock \doi{10.1063/5.0234014}.
\newblock URL \url{http://dx.doi.org/10.1063/5.0234014}.

\bibitem[Severac and Serre(2007)]{Severac_2007}
E.~Severac and E.~Serre.
\newblock A spectral vanishing viscosity for the les of turbulent flows within rotating cavities.
\newblock \emph{Journal of Computational Physics}, 226\penalty0 (2):\penalty0 1234–1255, 2007.
\newblock ISSN 0021-9991.
\newblock \doi{10.1016/j.jcp.2007.05.023}.
\newblock URL \url{http://dx.doi.org/10.1016/j.jcp.2007.05.023}.

\bibitem[Shahbazi et~al.(2007)Shahbazi, Fischer, and Ethier]{Shahbazi_2007}
K.~Shahbazi, P.~F. Fischer, and C.~R. Ethier.
\newblock A high-order discontinuous galerkin method for the unsteady incompressible navier–stokes equations.
\newblock \emph{Journal of Computational Physics}, 222\penalty0 (1):\penalty0 391–407, 2007.
\newblock ISSN 0021-9991.
\newblock \doi{10.1016/j.jcp.2006.07.029}.
\newblock URL \url{http://dx.doi.org/10.1016/j.jcp.2006.07.029}.

\bibitem[Skrbek and Stalp(2000)]{skrbek2000}
L.~Skrbek and S.~R. Stalp.
\newblock On the decay of homogeneous isotropic turbulence.
\newblock \emph{Physics of fluids}, 12\penalty0 (8):\penalty0 1997--2019, 2000.

\bibitem[Spalart et~al.(1991)Spalart, Moser, and Rogers]{SPALART1991}
P.~R. Spalart, R.~D. Moser, and M.~M. Rogers.
\newblock Spectral methods for the navier-stokes equations with one infinite and two periodic directions.
\newblock \emph{Journal of Computational Physics}, 96\penalty0 (2):\penalty0 297--324, 1991.
\newblock ISSN 0021-9991.
\newblock \doi{https://doi.org/10.1016/0021-9991(91)90238-G}.
\newblock URL \url{https://www.sciencedirect.com/science/article/pii/002199919190238G}.

\bibitem[Tadmor(1989)]{Tadmor_1989}
E.~Tadmor.
\newblock Convergence of spectral methods for nonlinear conservation laws.
\newblock \emph{SIAM Journal on Numerical Analysis}, 26\penalty0 (1):\penalty0 30–44, 1989.
\newblock ISSN 1095-7170.
\newblock \doi{10.1137/0726003}.
\newblock URL \url{http://dx.doi.org/10.1137/0726003}.

\bibitem[Tadmor(1994)]{Tadmor_1994}
E.~Tadmor.
\newblock \emph{Super-viscosity and spectral approximations of nonlinear conservation laws}, page 69–82.
\newblock Oxford University PressOxford, 1994.
\newblock ISBN 9781383026016.
\newblock \doi{10.1093/oso/9780198536963.003.0005}.
\newblock URL \url{http://dx.doi.org/10.1093/oso/9780198536963.003.0005}.

\bibitem[Taylor and Green(1937)]{taylor_1937}
G.~I. Taylor and A.~E. Green.
\newblock Mechanism of the production of small eddies from large ones.
\newblock \emph{Proceedings of the Royal Society of London. Series A-Mathematical and Physical Sciences}, 158\penalty0 (895):\penalty0 499--521, 1937.
\newblock ISSN 2053-9169.
\newblock \doi{10.1098/rspa.1937.0036}.
\newblock URL \url{http://dx.doi.org/10.1098/rspa.1937.0036}.

\bibitem[Tikhomirov(1991)]{Tikhomirov_1991}
V.~M. Tikhomirov.
\newblock \emph{Local structure Of turbulence in an incompressible viscous fluid at very large Reynolds numbers}, page 312–318.
\newblock Springer Netherlands, 1991.
\newblock ISBN 9789401130301.
\newblock \doi{10.1007/978-94-011-3030-1_45}.
\newblock URL \url{http://dx.doi.org/10.1007/978-94-011-3030-1_45}.

\bibitem[Tlales et~al.(2024)Tlales, Otmani, Ntoukas, Rubio, and Ferrer]{Tlales_2024}
K.~Tlales, K.-E. Otmani, G.~Ntoukas, G.~Rubio, and E.~Ferrer.
\newblock Machine learning mesh-adaptation for laminar and turbulent flows: applications to high-order discontinuous galerkin solvers.
\newblock \emph{Engineering with Computers}, 40\penalty0 (5):\penalty0 2947–2969, 2024.
\newblock ISSN 1435-5663.
\newblock \doi{10.1007/s00366-024-01950-y}.
\newblock URL \url{http://dx.doi.org/10.1007/s00366-024-01950-y}.

\bibitem[Wang et~al.(2013)Wang, Fidkowski, Abgrall, Bassi, Caraeni, Cary, Deconinck, Hartmann, Hillewaert, Huynh, Kroll, May, Persson, van Leer, and Visbal]{Wang_2013}
Z.~J. Wang, K.~Fidkowski, R.~Abgrall, F.~Bassi, D.~Caraeni, A.~Cary, H.~Deconinck, R.~Hartmann, K.~Hillewaert, H.~T. Huynh, N.~Kroll, G.~May, P.‐O. Persson, B.~van Leer, and M.~Visbal.
\newblock High‐order cfd methods: current status and perspective.
\newblock \emph{International Journal for Numerical Methods in Fluids}, 72\penalty0 (8):\penalty0 811–845, 2013.
\newblock ISSN 1097-0363.
\newblock \doi{10.1002/fld.3767}.
\newblock URL \url{http://dx.doi.org/10.1002/fld.3767}.

\bibitem[Williamson(1980)]{Williamson_1980}
J.~H. Williamson.
\newblock Low-storage runge-kutta schemes.
\newblock \emph{Journal of Computational Physics}, 35\penalty0 (1):\penalty0 48–56, 1980.
\newblock ISSN 0021-9991.
\newblock \doi{10.1016/0021-9991(80)90033-9}.
\newblock URL \url{http://dx.doi.org/10.1016/0021-9991(80)90033-9}.

\bibitem[Winters et~al.(2018)Winters, Moura, Mengaldo, Gassner, Walch, Peiro, and Sherwin]{Winters_2018}
A.~R. Winters, R.~C. Moura, G.~Mengaldo, G.~J. Gassner, S.~Walch, J.~Peiro, and S.~J. Sherwin.
\newblock A comparative study on polynomial dealiasing and split form discontinuous galerkin schemes for under-resolved turbulence computations.
\newblock \emph{Journal of Computational Physics}, 372:\penalty0 1–21, 2018.
\newblock ISSN 0021-9991.
\newblock \doi{10.1016/j.jcp.2018.06.016}.
\newblock URL \url{http://dx.doi.org/10.1016/j.jcp.2018.06.016}.

\bibitem[Xi and Ji(2025)]{XI2025}
Y.~Xi and X.~Ji.
\newblock A finite element contour integral method for computing the scattering resonances of fluid-solid interaction problem.
\newblock \emph{Journal of Computational Physics}, 521:\penalty0 113539, 2025.
\newblock ISSN 0021-9991.
\newblock URL \url{https://www.sciencedirect.com/science/article/pii/S0021999124007873}.

\bibitem[Zhang and Liu(2022)]{Zhang_2022}
F.~Zhang and T.~Liu.
\newblock A high-order direct discontinuous galerkin mthod for variable density incompressible flows.
\newblock \emph{Communications in Computational Physics}, 32\penalty0 (3):\penalty0 850–877, June 2022.
\newblock ISSN 1991-7120.
\newblock \doi{10.4208/cicp.oa-2022-0064}.
\newblock URL \url{http://dx.doi.org/10.4208/cicp.oa-2022-0064}.

\bibitem[Zhang and Peet(2023)]{Zhang_2023}
F.~Zhang and Y.~T. Peet.
\newblock Discontinuous galerkin spectral element method for shock capturing with summation by parts properties.
\newblock \emph{Journal of Computational Physics: X}, 17:\penalty0 100123, November 2023.
\newblock ISSN 2590-0552.
\newblock \doi{10.1016/j.jcpx.2023.100123}.
\newblock URL \url{http://dx.doi.org/10.1016/j.jcpx.2023.100123}.

\bibitem[Zhou et~al.(2015)Zhou, Nagata, Sakai, Ito, and Hayase]{Zhou_2015}
Y.~Zhou, K.~Nagata, Y.~Sakai, Y.~Ito, and T.~Hayase.
\newblock On the evolution of the invariants of the velocity gradient tensor in single-square-grid-generated turbulence.
\newblock \emph{Physics of Fluids}, 27\penalty0 (7), July 2015.
\newblock ISSN 1089-7666.
\newblock \doi{10.1063/1.4926472}.
\newblock URL \url{http://dx.doi.org/10.1063/1.4926472}.

\end{thebibliography}
